\documentclass[nofootinbib,twocolumn,amsmath,amssymb,aps,prd,preprintnumbers,superscriptaddress]{revtex4-2}

\usepackage{graphicx}% Include figure files
\usepackage{dcolumn}% Align table columns on decimal point
\usepackage{bm}% bold math
\usepackage{multirow}
\usepackage{tabularx}
\usepackage{subfig}
\usepackage{verbatim}
\usepackage{appendix}

\usepackage[T1]{fontenc} % if needed
\usepackage{color}

\newcommand{\aap}{Astron. Astropart. Phys.}
\newcommand{\apss }{Astrophys. Sp. Sc.}
\newcommand{\apjl}{Astrophys. J. Lett.}

\newcommand{\jcap}{J. Cosmol. Astropart. Phys.}
\newcommand{\mnras}{Mon. Not. R. Astron. Soc.}
\newcommand{\pasa}{Publ.  Astr. Soc. Au.}
\newcommand{\til}{~}

\begin{document}

	\title{Exploiting the Einstein Telescope to solve the Hubble tension}

	\author{Matteo Califano}
	\email{matteo.califano@unina.it}
	\affiliation{Scuola Superiore Meridionale, Largo San Marcellino 10, I-80138, Napoli, Italy}
	\affiliation{INFN Sezione di Napoli, Compl. Univ. di
		Monte S. Angelo, Edificio G, Via Cinthia, I-80126, Napoli, Italy}
	
	\author{Ivan de Martino}
	\email{ivan.demartino@usal.es}
	\affiliation{Universidad de Salamanca, Departamento de Fisica Fundamental, P. de la Merced S/N, Salamanca, ES}
	
	\author{Daniele Vernieri}
	\email{daniele.vernieri@unina.it}
	\affiliation{Dipartimento di Fisica, Universit\`a di Napoli ``Federico II'', Compl. Univ. di Monte S. Angelo, Edificio G, Via Cinthia, I-80126, Napoli, Italy}
	\affiliation{Scuola Superiore Meridionale, Largo San Marcellino 10, I-80138, Napoli, Italy}
	\affiliation{INFN Sezione di Napoli, Compl. Univ. di
		Monte S. Angelo, Edificio G, Via Cinthia, I-80126, Napoli, Italy}
	
	\author{Salvatore Capozziello}
	\email{capozziello@unina.it}
	\affiliation{Dipartimento di Fisica, Universit\`a
		di Napoli ``Federico II'', Compl. Univ. di
		Monte S. Angelo, Edificio G, Via Cinthia, I-80126, Napoli, Italy}
	\affiliation{Scuola Superiore Meridionale, Largo San Marcellino 10, I-80138, Napoli, Italy}
	\affiliation{INFN Sezione di Napoli, Compl. Univ. di
		Monte S. Angelo, Edificio G, Via Cinthia, I-80126, Napoli, Italy}
	
	\date{\today}

	\begin{abstract}
		We probe four cosmological models which, potentially, can solve the Hubble tension according to  the dark energy equation of state. In this context, we demonstrate that the Einstein Telescope is capable of achieving a relative accuracy below $1\%$ on the Hubble constant independently of the specific dark energy model. We  firstly build  mock catalogs containing gravitational wave events for one, five and ten years of observations, and above Signal-to-Noise Ratio equal to nine. From these catalogs, we extract the events which are most likely associated with possible electromagnetic counterpart detected by THESEUS. Finally, we select four dark energy models, namely a non-flat $\omega$CDM, an interacting dark energy, an emergent dark energy, and a time varying gravitational constant model, to forecast the precision down to which the Einstein Telescope can bound the corresponding cosmological parameters. We foresee that the Hubble constant is always constrained with less than $1\%$ uncertainty, thereby offering a potential solution to the Hubble tension. The accuracy on the other cosmological parameters is at most comparable with the one currently obtained using multiple probes, except for the emergent dark energy model for which the Einstein Telescope alone will be able to improve the current limits by more than one order of magnitude.
	\end{abstract}

	\preprint{ET-0188A-22}
	\maketitle

	\section{Introduction}\label{sec:intro}
	The detection of Gravitational Waves (GWs) from the coalescence of merging Binary Black Holes (BBH) and Binary Neutron Stars (BNS)\til\cite{Abbott2016,GW170817} opened a new window to test the General Relativity, relativistic astrophysics, and cosmology\til\cite{GW170817,Abbott2016b,Ezquiaga2017}. As it is well known, the GWs bring direct information on the luminosity distance of  sources and, therefore, they can be used as rulers to measure distances in the Universe. Indeed, they are usually called {\em standard sirens}\til\cite{Schutz1986,Holz2005}, and are fully complementary to the  {\em standard candles},  such as Cepheids and Supernovae Type Ia (SNeIa) among the others, that are instead based on the detection of their electromagnetic emission and need to be calibrated on closer sources in order to get a measure of their luminosity distance.
	Although GWs offer an alternative method to obtain distances in cosmology and are not affected by calibration problems, they are not free of issues. Indeed, the GWs waveform encodes both information on the systems, such as masses, spin, and inclination angle among others, and information on the cosmology such as the distance. Furthermore, they  encode information on a given theory of gravity and, potentially, can probe 
	it\til\cite{Bogdanos:2009tn,Oikonomou:2022xoq,Odintsov:2022cbm}.
	However, information on  masses and  redshift is completely degenerate, and the only way to break such degeneracy is to have prior information on the redshift from an electromagnetic counterpart.
	There are several ways to get  accurate information on the redshift. For instance, one can assign to the GWs source the redshift of the host galaxy\til\cite{Schutz1986,Holz2005,Chen2018} or, alternatively, looking at the electromagnetic emission following the GWs such as 
	short Gamma-Ray Burst (GRB)\til\cite{Capozziello2011,GW170817} and kilonovae\til\cite{GW170817}. However, the host galaxy can be accurately detected only up to redshift below one\til\cite{Gray2020}, and kilonovae will be detected up to redshift $z\sim 1$\til\cite{Chase2022}.  On the contrary, short GRBs may be detected using forthcoming satellites, such as Transient High Energy Sources and Early Universe Surveyor (THESEUS), up to redshift $z\sim 8$\til\cite{THESEUS:2017wvz,Amati2021,Stratta2022}. Such high redshift detections can allow also to deeply test the cosmological evolution\til\cite{Rosati2021,Tanvir2021,Dainotti2021,Dainotti2022}. Furthermore, a complementary avenue to obtain the redshift information is represented by  the observation of tidal deformation in BNS mergers\til\cite{messenger:Read,Chatterjee2021}. Indeed, it may supply redshift estimation with an accuracy ranging from  $8$\% to $40$ \% depending on the choice of the equation of state (EoS).
	
	Nowadays, the LIGO/Virgo/KAGRA collaborations have explored several ways to constrain  cosmological models from  GWs events.  A turning point was the event GW170817, {\em i.e.} the first merger of a BNS with the simultaneous detection of the GRB 170817A\til\cite{GW170817}, yielding to the first estimation of the Hubble constant with GWs, $H_0= 70_{-8}^{+12}$ km s$^{-1}$ Mpc$^{-1}$ at 68\% of confidence level\til\cite{LIGO_H0_2017}. Afterward, the LIGO/Virgo/KAGRA collaborations have explored the possibility to constrain $H_0$ by analyzing the population distribution of BBH mergers, and searching for the host galaxy identification\til\cite{LVK_H0_2021}.  However, these new measurements of the Hubble constant are in agreement with both the late-time and the early-time measurements of $H_0$ and, therefore, do not help to solve the so-called Hubble tension\til\cite{DiValentino2021,Abdalla2022,Krishnan2021,Colgain2022,Colgain2022b} {which may, however, be alleviated to a 2.1$\sigma$ tension if the errors on the Hubble constant were underestimated\til\cite{Lopez2022}. Other examples of solutions to the Hubble tension may rely on modifications of fundamental laws\til\cite{Capozziello:2020nyq,Spallicci:2021kye,Capozziello2023PDU}.} Nevertheless, the next generation of GW detectors, {\em e.g.} the Einstein Telescope (ET), can strongly improve the accuracy on the Hubble constant reducing it below 1\%\til\cite{Maggiore2020,Branchesi2023}, and promise to offer a solution to such a tension pointing out its correct value. Therefore, there is an important need to study also the theoretical framework related to the Hubble tension. Let us remember that the Hubble tension is  $4.2\sigma$ discrepancy between the measurements of  $H_0$ obtained by fitting the CMB power spectra\til\cite{Planck2020}, and using the {\em standard candles} such as Cepheids\til\cite{Riess2019} both in the framework of the {\em concordance} cosmological model, also known as $\Lambda$ Cold Dark  Matter ($\Lambda$CDM) model.  
	
	Since the nature of Dark Energy (DE) is still a puzzle (for a comprehensive review see Ref.\til\cite{Abdalla2022}), there are many attempts to explain the $H_0$ tension modifying the DE EoS\til\cite{Zhang2019,Belgacem:2019tbw,DiValentino2021,Abdalla2022,Jin2022b,Yu2021}, or the underlying theory of gravity\til\cite{Belgacem2019,Belgacem2019b,Abdalla2022,Ferreira2022}. Several analyses have been carried out to measure the Hubble constant and DE EoS with GWs  making use of different techniques such as the identification of the electromagnetic counterpart of GW sources\til\cite{Bachega2020,Caprini2016,Matos2022}, the cross-correlation between GW sources and galaxies\til\cite{Mukherjee2021MNRAS}, the statistical host identification techniques with galaxies\til\cite{Gray2020,Laghi2021,Muttoni2022,Zhu2022,zhu2022b}, the hierarchical inference without galaxy surveys\til\cite{Sturani2022,Ye:2021klk,Ding:2018zrk}, the study of lensed events\til\cite{Liu2019,Balaudo2022} and the analysis of NS EoS\til\cite{Delpozzo2017,ghosh2022,Jin2022a}. 
	
	{ It is worth noting that the $\Lambda$CDM model has its own issues such as, for instance, the well-known small-scale problem of the CDM paradigm and the extremely small value of the cosmological constant compared to the expectation from quantum field theory\til\cite{deMartino2020}. Therefore, alternative approaches have been considered to overcome both the lack of experimental detection of the dark ingredients and the small-scale issues of CDM as well. For instance, the modification of the underlying theory of gravity can explain the acceleration of the Universe by means of extra degrees of freedom arising from higher-order curvature invariants or extra scalar fields\til\cite{Sotiriou2010,Kase2019}. Alternatively, models of non-homogeneous universes or violations of the Copernican principle can also account for the accelerated expansion\til\cite{Dam2017, Colin2019}. Nevertheless, the debate on whether there is or not a real need to shift from the $\Lambda$CDM to a more complicated model is still under debate\til\cite{Pardo2020}.}
	
	Here, we will focus on a set of models which modify the DE EoS in view of solving the Hubble tension\til\cite{DiValentino2021,Abdalla2022}. For each model, we will predict the accuracy down to which ET will be able to detect departures from the $\Lambda$CDM offering in such a way a theoretical framework of DE capable of solving the Hubble tension. In Sec.\til\ref{sec:models},  we will briefly introduce the DE models. In Sec.\til\ref{sec:mockdata}, we will summarize the procedure adopted to build mock data that will mimic the ET observations of the luminosity distance. In Sec.\til\ref{sec:stats}, we will give  details of our statistical analysis, while in Sec.\til\ref{sec:results} we will show  results. Finally, in Sec.\til\ref{sec:conclusion} we will give our final discussion and conclusions.

	\section{Dark Energy models}\label{sec:models}
	
	We focus on four DE models which may help in solving the Hubble tension\til\cite{DiValentino2021,Abdalla2022,Dagostino2019,Capozziello:2021xjw}, and differ from each other in the way they affect the DE EoS leading to a modification of the luminosity distance.
	In the context of General Relativity, and of the Friedman--Lema\^{i}tre--Robertson--Walker (FLRW) cosmology,  the luminosity distance is defined as
	\begin{equation}\label{luminosty_distance}
		d_{L}(z) = 
		\begin{cases}
			\frac{c(1+z)}{H_0}\frac{1}{\sqrt{\Omega_k}}\sinh{\left[ \sqrt{\Omega_k} \int_{0}^{z}\frac{dz'}{E(z')}\right]}\quad &\mbox{for}\ \Omega_k >0,\\
			\frac{c(1+z)}{H_0} \int_{0}^{z}\frac{dz'}{E(z')}\qquad &\mbox{for}\ \Omega_k =0,\\
			\frac{c(1+z)}{H_0}\frac{1}{\sqrt{|\Omega_k|}}\sin{\left[ \sqrt{|\Omega_k|} \int_{0}^{z}\frac{dz'}{E(z')}\right]}\quad &\mbox{for}\ \Omega_k <0,
		\end{cases}
	\end{equation}
	where $z$ is the redshift, $c$ is the speed of light, $H_0$ is the Hubble constant, $\Omega_k,0$ is the value  of the curvature parameter at $z=0$, and 
	\begin{equation}\label{E_z}
		E^{2}(z)=\Omega_{m,0}(1+z)^{3} + \Omega_{k,0}(1+z)^{2} + \Omega_{DE}(z)\,,
	\end{equation}
	where $\Omega_{m,0}$ is the value  of the matter density parameter at $z=0$, and $\Omega_{DE}(z)$ is the DE density parameter as a function of the redshift. In the $\Lambda$CDM model, $\Omega_{DE}(z)=\Omega_{\Lambda,0}$.
	Since Eq.\til\eqref{E_z} is strictly related to the Friedman equations and to the DE EoS, changing the DE model leads to different expressions of the function $E(z)$\til\cite{Capozziello:2019cav}. To study the proprieties of  DE component in a general framework, one can consider the ratio of the DE pressure to its energy density as a function of redshift\til\cite{Chevallier2001,Linder2003}:
	\begin{equation}
		\omega_{DE} (z) = \frac{p_{DE}(z)}{\rho_{DE}(z)}\,.
	\end{equation}
	Again, $\Lambda$CDM model is recovered for $\omega_{DE} (z)=-1$.
	In the next subsections, we will discuss four DE models whose modifications of the Eq.\til\eqref{E_z} may serve to solve the Hubble tension. Moreover, we will also discuss the limits in which such models recover the $\Lambda$CDM cosmology.

	\subsection{Non-flat $\omega$CDM}\label{subsec:omCDM}
	
	We focus on the simplest extension of the $\Lambda$CDM model in which $\omega_{DE} (z)$ is a constant, but it can assume values different from $\omega_{DE} = -1$, {\em i.e.} the cosmological constant. Hence, the modification to Eq.\til\eqref{E_z} appears as\til\cite{Copeland2006}
	\begin{equation}\label{eq:Ez}
		E^2 (z)  =  \Omega_{m,0}(1+z)^{3} + \Omega_{k,0}(1+z)^{2} + \Omega_{\Lambda,0}(1+z)^{3(1+\omega_{DE})}\,.
	\end{equation}
	
	In\til\cite{Gao:2021}, it was shown using the CMB + BAO + SN +$H_0$ dataset observations, the aforementioned model may solve the Hubble tension at 95\% Confidence Level (CL). The best-fit values are: $H_0 = 69.88_{-0.76}^{+0.77}$ km s$^{-1}$ Mpc$^{-1}$ and $\omega_{DE} = -1.08\pm 0.03$.
	In\til\cite{Belgacem:2019tbw}, generating a mock dataset of combined events  by ET and THESEUS, they obtained the following accuracy on the cosmological parameter $\omega_{DE}$:  $\sigma_{\omega_{DE}} =0.3$
	
	Since,  we are considering a non-flat $\omega$CDM model, we can recast $\Omega_{m,0}$  as $1- \Omega_{k,0}-\Omega_{\Lambda,0}$, and when $\omega_{DE}$ assumes values different from $-1$ the model departs from the  standard cosmological constant. For instance, the case $\omega_{DE} > -1$ is usually referred to as “quintessence”\til\cite{Copeland2006}, while the case with $\omega_{DE} < -1$ as “phantom”\til\cite{Bamba2012}.

	\subsection{Interacting Dark Energy}\label{subsec:Interacting Dark Energy}
	
	Another scenario capable of solving the Hubble tension considers that Dark Matter 
	(DM) and DE interact not only gravitationally. This is the so-called {\em Interacting Dark Energy} (IDE) model. 
	Following\til\cite{Valiviita2008,Gavela2009}, one can parameterize the interaction between DM and DE as follows 
	\begin{align}
		\nabla_{\mu} T^{(DM)\mu}_{\phantom{(DM)\mu}\nu} &= Q u^{(DM)}_{\nu}/a\,,\\
		\nabla_{\mu} T^{(DE)\mu}_{\phantom{(DE)\mu} \nu} &=- Q u^{(DM)}_{\nu}/a\,,
	\end{align}
	where $T^{(DM)\mu}_{\phantom{(DM)\mu}\nu}$ and $T^{(DE)\mu}_{\phantom{(DE)\mu} \nu}$  are the energy-momentum tensors for the DM and DE, respectively.  The coefficient $Q$ encodes the  coupling between the two dark components. Although different functional forms of $Q$ have been explored\til\cite{Gavela2009,Wang2016,Yang2019},  we select the following coupled model: $Q= \xi H(z) \rho_{DE}$; because a generic interaction coupling might have several instabilities, while the model we choose  can avoid them under some suitable conditions on $\xi$ and $\omega$\til\cite{Gavela2009,Wang2016}. Hence, the DM and DE background evolve with respect to the cosmic time as\til\cite{Gavela2009}
	\begin{align}
		\label{IDE_DM}    
		\dot{\rho}_{DM} +3H(z)\rho_{DM} &= \xi H(z)\rho_{DE},\\
		\label{IDE_DE}
		\dot{\rho}_{DE} +3H(z)\rho_{DE}(1 + \omega_{DE}) &= -\xi H(z)\rho_{DE}.
	\end{align}
	Since the DM density must be positive along the cosmic evolution if  $\omega_{DE} \ < 0$ and $\xi \ >0$, we need to impose the following condition: $\xi < - \omega_{DE}$\til\cite{Gavela2009}.
	
	Solving the Eqs.\til\eqref{IDE_DM} and \eqref{IDE_DE}, one can rewrite the Eq. \eqref{E_z} for the case of an IDE model as 
	\begin{equation}\label{eq:Ez:interacting}
		\begin{aligned}
			E^2 (z)  =  \Omega_{m,0}(1+z)^{3}  + \Omega_{\Lambda,0}\left[(1+z)^{3 \left(1+\omega_{DE}^{eff}\right)}\ \right.\\
			\left.+ \frac{\xi}{3\omega_{DE}^{eff}}\left(1-(1+z)^{3 \omega_{DE}^{eff}}\right)(1+z)^{3}\right]\,,
		\end{aligned}
	\end{equation}
	where $\omega_{DE}^{eff}= \omega_{DE} +\frac{\xi}{3}$. In order to avoid the early time instability, the quantities $(1 + \omega_{DE})$ and $\xi$ must have opposite sign\til\cite{Gavela2009}. It is worth  noticing that the $\Lambda$CDM is recovered by setting $\omega_{DE} = -1$ and $\xi = 0$.
	In our analysis, we will consider two cases: (i) $\omega_{DE}$ is fixed  to $-1$ and $\xi$ is a free parameter, (ii) $\omega_{DE}$ and $\xi$ are both free parameters.
	
	Using the CMB dataset, it has been shown, for case (i), that the IDE model is capable of solving the Hubble tension making early and late time measurements of $H_0$ agree at 68\% CL\til\cite{DiValentino2020a,Divalentino2020b,Pan2019}. The best-fit values are: $H_0 = 72.8_{-1.6}^{+3.0}$ km s$^{-1}$ Mpc$^{-1}$ and $\xi = -0.51_{-0.29}^{+0.12}$. In the case (ii), using CMB+Cepehids, the best-fit values are: $H_0 = 73.3_{-1.0}^{+1.2}$ km s$^{-1}$ Mpc$^{-1}$,  $\omega_{DE} = -0.95_{-0.05}^{+0.01}$ and $\xi = -0.73_{-0.10}^{+0.05}$.

	\subsection{Emergent Dark Energy}\label{subsec:Emergent Dark Energy}
	
	Another solution to the Hubble constant is that  DE contributes to the total energy density budget of the  Universe only at late time\til\cite{Li2019,Pan2020}. In such a case,  the Eq.\til\eqref{E_z} can be re-written as follows
	\begin{equation}\label{eq:Ez:emergent}
		E^2 (z)  =  \Omega_{m,0}(1+z)^{3}  + \tilde{\Omega}_{DE}(z)\,.
	\end{equation}
	
	In the simplest parameterization,  the DE evolves as\til\cite{Li2019}
	\begin{equation}
		\tilde{\Omega}_{DE}(z) =   \Omega_{\Lambda,0}\left[ 1- \tanh{\left( \log_{10}\left(1+z \right) \right) } \right]\,.
	\end{equation}
	In this parameterization, there are the same degrees of freedom of the $\Lambda $CDM model. Indeed, there is only one free parameter, namely $\Omega_{\Lambda,0}$. Despite this is not a sever modification of the parameter space, the statistical analysis of the temperature fluctuations of the CMB data data\til\cite{Planck2020} provides a higher value of the Hubble constant, $H_0 = 72.35_{-0.79}^{+0.78}\ \mbox{km}\ \mbox{s}^{-1}\ \mbox{Mpc}^{-1}$\til\cite{Yang2020}, with respect to the $\Lambda$CDM cosmology, which results to agree with the late-time measurements of $H_0$ at 68\% CL.
	
	We will focus on a generalization of the aforementioned model where the DE contribution arises at a specific transition redshift $z_t$. In such a model the DE critical density can be written as\til\cite{Li:2020}
	\begin{equation} \label{eq:EME}
		\tilde{\Omega}_{DE}(z) =   \Omega_{\Lambda,0}\left[ \frac{1- \tanh{\left(\Delta \log_{10}\left(\frac{1+z}{1+z_t} \right) \right)} }{1+ \tanh{\left(\Delta \log_{10}\left(1+z_t\right)\right)}} \right]\,,
	\end{equation}
	where $\Delta$ is a free parameter and $z_t$ is the epoch where the matter-energy density and the DE density are equal. More precisely,  $z_t$ is defined by the following equality:
	\begin{equation}
		\Omega_{m,0}(1+z_t)^{3}\ =\   \frac{\Omega_{\Lambda,0}}{1+ \tanh{\left(\Delta \log_{10}\left(1+z_t\right)\right)}} \ .
	\end{equation}
	In this case, there is only one extra free parameter, $\Delta$, which can discriminate between the $\Lambda$CDM model, which is recovered for $\Delta = 0$, and the emergent DE parameterization given in Eqs. \eqref{eq:Ez:emergent}  and \eqref{eq:EME}, {with  $\Delta \neq 0$}. Under the parameterization in Eq. \eqref{eq:EME}, it has been shown using CMB+BAO+Cepheids that the $H_0$ tension reduces to $1.8\ \sigma$ with the best-fit values of $H_0 = 71.0_{-1.3}^{+1.4}\ \mbox{km}\ \mbox{s}^{-1}\ \mbox{Mpc}^{-1}$ and  $\Delta = 0.85_{-0.41}^{+0.44} $\til\cite{Yang2021}.
	
	Finally, we focus on the second parameterization in Eq. \eqref{eq:EME}  because it admits a direct limit to the $\Lambda$CDM cosmological model and, therefore, allows us to predict the accuracy down to which departure from  the $\Lambda$CDM  model may be detected by future experiments.

	\subsection{Time-Varying Gravitational Constant}\label{subsec:time-Varying}
	
	Alternatively to the previous models, one can investigate the case in which gravitational coupling is a function of the redshift through some scalar field\til\cite{Mota:2011iw}. Starting from  an effective quantum theory of gravity that is asymptotically safe, one can obtain $G_N (z) = G_{N,0}(1 + z)^{-\delta_{G}}$\til\cite{Weinberg1976,Weinberg2010}. The term $G_{N,0}$ refers to the values of gravitational constant at $z=0$, and $\delta_{G}$ parameterizes its evolution with redshift. Indeed, $\delta_{G}=0$ means that the gravitation constant is the Newtonian one, and no evolution with redshift is considered.
	Since the gravitational coupling is no longer a constant, the cosmological constant would also be redshift-dependent: $\Lambda(z) = \Lambda_{0} (1 + z)^{\delta_{\Lambda}}$\til\cite{Xue2015}. The density of matter and of DE evolve according to the following equations:
	\begin{align}
		\left(\frac{G_N}{G_{N,0}}\right) \rho_{m} &= \rho_{m,0}(1+z)^{(3-\delta_{G})}\,,\\
		\left(\frac{G_N}{G_{N,0}}\right) \rho_{\Lambda} &= \rho_{\Lambda,0}(1+z)^{\delta_{\Lambda}}.
	\end{align}
	Therefore, the Eq. \eqref{E_z} can be recast in the following form
	\begin{equation}\label{eq:Ez:Gvar}
		E^2 (z)  =  \Omega_{m,0}(1+z)^{(3-\delta_{G})}  + \Omega_{\Lambda,0}(1+z)^{\delta_{\Lambda}}\,,
	\end{equation}
	The requirement for a flat Universe leads to the relation\til\cite{Xue2015}
	\begin{equation}\label{delta_relation}
		\delta_{\Lambda}=\delta_{G}\frac{\Omega_{m,0}}{\Omega_{\Lambda,0}}.
	\end{equation}
	Let us notice that the $\Lambda$CDM cosmology is recovered by setting $\delta_{G}=0$.
	Using the CMB + BAO + SN +$H_0$ dataset, it was shown that the model mitigates the tension in the Hubble constant reducing it at $2\sigma$\til\cite{Gao:2021}. In their analysis, the best-fit values are: $H_0 = 70.69_{-1.08}^{+1.06}\ \mbox{km}\ \mbox{s}^{-1}\ \mbox{Mpc}^{-1}$ and  $\delta_{G} = -0.0062_{-0.0023}^{+0.0025} $.
	
	\subsection{Comparing the dark energy models with the $\Lambda$CDM cosmology}
\begin{figure}
		\centering
		\includegraphics[width=0.48\textwidth]{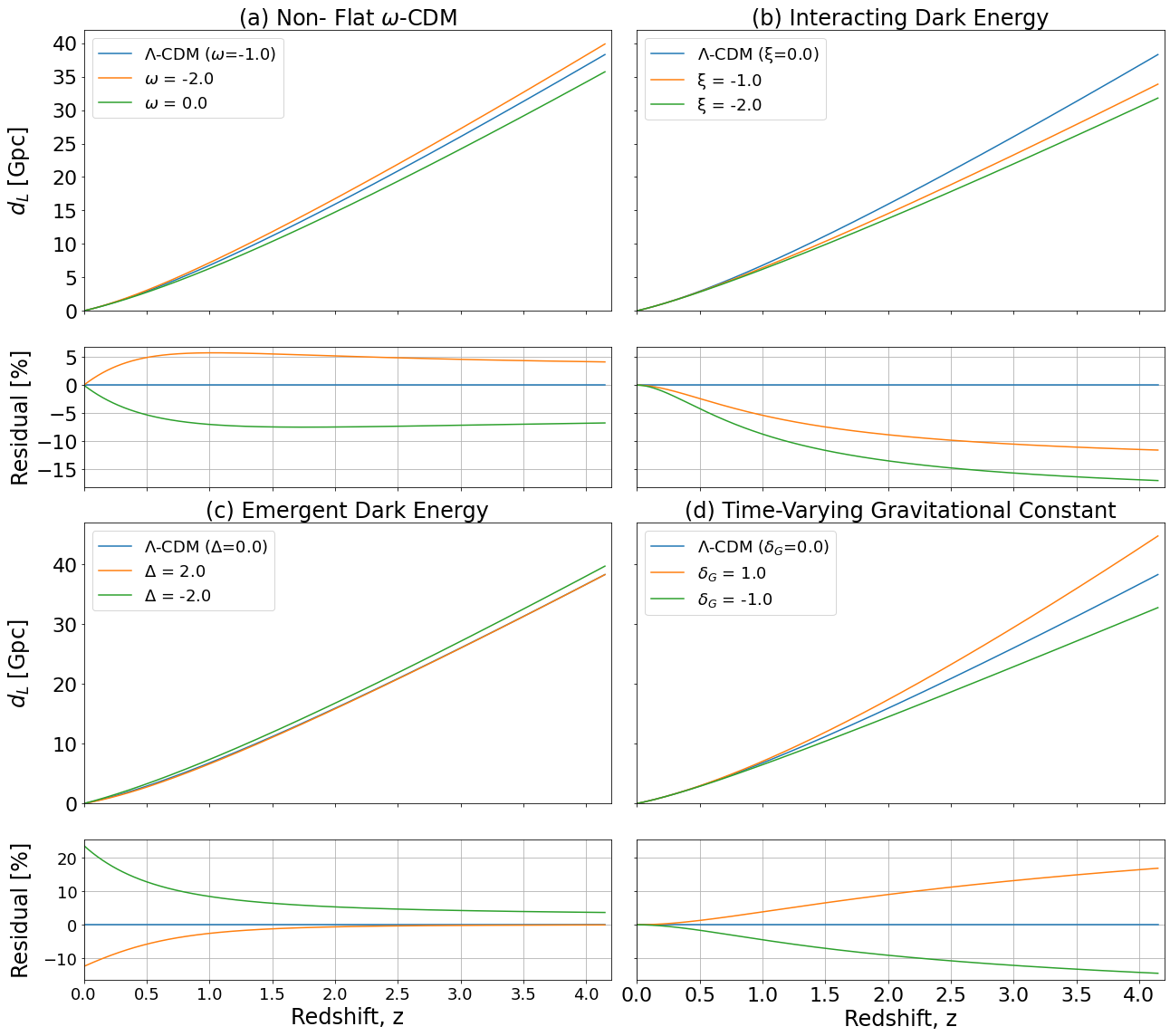}
		\caption{The panels report the predicted luminosity distance as a function of redshift for the non-flat $\omega$CDM (upper left panel), IDE model (upper right panel), emergent DE model (lower left panel), and time-varying gravitational constant model (lower right panel). In each panel, we depict our {\em fiducial} model (blue solid line), and the DE models where $[H_0,\, \Omega_{m,0},\, \Omega_{k,0},\, \Omega_{\Lambda,0}]$ are set to their fiducial values, and the extra-parameters are varied. For each model, we also show the residuals with respect to the $\Lambda$CDM model.}
		\label{fig:models}  
	\end{figure}

	In Fig.\til\ref{fig:models}, we illustrate the impact of the  DE parameters on the luminosity distance  for each model. In the upper left panel, we depict the non-flat $\omega$-CDM model for the value of $\omega=[-2; 0]$ as red and green solid lines, respectively. In the upper right panel, we show the IDE model for the value of $\xi=[-1; -2]$ as red and green solid lines, respectively. In the lower-left panel, we depict the emergent DE model for the value of $\Delta=[-2; 2]$ as red and green solid lines, respectively. And, finally, in the lower right panel, we show the time-varying gravitational constant model for the value of $\delta_G=[-1; 1]$ as red and green solid lines, respectively. In all panels, we also report, for comparison,
	our {\em fiducial} cosmological model, as a blue solid line, which is a flat-$\Lambda$CDM with the following values of cosmological parameters\til\cite{Planck2020}:
	\begin{equation}\label{baseline_model}
		\begin{aligned}
			H_0  =  67.66\ \mbox{km}\ \mbox{s}^{-1}\mbox{Mpc}^{-1} ,\  \Omega_{m,0}  =  0.3111,\\
			\Omega_{\Lambda,0}  =  0.6889\ \mbox{and}\ \Omega_{k,0}  =  0.00\,.\end{aligned}
	\end{equation}
	
	Below each panel, we report the residuals to illustrate the level of departure expected from the $\Lambda$CDM model. The maximum departure in the case of  non-flat $\omega$-CDM model is $\sim 7 \%$, while it reaches $\sim 18 \%$
	at $z\approx 4$ for the IDE model. The emergent DE model reaches a maximum departure when $z\sim 0$ as the model claims to solve the Hubble tension with a modification of the DE contribution at a late-time. Finally, the departure of the time-varying gravitational constant model from the {\em fiducial} model reaches $\sim 15 \%$ at $z\approx 4$.
	
	\section{Mock Data}\label{sec:mockdata}
	
	Here we briefly summarize the procedure adopted to build up the mock catalogs. We closely follow the recipe given in\til\cite{Califano2022}, and assign the redshift to the GWs sources extracting it from  the following redshift probability distribution\til\cite{Regimbau:2012,Cai:2017,Belgacem:2019tbw}
	\begin{equation}\label{rate:unit_of_redshift}
		p(z) = \mathcal{N}\frac{R_m (z)}{1+z}\frac{dV(z)}{dz},
	\end{equation} 
	where $\mathcal{N}$ is a normalization factor, $dV(z)/dz$ is the comoving volume element, and $R_m (z)$ is the merger rate per unit of volume in the source frame. The latter takes the form\til\cite{RegimbauHughes:2009,Meacher:2016,Regimbau:2017}
	\begin{equation}\label{merger_rate}
		R_{m} (z) = R_{m,0} \int_{t_{min}}^{t_{max}} R_f[t(z)-t_d] P(t_d) d t_d \ ,
	\end{equation}
	where $R_f[t(z)-t_d]$ is the Star Formation Rate (SFR), and $P(t_d)$ is the time delay distribution. We assume, for the SFR, the model proposed in\til\cite{Vangioni:2014} and for the time delay distribution a power law functional form, $P(t_d)\propto t_{d}^{-1}$, as suggested by population synthesis models\til\cite{Tutukov:1994,Lipunov:1995,Pacheco:2006,Belczynski:2006,Shaughnessy:2008}. Nevertheless, it is worth noticing that setting the SFR and the time delay distribution to other models do not affect the accuracy of final results (for more details we refer to Sec.\til 5.3 in\til\cite{Califano2022}).
	We integrate Eq.\til\eqref{merger_rate} between a minimum time delay of $20$ Myr and the maximum fixed to the Hubble time. Furthermore, the quantity $R_{m,0}$ is  the  normalization  of the merger rate at $z=0$. We, therefore, set it to the best-fit value obtained by the LIGO/Virgo/KAGRA collaboration:  $R_{m}(z=0)=105.5^{+190.2}_{-83.9}\ \mbox{Gpc}^{-3} \mbox{yr}^{-1}$\til\cite{LIGO2021:population}. Once the redshift is extracted from the probability distribution in Eq. \eqref{rate:unit_of_redshift}, we can assign a {\em fiducial} luminosity distance, $d_L^{fid}(z)$, based on 
	our {\em fiducial} cosmological model in Eq. \eqref{baseline_model}.

	The ET will have three independent interferometers and, hence, the combined SNR is $\rho=\left(\sum\limits_{i=1}^{3}(\rho_{(i)})^2\right)^{1/2}$. The SNR of the single interferometer, $\rho_{(i)}$, %detected by matched filtering with an optimum filter 
	in the ideal case of Gaussian noise, is:
	\begin{equation}\label{eq:SNR}
		\rho^{2}_{(i)}=4 \int_{f_{\rm lower}}^{f_{\rm upper}} \frac{|F_{+,i}\tilde{h}(f)_{+}+F_{\times,i}\tilde{h}(f)_{\times}|}{S_{h,i}(f)} df.
	\end{equation}
	
\begin{figure}
		\centering
		\includegraphics[width=0.48\textwidth]{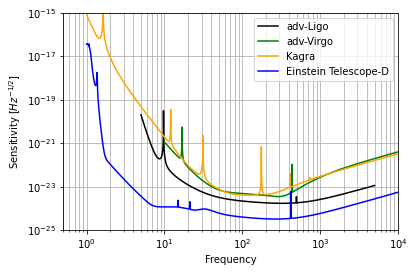}
		\caption{The sensitivities of advanced LIGO and Virgo and of the latest sensitivity curve made available by for the ET in the ET-D configuration.}
		\label{fig:sems_curv}
	\end{figure}

	In the previous definition, $S_{h,i}(f)$ is the one-sided noise power spectral density of $i$-th interferometer, $\tilde{h}_{+}$ and $\tilde{h}_{\times}$ are the GW strain amplitudes of $+$ and $\times$ polarizations, and $F_{+,i}(\psi,\theta,\phi)$ and $F_{\times,i}(\psi,\theta,\phi)$ are the so-called beam pattern functions\til\cite{FinnChernoff:1993}. The whole sensitivity function\footnote{The latest power spectral density $S_h(f)$ can be downloaded at \url{https://apps.et-gw.eu/tds/?content=3&r=14065}.} $S_h(f)$ is depicted in in Fig.\til\ref{fig:sems_curv} . To integrate Eq. \eqref{eq:SNR}, we set a lower cutoff, $f_{lower}$, at $1$ Hz\til\cite{Sensitivity:2011} and the upper one to $f_{upper}=\frac{c^3}{(6\sqrt{6}\pi M_{obs})G}$. 
	We can compute the total number of observable BNS mergers, $N$, from the equation
	\begin{equation}
		N = T_{obs}\ \Theta \int_{0}^{10} \frac{R_m (z)}{1+z}\frac{dV(z)}{dz}dz\,,
	\end{equation}
	where $\Theta$ is the duty cycle and $T_{obs}$ is the observation time. In order to generate the catalog and select the events above a certain threshold of the signal-to-noise ratio (SNR) $\geq 9$, we assume an isotropic distribution for the sky angles $\theta$ and $\phi$,  and an uniform distribution for the  orientation angle $\cos i$ and the polarization $\psi$. Moreover,  to generate the synthetic signal self-consistently with our choice of $R_{m,0}$, we  follow the LIGO/Virgo/KAGRA collaboration and set a uniform NS mass range in the interval $[1, 2.5]\ M_\odot$. 
	Thus, we obtain a rate of $\sim 3\times 10^4$ events per year, assuming a duty cycle of 80\%. In our analysis, we will consider 1, 5, and 10 years of observations.
	
	Once the {\em fiducial} luminosity distances are generated, we add a  Gaussian noise component, $\mathcal{N}(d_{L}^{fid},\sigma_{ d_{L}})$, to them in order generate our mock observations. 
	The variance $\sigma^2_{ d_{L}}$ includes the contributions due to the instrument, $\sigma^2_{inst}$,  the lensing, $\sigma^2_{lens}$, and  the peculiar velocity of the host galaxy, $\sigma^2_{pec}$. Therefore, the total variance will be:
	\begin{equation}\label{sigma_dl}
		\sigma_{d_L}^2={\sigma_{inst}^2+\sigma_{lens}^2 +\sigma_{pec}^2}\,.
	\end{equation}
	The contribution due to the instrumental noise component  $\sigma_{inst}$ is\til\cite{Cutler1994,Dalal:2006}
	\begin{equation}
		\sigma_{inst}=\frac{2}{\rho}d_L(z),
	\end{equation}
	where factor two accounts for the degeneration between $\rho$ and the inclination angle, which may differ for each event.
	The contribution due to the weak lensing distortions, $\sigma_{lens}$, is given by\til\cite{Hirata:2010,Tamanini:2016}
	\begin{equation}
		\sigma_{lens}=0.066\left(\frac{1-(1+z)^{-0.25}}{0.25}\right)^{1.8}d_L(z)F_{delens}(z),
	\end{equation}
	where $F_{delens}(z)= 1- \frac{0.3}{\pi /2}\arctan{\frac{z}{z_*}}$, with $z_*=0.073$\til\cite{Tamanini:2016}. The latter factor takes into account the possibility to reduce the uncertainty due to weak lensing with the future detectors such as the Extremely Large Telescope\til\cite{Speri:2021}.
	Finally, $\sigma_{pec}$ is related to the peculiar velocities\til\cite{Hjorth2017,Howlett2020,Mukherjee2021} and can be approximated with the following fitting formula\til\cite{Kocsis:2006}
	\begin{equation}
		\sigma_{pec}=\left[ 1+\frac{c(1+z)^2}{H(z)d_L (z)}\right]\frac{\sqrt{\langle v^2\rangle}}{c}d_L (z)\,,
	\end{equation}
	where we set the averaged peculiar velocity $\langle v^2\rangle$ to  $500$ km/s, in agreement with the observed values in galaxy catalogs\til\cite{Cen2000}. { To build such a mock data set we are relying on the somehow simplistic assumption that there are no correlations to take into account with the source parameters of each system. Of course, in reality, this will be not the case. However, following\til\cite{Cutler1994,Dalal:2006,Cai:2017,Belgacem:2019tbw,Zhang2019,Dagostino2019,Fu2020,Yang:2021qge,Mitra2021}, we are assuming that the luminosity distance is measured with a statistical error that depends on the observational features of the Einstein Telescope. Moreover,  it is worth remarking that our estimation of the error bars is, on average, a factor $1.7$ larger than the one expected for the Einstein Telescope \cite{Maggiore2020} providing, therefore, a conservative estimation of the luminosity distance.}

	\subsection{Electromagnetic counterpart}
	
\begin{figure}
		\centering
		\includegraphics[width=0.48\textwidth]{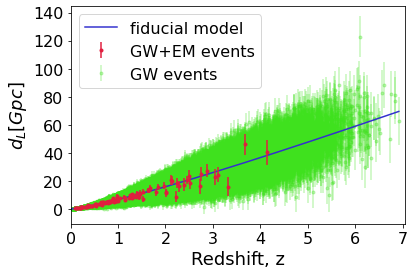}
		\caption{The catalog of all GW events after 10 years of observation. In red, we highlight the 110 events with the electromagnetic counterpart. The blue line is the luminosity distance in the fiducial cosmological model.}
		\label{fig:catalog}
	\end{figure}

	The predicted outcomes for a BNS merger are a relativistic outflow, which is highly anisotropic and can produce an observable high energy transient; a thermal and radioactive source emitting most of its energy at ultraviolet, optical, and near-infrared wavelengths; and  a burst of MeV neutrinos\til\cite{Pian2021}. 
	The neutrino burst  is hard to detect. Indeed, with current instruments such as the IceCube Neutrino Observatory\til\cite{IceCube2017}, we can detect a neutrino counterpart  only for events located at redshift below $0.1$\til\cite{Aartsen2020}.
	The thermal sources, {\em i.e.} kilonova, produced by the radioactive decay of unstable heavy elements synthesized during the coalescence can be detected up to $z\sim 1$ with the current and forthcoming telescopes such as the Roman Space Telescope\til\cite{Spergel2015,Hounsell2018,Chase2022,Alfradique2022}. Since, we are interested to study the accuracy on the cosmological parameters and, more specifically, constraining DE models, we only focus on the first kind of outcomes: the relativistic outflows. In particular, we study the case of short GRBs because  forthcoming Gamma-Ray and X-Ray satellites will detect electromagnetic counterparts at $z\sim 8$\til\cite{THESEUS:2017wvz}. 
	In particular, we consider the THESEUS satellite that could overlap with ET and provides the electromagnetic counterpart of the GWs events\til\cite{THESEUS:2017qvx,THESEUS:2017wvz,Amati2021,Ciolfi2021,Ghirlanda2021,Rosati2021}. Again we closely follow Ref.\til\cite{Califano2022} and simulate the observed photon flux of the GRB events associated with GW events through the luminosity distance by sampling the luminosity  probability distribution $\phi(L)$\til\cite{Yang:2021qge,Califano2022}. We assume $\phi(L)$ be a standard broken power law distribution\til\cite{Wanderman:2014eza} and the jet profile be Gaussian\til\cite{Resmi2018,Howell:2019} { (more details can be found in Sec. 3.4 of \cite{Califano2022})}.
	Once we have extracted the flux from the relation flux-luminosity, we select only the events which are above the flux threshold of $0.2\ \mbox{photon}\ \mbox{cm}^{-2} \ \mbox{s}^{-1}$. 
	{  To obtain the number of combined events, we set the duty cycle of the THESEUS satellite to 80\%\til\cite{THESEUS:2017wvz} mainly due to a reduction of observing time owing to the passage through the Southern Atlantic Anomaly. The THESEUS field of view is $\sim 2\pi$ sr,  in order to have accuracy on the localization of  about 5 arcmins the source must be within the central 2 sr of its field of view. This feature reduces the total number of combined events of a factor $2/(2\pi)\sim \ 1/3$ \til\cite{Belgacem:2019tbw}. }
	We estimate a rate of $\sim 11$ events per year. To show the effectiveness of the procedure, in Fig.\til\ref{fig:catalog}, we depict all GW events recorded after 10 years of observations with the corresponding error bars (green points), the ones with the electromagnetic counterpart (red points), and the {\em fiducial} cosmological model (blu solid line).
	
	\section{Statistical Analysis}\label{sec:stats}
	
	We carry out a Monte Carlo Markov Chain (MCMC) analysis to estimate the accuracy down to which each DE model parameter (that depends on the specific choice of the DE EoS) can be constrained with the future observation from ET. Our mock data are built using  the flat-$\Lambda$CDM cosmology in Eqs.\til\eqref{E_z} and \eqref{baseline_model} as our {\em fiducial} model. Then, 
	we expect the posterior distributions of the parameters of the DE models introduced in Sec.\til\ref{sec:models} to be centered around the {\em fiducial} model. Therefore, the error on the model parameters will indicate the accuracy that we will be able to reach with ET. To this aim, we will run our MCMC pipeline on both bright and dark sirens, {\em i.e.} events whose electromagnetic counterpart has been and has not been detected, respectively, and we will point out the main differences in the results.
	
	Our MCMC is based on the {\texttt{emcee}} package\til\cite{emcee}, and will employ the likelihood of all GWs events
	defined as the product of the single event likelihood, $ p(\textbf{d}|\bm{\lambda}) = \prod_{i=1}^{N} p(d_i|\bm{\lambda})$. Here,  $\bm{\lambda}$ are the cosmological parameters of interest for the specific model, and \textbf{d}$\equiv\lbrace d_i \rbrace_{i=1}^{N}$ is the mock dataset with $N$ equal to the number of observations. In order to write down the single event likelihood,  one must distinguish between the run with the bright and dark sirens\til\cite{Califano2022}. When using bright sirens, the redshift information is assumed known from the detection of an electromagnetic counterpart which is, in our case, a short GRB. In such a case, the single event likelihood can be written as\til\cite{ Mandel:2018mve, Ye:2021klk}
	\begin{equation}\label{likelihood_bright}
		p(d_i | \bm{\lambda})= \frac{\int  p(d_i|D_L)p_{pop}(D_L |z_i , \bm{\lambda})d D_L}{\int p_{det}(D_L)p_{pop}(D_L|z_i , \bm{\lambda}) dD_L}\,,
	\end{equation} 
	where $p_{pop}(D_L | \bm{\lambda})=\delta(D_L - d_{L}^{th}(z_i,\bm{\lambda}))$\til\cite{DelPozzo:2011vcw}.
	In the Eq.\til\eqref{likelihood_bright}, the denominator is a normalization factor that takes into account the selection effects\til\cite{Mandel:2018mve,Vitale:2020aaz}. 
	
	Instead, when using {dark sirens}, we assume that   the redshift distribution of the BNS population is known, and marginalized over the redshift\til\cite{Ding:2018zrk,Ye:2021klk}. In such a case, the probability of detecting an event $d_i$ in a specified cosmological model is given by
	\begin{equation}
		\begin{aligned}
			p(d_i|\bm{\lambda}) &= \int_{0}^{z_{max}} p(d_i , z_i |\bm{\lambda}) dz_i \\
			&= \int_{0}^{z_{max}} p(d_i|d_L^{th}(z_i,\bm{\lambda})) p_{obs}(z_i|\bm{\lambda})dz_i\,,
		\end{aligned}
	\end{equation}
	where the probability prior distribution of the redshift, $p_{obs}(z_i|\bm{\lambda})$, is obtained from the observed events and already includes detector selection effects\til\cite{Ding:2018zrk}. 
	
	{ Finally, in our analysis, we neglect {\em (i)} the contribution of the spin of the source to the  amplitude of the signal\til\cite{Poisson:1995ef,Baird:2012cu}, {\em (ii)} assume a flat uniform prior on the cosmological parameters of interest as reported in Table\til\ref{tab:prior}, and {\em (iii)} we neglect the impact of the merger rate and of the time-delay distribution for being negligible within this dataset (we refer the reader to Sec. 5.3 in \cite{Califano2022}).}

	\begin{table}[!ht]
		\centering
		\begin{tabular}{cc||cc}
			\hline
			\hline
			Parameters  &  Prior & Parameters  &  Prior\\
			\hline
			$H_0$ & $\mathcal{U}(35,85)$ & $\omega_{DE}$ & $\mathcal{U}(-3,0)$\\
			$\Omega_{m,0}$ & $\mathcal{U}(0,1)$ & $\xi$ & $\mathcal{U}(-3,3)$ \\
			$\Omega_{\Lambda,0}$ & $\mathcal{U}(0,1)$ & $\Delta$ & $\mathcal{U}(-2,2)$\\
			$\Omega_{k,0}$ & $\mathcal{U}(-1,1)$ & $\delta_{G}$ & $\mathcal{U}(-3,3)$ \\
			\hline
			\hline
		\end{tabular}
		\caption{Uniform priors on the cosmological parameters involved in the DE models explained in Sec.\til\ref{sec:models}.}
		\label{tab:prior}
	\end{table}

	\begin{table*}[!ht]
		\centering
		\begin{tabular}{c|c|c|c|c|c|c|c|c}
			\hline
			\hline
			\multicolumn{9}{c}{{\bf $\omega$CDM}}\\
			\hline
			\multirow{2}{2.5em}{{\bf years}}&\multicolumn{4}{c|}{{\bf Bright Sirens}}&\multicolumn{4}{c}{{\bf Dark Sirens}} \\
			\cline{2-9}
			& ${\bf H_0}$ & ${\bf \Omega_{k,0}}$  & ${\bf \Omega_{\Lambda,0}}$ & $ {\bf \omega_{DE}}$ & ${\bf H_0}$ & ${\bf \Omega_{k,0}}$  & ${\bf \Omega_{\Lambda,0}}$ & $ {\bf \omega_{DE}}$ \\
			\hline
			1  & $66.70_{-2.24}^{+2.50}$ &$-0.08_{-0.28}^{+0.38}$ &$0.70_{-0.34}^{+0.21}$ &$-1.56_{-0.97}^{+1.39}$ &$67.52_{-0.17}^{+0.19}$ &$0.02_{-0.03}^{+0.04}$ &$0.68_{-0.04}^{+0.04}$ &$-1.20_{-0.36}^{+0.28}$ \\
			5  & $67.80_{-1.02}^{+0.99}$ &$0.13_{-0.22}^{+0.22}$ &$0.57_{-0.18}^{+0.22}$ &$-1.63_{-0.97}^{+1.08}$ &$67.70_{-0.07}^{+0.07}$ &$-0.02_{-0.02}^{+0.02}$ &$0.67_{-0.03}^{+0.04}$ &$-0.97_{-0.15}^{+0.15}$  \\
			10  & $67.55_{-1.03}^{+1.02}$ &$-0.05_{-0.17}^{+0.19}$ &$0.66_{-0.16}^{+0.20}$ &$-1.35_{-0.98}^{+0.84}$ & $67.68_{-0.05}^{+0.06}$ &$-0.01_{-0.02}^{+0.02}$ &$0.68_{-0.03}^{+0.03}$ &$-0.95_{-0.11}^{+0.09}$\\
			\hline
			\hline
			\multicolumn{9}{c}{{\bf Interacting Dark Energy ($\omega_{DE}$-fixed)}}\\
			\hline
			\multirow{2}{2.5em}{{\bf years}}&\multicolumn{4}{c|}{{\bf Bright Sirens}}&\multicolumn{4}{c}{{\bf Dark Sirens}} \\
			\cline{2-9}
			& ${\bf H_0}$ & ${\bf \Omega_{m,0}}$  & ${\bf {\bf \omega_{DE}}}$ & $ {\bf \xi}$ & ${\bf H_0}$ & ${\bf \Omega_{m,0}}$  & $ {\bf \omega_{DE}}$ & $ {\bf \xi}$ \\
			\hline
			1  & $66.71_{-1.66}^{+1.35}$ &$0.25_{-0.17}^{+0.25}$ & -- &$-0.98_{-1.11}^{+1.50}$  &  $67.72_{-0.24}^{+0.27}$ &$0.30_{-0.03}^{+0.03}$ & --&$0.13_{-0.20}^{+0.20}$ \\
			5  & $68.20_{-0.99}^{+0.91}$ &$0.22_{-0.13}^{+0.16}$ & --&$-0.65_{-0.84}^{+1.10}$& $67.68_{-0.12}^{+0.14}$ &$0.33_{-0.02}^{+0.02}$ & -- &$0.01_{-0.11}^{+0.12}$  \\
			10  & $67.81_{-0.93}^{+0.97}$ &$0.24_{-0.14}^{+0.13}$ & -- &$-0.76_{-0.92}^{+0.83}$ &$67.70_{-0.05}^{+0.05}$ &$0.32_{-0.01}^{+0.01}$ & -- &$-0.02_{-0.06}^{+0.06}$  \\
			\hline
			\hline
			\multicolumn{9}{c}{{\bf Interacting Dark Energy ($\omega_{DE}$-variable)}}\\
			\hline
			1  &  $67.86_{-2.15}^{+2.50}$ &$0.51_{-0.15}^{+0.11}$ &$-2.08_{-0.66}^{+1.05}$ &$0.96_{-0.95}^{+1.18}$&  $67.51_{-0.19}^{+0.19}$ &$0.37_{-0.05}^{+0.04}$ &$-1.14_{-0.21}^{+1.16}$ &$0.79_{-0.50}^{+0.59}$\\
			5  & $68.64_{-1.17}^{+1.22}$ &$0.42_{-0.25}^{+0.16}$ &$-1.51_{-0.63}^{+0.52}$ &$0.37_{-1.22}^{+1.72}$ &$67.70_{-0.10}^{+0.11}$ &$0.28_{-0.18}^{+0.17}$ &$-0.97_{-0.09}^{+0.09}$ &$-0.25_{-0.26}^{+0.27}$\\
			%\cline{2-5}\cline{6-8}
			%\cline{2-8}
			10  & $67.99_{-1.04}^{+1.35}$ &$0.43_{-0.26}^{+0.16}$ &$-1.38_{-0.64}^{+0.43}$ &$0.15_{-1.33}^{+1.68}$& $67.63_{-0.04}^{+0.04}$ &$0.28_{-0.16}^{+0.17}$ &$-0.92_{-0.08}^{+0.08}$ &$-0.07_{-0.18}^{+0.16}$ \\
			\hline
			\hline
			\multicolumn{9}{c}{{\bf Emergent Dark Energy}}\\
			\hline
			\multirow{2}{2.5em}{{\bf years}}&\multicolumn{4}{c|}{{\bf Bright Sirens}}&\multicolumn{4}{c}{{\bf Dark Sirens}} \\
			\cline{2-9}
			& ${\bf H_0}$ & ${\bf \Omega_{m,0}}$  & ${\bf {\bf \Delta}}$ & -- & ${\bf H_0}$ & ${\bf \Omega_{m,0}}$  & $ {\bf \Delta}$ & -- \\
			\hline
			1  & $66.46_{-1.35}^{+4.16}$ &$0.35_{-0.10}^{+0.09}$ &$-0.06_{-0.91}^{+1.09}$& --& $67.86_{-0.24}^{+0.34}$ &$0.31_{-0.01}^{+0.01}$ &$0.21_{-0.34}^{+0.28}$& --\\
			5  & $67.30_{-0.82}^{+2.71}$ &$0.32_{-0.06}^{+0.05}$ &$0.26_{-0.77}^{+0.80}$& --&$67.60_{-0.08}^{+0.09}$ &$0.31_{-0.01}^{+0.01}$ &$-0.02_{-0.06}^{+0.06}$& --\\
			10  & $66.92_{-0.68}^{+2.17}$ &$0.36_{-0.06}^{+0.05}$ &$0.21_{-0.83}^{+0.89}$& --  & $67.66_{-0.03}^{+0.03}$ &$0.310_{-0.002}^{+0.002}$ &$0.00_{-0.01}^{+0.01}$& --  \\
			\hline
			\hline
			\multicolumn{9}{c}{{\bf Time-Varying Gravitational Constant}}\\
			\hline
			\multirow{2}{2.5em}{{\bf years}}&\multicolumn{4}{c|}{{\bf Bright Sirens}}&\multicolumn{4}{c}{{\bf Dark Sirens}} \\
			\cline{2-9}
			& ${\bf H_0}$ & ${\bf \Omega_{m,0}}$  & ${\bf {\bf \delta_{G}}}$ & -- & ${\bf H_0}$ & ${\bf \Omega_{m,0}}$  & $ {\bf \delta_{G}}$ & -- \\
			\hline
			1  &$66.92_{-1.70}^{+1.30}$ &$0.26_{-0.17}^{+0.26}$ &$-0.49_{-1.52}^{+1.63}$ & --  &$67.64_{-0.08}^{+0.07}$ &$0.31_{-0.01}^{+0.01}$ &$-0.03_{-0.04}^{+0.05}$& -- \\
			5  &$67.49_{-0.89}^{+0.87}$ &$0.35_{-0.12}^{+0.12}$ &$0.22_{-0.60}^{+1.21}$&--&$67.68_{-0.04}^{+0.05}$ &$0.32_{-0.01}^{+0.01}$ &$-0.01_{-0.03}^{+0.02}$& --  \\
			10  & $67.51_{-0.92}^{+0.81}$ &$0.29_{-0.07}^{+0.10}$ &$-0.26_{-0.46}^{+0.42}$& -- &$67.65_{-0.04}^{+0.04}$ &$0.31_{-0.01}^{+0.01}$ &$-0.02_{-0.02}^{+0.02}$ & -- \\
			\hline
			\hline
		\end{tabular}  
		\caption{The table lists the best fist values and the $1\sigma$ uncertainty on the cosmological parameters of interest for each DE model presented in Sec.\til\ref{sec:models}.}\label{tab:results}
	\end{table*}

	\section{Results}\label{sec:results}
	
\begin{figure*}[!ht]
		\centering
		\includegraphics[width=0.48\textwidth]{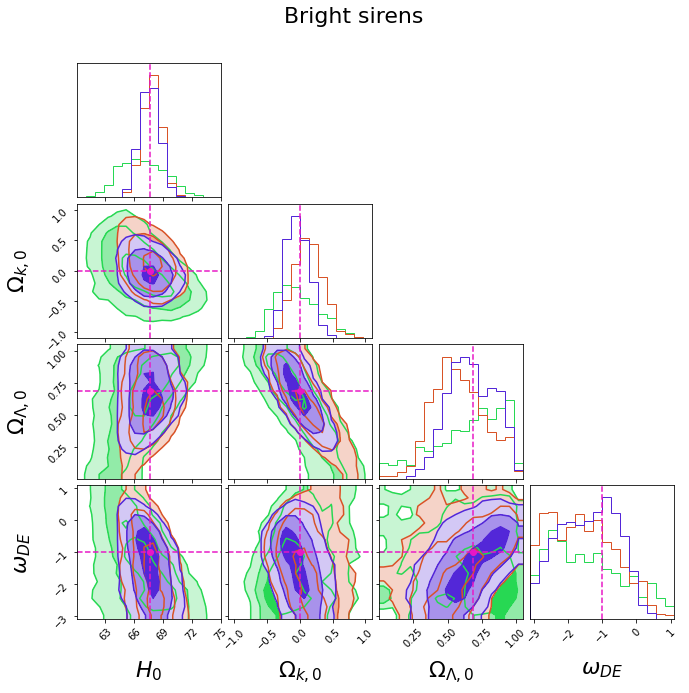}
		\includegraphics[width=0.48\textwidth]{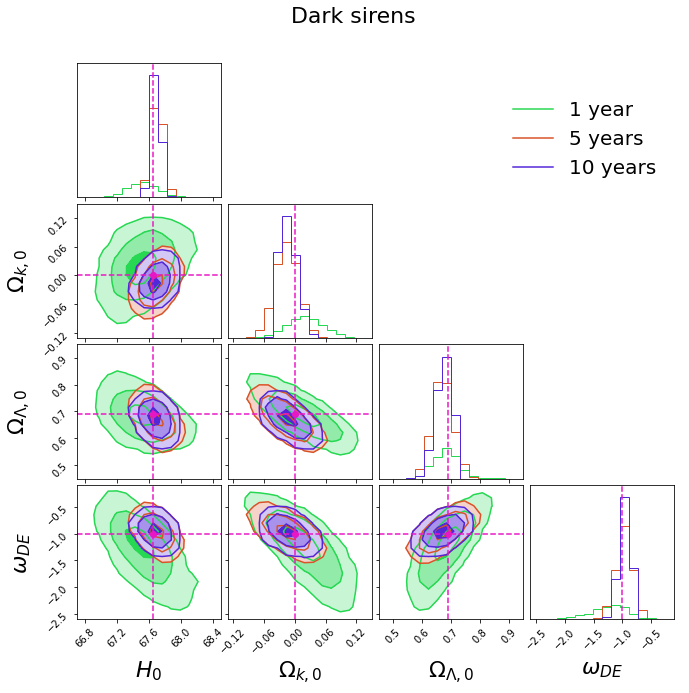}
		\caption{The figure illustrates the posterior distribution for the $\omega$CDM model  obtained from the bright and dark sirens, depicted on the left and right side respectively. In each contour plot, we represent with green, orange, and blue histograms and filled areas the results from one, five, and ten years of observation, respectively. The different level of the transparency of the contours corresponding to a specific total number of years of observations depicts the 68\%, 95\% and 99\% CL from the darkest to the lightest color, respectively. Finally, the vertical dashed red line indicate the value of the {\em fiducial} cosmological parameters.}
		\label{fig:contour_interacting_wcdm}
	\end{figure*}

\begin{figure*}[!ht]
		\centering
		\includegraphics[width=0.48\textwidth]{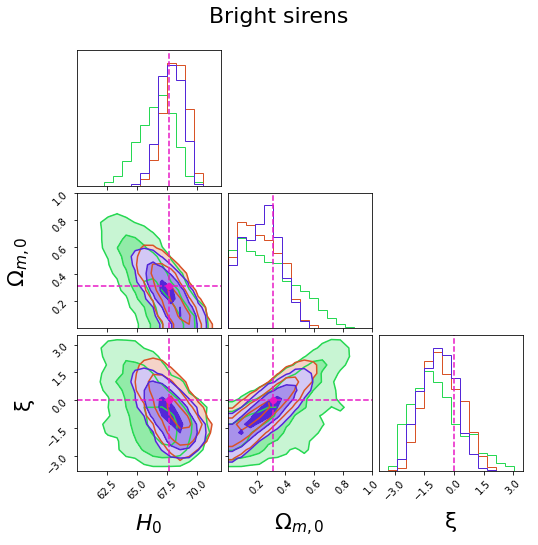}
		\includegraphics[width=0.48\textwidth]{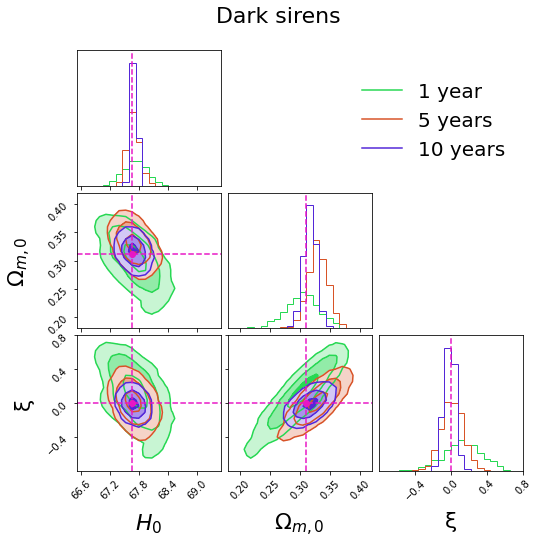}
		\caption{The same of Fig.\til\ref{fig:contour_interacting_wcdm} for Interacting Dark Energy model  with $\omega_{DE}$-fixed }
		\label{fig:contour_interacting_1p}
	\end{figure*}

\begin{figure*}[!ht]
		\centering
		\includegraphics[width=0.48\textwidth]{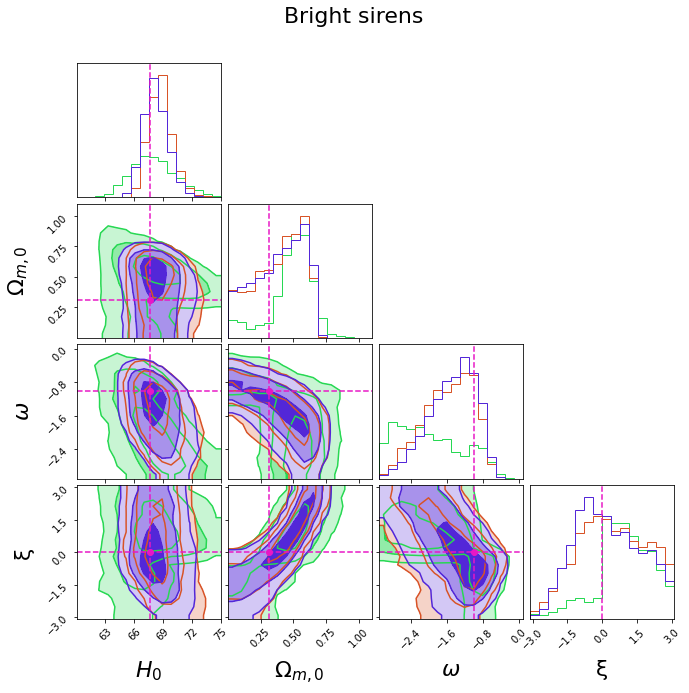}
		\includegraphics[width=0.48\textwidth]{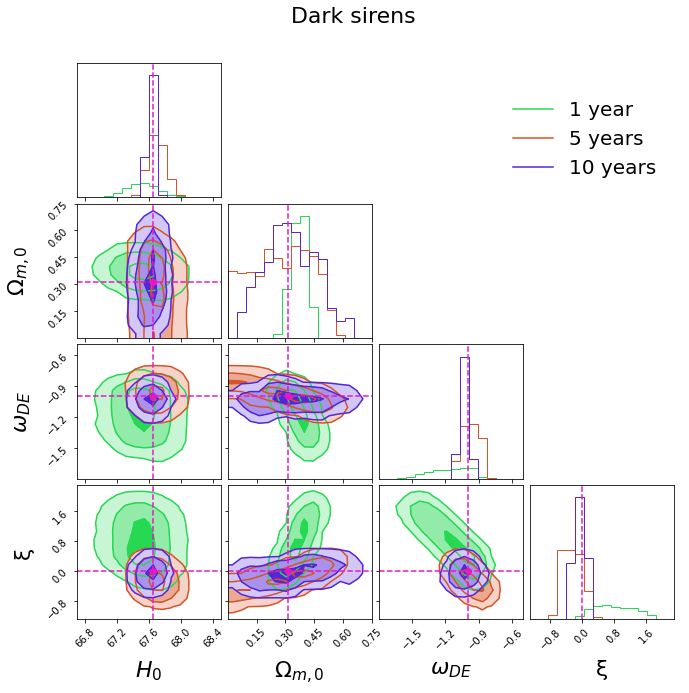}
		\caption{The same of Fig.\til\ref{fig:contour_interacting_wcdm} for Interacting Dark Energy model with $\omega_{DE}$-variable }
		\label{fig:contour_interacting_2p}
	\end{figure*}

\begin{figure*}[!ht]
		\centering
		\includegraphics[width=0.48\textwidth]{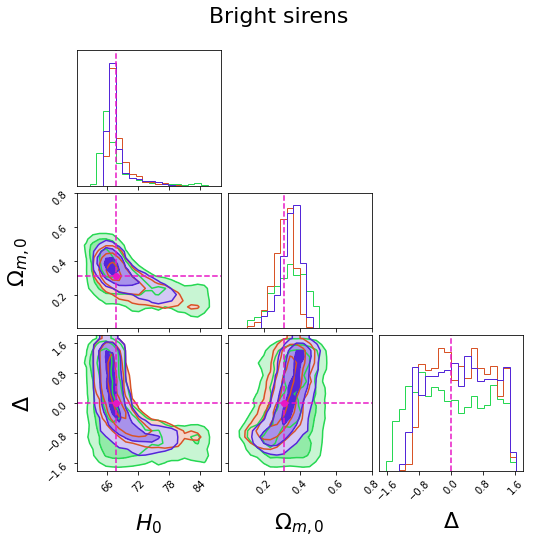}
		\includegraphics[width=0.48\textwidth]{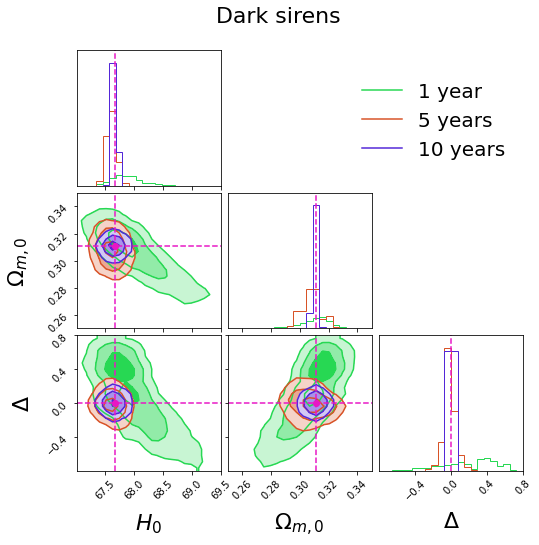}
		\caption{The same of Fig.\til\ref{fig:contour_interacting_wcdm} for Emergent Dark Energy model }
		\label{fig:contour_emergent}
	\end{figure*}

\begin{figure*}[!ht]
		\centering
		\includegraphics[width=0.48\textwidth]{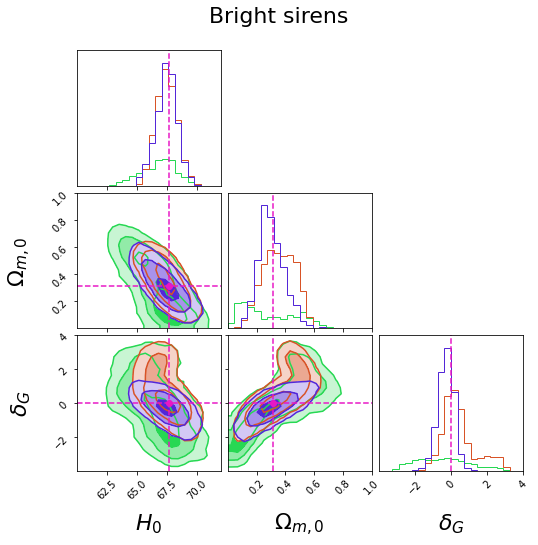}
		\includegraphics[width=0.48\textwidth]{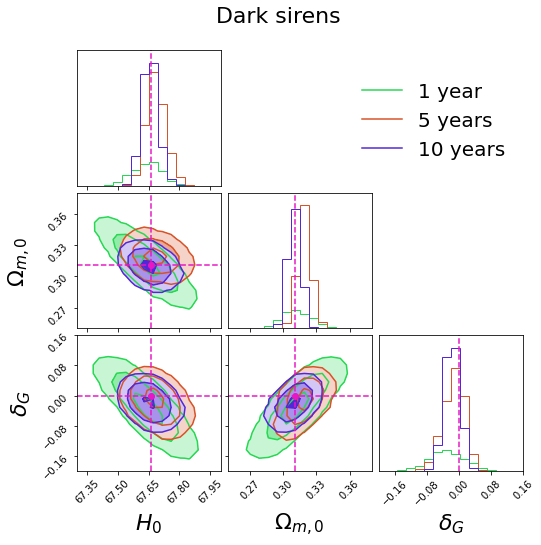}
		\caption{The same of Fig.\til\ref{fig:contour_interacting_wcdm} for Time-Varying Gravitational Constant }
		\label{fig:time_varying}
	\end{figure*}

	We carried out a MCMC run for each mock catalog and for each DE model. We built six mock catalogs: three of them reporting all the GW events at one, five, and ten years of observations but without having prior redshift information (dark sirens); and the other three containing those GW events with a detected electromagnetic counterpart and, therefore, having prior redshift information (bright sirens) from a X-ray telescope, such as THESEUS. 
	We use these mock catalogs to constrain the four DE models mentioned in Sec.\til\ref{sec:models}. All results of our run are reported in Table\til\ref{tab:results}. For each DE model we also show the corner plot of the posterior distributions of the cosmological parameters of interest, see Figs.\til\ref{fig:contour_interacting_wcdm},\til\ref{fig:contour_interacting_1p},\til\ref{fig:contour_interacting_2p},\til\ref{fig:contour_emergent}, and\til\ref{fig:time_varying}. 
	In each Figure, we report on the left and right sides the posterior distribution obtained from the bright and dark sirens, respectively. In each contour plot, we depict with green, orange, and blue histograms and filled areas the results from one, five, and ten years of observation, respectively. The different level of the transparency of the contours corresponding to a specific total number of years of observations depicts the 68\%, 95\%, and 99\% CL from the darkest to the lightest color, respectively. Finally, the vertical dashed red line indicates the value of the {\em fiducial} cosmological parameters. Let us now discuss in detail the results for each DE model comparing them with the current observational results. In the discussion, we will always refer to the results obtained after ten years of observations.
	
	{\bf \em $\omega$CDM}: we report the results in Fig.\til\ref{fig:contour_interacting_wcdm}. 
	As reported in Table\til\ref{tab:results},  we may constrain the cosmological parameters with an accuracy of $[\sigma_{H_0}, \sigma_{\Omega_{k,0}}, \sigma_{\Omega_{\Lambda,0}}, \sigma_{\omega_{DE}}]= [1.02, 0.18, 0.18, 0.93]$ and $[0.06, 0.02,0.03,0.10]$ in the case of bright and dark sirens, respectively.  These results correspond to a relative accuracy of the Hubble constant and the $\omega_{DE}$ of $[\sim1.5\%,\sim 69\%]$ and  $[\sim 0.1\%,\sim 10\%]$, in the case of bright and dark sirens, respectively. As a reference,  the current accuracy on $H_0$ is at a level of 1\%,  and on $\omega_{DE}$ is at a level of 3\%\til\cite{Gao:2021}. However, it is important to remark that those bounds were obtained by using not only luminosity distances but also the distance prior data obtained by {\em Planck} satellite\til\cite{Gao:2021}. Therefore, we show that by using bright sirens, ET will bound the Hubble constant with the same level of accuracy, but it will be capable of improving it   by one order of magnitude using dark sirens. Nevertheless, ET alone will not be capable of improving the accuracy on the  $\omega_{DE}$ parameter not even with the dark sirens catalog.

	{ \bf \em Interacting Dark Energy ($\omega_{DE}$-fixed)}:  we report the posterior distributions in Fig.\til\ref{fig:contour_interacting_1p}, and the best fit values  of the cosmological parameters in Table\til\ref{tab:results}. The 68\% uncertainties are:  $[\sigma_{H_0}, \sigma_{\Omega_{m,0}}, \sigma_{\xi}]= [0.95, 0.13, 0.88]$ and $[0.05, 0.01, 0.06]$ in the case of bright and dark sirens, respectively. These results translate   in accuracy on the Hubble constant of $\sim 1.4\%$ and  $\sim 0.1\%$, which is always better than current constraints shown in\til\cite{DiValentino2020a,Divalentino2020b,Pan2019} by a factor of $\sim 2.4$ and $\sim 46$. On the contrary, the accuracy on the parameter $\xi$ is improved only with dark sirens by a factor of $\sim 3.3$.
	
	{\bf \em Interacting Dark Energy ($\omega_{DE}$-variable)}: the results of the MCMC algorithm are shown in Fig.\til\ref{fig:contour_interacting_2p}  and listed in Table\til\ref{tab:results}. The cosmological parameters are constrained with an accuracy of $[\sigma_{H_0}, \sigma_{\Omega_{m,0}}, \sigma_{\omega_{DE}},\sigma_{\xi}]= [1.19, 0.21, 0.53, 1.5]$ and $[0.04, 0.16,0.08, 0.17]$ which corresponds to a relative accuracy on the Hubble constant and the $\omega_{DE}$ of $[\sim1.7\%,\sim 38\%]$ and  $[\sim 0.1\%,\sim 9\%]$, in the case of bright and dark sirens, respectively.
	Using only bright sirens, the uncertainty and the relative accuracy on $H_0$, $\omega_{DE}$ and $\xi$ are not comparable with the ones obtained in\til\cite{DiValentino2020a,Divalentino2020b,Pan2019}. Nevertheless, when we use dark sirens the accuracy on $H_0$ improves of a factor $\sim 27.5$ while the constraints on $\omega_{DE}$ and $\xi$ do not still improve results from\til\cite{DiValentino2020a,Divalentino2020b,Pan2019}. It is worth noticing that previous analyses are based on multiple datasets, such as CMB and Cepheids, while we are only focusing on studying the capability of ET.
	
	{\bf \em Emergent Dark Energy}:  the results are reported in Fig.\til\ref{fig:contour_emergent} and  in Table\til\ref{tab:results}.  We constrain the cosmological parameters with the following  accuracy:  $[\sigma_{H_0}, \sigma_{\Omega_{m,0}}, \sigma_{\Delta}]= [1.43, 0.06, 0.86]$ and $[0.03, 0.002, 0.01]$ in the case of bright and dark sirens, respectively.  These results provide us with a relative accuracy on $H_0$  of $\sim 2.1\%$ and  $\sim 0.04\%$.   Using bright sirens allows us to obtain bounds on the Hubble constant comparable with current constraints shown in\til\cite{Yang2021}. Instead, using dark sirens, we improve such a constraint by a factor $\sim 46$. The parameter $\Delta$ is constrained with a better accuracy only when dark sirens are taken into account, improving the bounds in\til\cite{Yang2021} of a factor $\sim 40$.

	{\bf \em Time-Varying Gravitational Constant}:  we report the posterior distributions in Fig.\til\ref{fig:time_varying},  while the constraints on the cosmological parameters are reported  in Table\til\ref{tab:results}.  We show that, in the framework of a time-varying gravitational constant,  ET will be capable of bounding the cosmological parameters with an accuracy of $[\sigma_{H_0}, \sigma_{\Omega_{m,0}}, \sigma_{\delta_G}]= [0.86, 0.08, 0.44]$ and $[0.04, 0.01, 0.02]$, in the case of bright and dark sirens, respectively. Hence,  the predicted relative accuracy on the Hubble constant is $\sim 1.2\%$ and  $\sim 0.06\%$.  Using the CMB + BAO + SN +$H_0$ dataset, the bounds on the Hubble constant are currently at the level of $\sim 1.5\%$, while the accuracy on  $\delta_G$ is of the order $0.002$\til\cite{Gao:2021}. Therefore, while using dark sirens ET will be capable of improving more than one order of magnitude of the relative error on $H_0$, it will not be capable alone of improving the bounds on $\delta_G$.

	\section{Discussion and  Conclusions}\label{sec:conclusion}
	
	The Hubble tension is one of the most important issues of modern cosmology\til\cite{DiValentino2021,Abdalla2022}. It is still not clear whether the solution to such a tension regards more the observational and statistical sector than the theoretical one with the possibility of some "{\it new physics}". Most of the solutions proposed up to date are focused on extending  the $\Lambda$CDM model\til\cite{Belgacem:2019tbw,DiValentino2021,Abdalla2022}, and on changing the underlying theory of gravity\til\cite{Belgacem2019,Belgacem2019b,Abdalla2022,Ferreira2022,Capozziello:2019cav}. Nowadays, this tension is established to be at $4.2\sigma$ and arises from a discrepancy in the value of the Hubble constant obtained from late-time observations, such as Cepheids, SNeIa, and BAO among the others,  and the observation of the CMB power spectrum at early-time. A dataset complementary to the usual late-time observations is represented by the estimation of the luminosity distance from the GWs. Since the latter is not based on the measurements of the photon flux, they must not be calibrated on the closer electromagnetic sources, such as Cepheids and  SNeIa. Therefore, they represent a potential way to solve the Hubble tension and may identify its cause  whether it is related to the observations at the late or early time, or to the theoretical limitations of the $\Lambda$CDM model. To this aim, the LIGO/Virgo/KAGRA collaborations have constrained the Hubble constant with GWs to be $H_0= 70_{-8}^{+12}$ km s$^{-1}$ Mpc$^{-1}$ at 68\% CL\til\cite{LIGO_H0_2017}. However, the accuracy reached is not enough to provide a definitive answer. 
	
	{ LISA and 3G detectors such as ET will provide  GW standard sirens at high redshifts, allowing us to  test different dark energy models and, possibly, the dark energy equation of state. The population of BNS mergers detected out to redshift equal to 3  by 3G detectors, and  the population of  black hole binaries mergers that will be detected up to redshift 10 by LISA, will allow to probe the cosmological principle, to map the dark matter distribution, to probe the equation of state of dark energy \cite{Bailes2021, Arun2022}.} The 3G GW detector ET promises to constrain the Hubble constant with sub-percent accuracy\til\cite{Maggiore2020},  offering a possible solution to the Hubble tension. Therefore, we have forecast the accuracy down to which ET may bound the cosmological parameters of four DE models which have the potential to solve the Hubble tension\til\cite{Abdalla2022}. Namely, we investigate the non-flat $\omega$CDM, the interacting dark energy, the emergent dark energy, and the time-varying gravitational constant models. We have predicted the luminosity distance expected in those models varying the cosmological parameters, and fit it to the mock data built to mimic the expected rate of observations and accuracy of ET, whose construction was explained in Sec.\til\ref{sec:mockdata}. Our fitting procedure is based on the MCMC algorithm explained in Sec.\til\ref{sec:stats}. The results are reported in Table\til\ref{tab:results}, and we also show the posterior distributions of the cosmological parameters of interest in  Figs.\til\ref{fig:contour_interacting_wcdm},\til\ref{fig:contour_interacting_1p},\til\ref{fig:contour_interacting_2p},\til\ref{fig:contour_emergent}, and\til\ref{fig:time_varying}. 
	
	Our results clearly indicate that ET will be capable of reaching an accuracy of $\sim 1\%$ with bright sirens, and go below $\sim 0.1\%$ with dark sirens, independently by the theoretical framework used in the statistical analysis. This accuracy will be adequate to solve the Hubble tension. Nevertheless, ET alone will not always be capable of improving current constraints on the additional cosmological parameters that depend on the specific choice of  DE model. For instance, in the non-flat $\omega$CDM and interacting DE models, the parameters $\omega_{DE}$ and $\xi$ will be constrained with an accuracy worse than current bounds\til\cite{DiValentino2020a,Divalentino2020b,Pan2019,Gao:2021}. In the case of the time-varying gravitational constant model, the accuracy reached by ET will be still one order of magnitude higher than current constraints\til\cite{Gao:2021}. On the contrary, in the emergent DE model, we show that ET will be also able to improve the bounds on the additional cosmological parameter $\Delta$ by a factor of 40 with respect to current analysis\til\cite{Yang2021}. These results show the huge capability of ET to solve the Hubble tension independently by the theoretical framework chosen but also point out that, to strongly constrain the DE models we have considered, ET will need to be complemented with other datasets.
	
	{ For instance, LISA will be able to obtain an accuracy on the dark energy cosmological parameters similar to the one we forecast for ET\til\cite{Tamanini:2016}. On the other hand, LSST will provide weak lensing and BAO  observations to set stringent constraints on the equation of state of dark energy\til\cite{Zhan2018RPPh}.  Indeed, using
		weak lensing and BAO separately, LSST will achieve accuracy on the curvature parameter of $10^{-3}$ that can be strongly improved by a joint analysis of these two cosmological probes\til\cite{Zhan2009}.  This represents an improvement in the accuracy of the curvature parameter  $\Omega_{k,0}$ of three orders of magnitude with respect to  our results. This is due to the nature of the cosmological observations provided by LSST that can probe curvature much better than the $d_L$-redshift relation. Finally, another interesting possibility will come from SKA which will allow testing cosmology up  to redshift $z=6$ using neutral hydrogen intensity mapping\til\cite{ska2020}. Indeed, it will allow us to bound the dark energy equation of state with an accuracy of $\sim 0.34\%$\til\cite{Xiao2022}.}
	
	\section*{Acknowledgments}
	{ MC, DV, and SC acknowledge the support of Istituto Nazionale di Fisica Nucleare (INFN) {\it iniziative specifiche} MOONLIGHT2, QGSKY, and TEONGRAV.
		IDM acknowledges support from Grant IJCI2018-036198-I  funded by MCIN/AEI/10.13039/501100011033 and, as appropriate, by “ESF Investing in your future” or by “European Union NextGenerationEU/PRTR”. IDM and RDM also acknowledge support from the  grant PID2021-122938NB-I00 funded by MCIN/AEI/10.13039/501100011033 and by “ERDF A way of making Europe”. 
		DV also acknowledges the FCT project with ref. number
		PTDC/FIS-AST/0054/2021.}

	\bibliographystyle{apsrev4-2}
	\bibliography{Biblio.bib}

\providecommand{\noopsort}[1]{}\providecommand{\singleletter}[1]{#1}%
\begin{thebibliography}{142}%
\makeatletter
\providecommand \@ifxundefined [1]{%
 \@ifx{#1\undefined}
}%
\providecommand \@ifnum [1]{%
 \ifnum #1\expandafter \@firstoftwo
 \else \expandafter \@secondoftwo
 \fi
}%
\providecommand \@ifx [1]{%
 \ifx #1\expandafter \@firstoftwo
 \else \expandafter \@secondoftwo
 \fi
}%
\providecommand \natexlab [1]{#1}%
\providecommand \enquote  [1]{``#1''}%
\providecommand \bibnamefont  [1]{#1}%
\providecommand \bibfnamefont [1]{#1}%
\providecommand \citenamefont [1]{#1}%
\providecommand \href@noop [0]{\@secondoftwo}%
\providecommand \href [0]{\begingroup \@sanitize@url \@href}%
\providecommand \@href[1]{\@@startlink{#1}\@@href}%
\providecommand \@@href[1]{\endgroup#1\@@endlink}%
\providecommand \@sanitize@url [0]{\catcode `\\12\catcode `\$12\catcode
  `\&12\catcode `\#12\catcode `\^12\catcode `\_12\catcode `\%12\relax}%
\providecommand \@@startlink[1]{}%
\providecommand \@@endlink[0]{}%
\providecommand \url  [0]{\begingroup\@sanitize@url \@url }%
\providecommand \@url [1]{\endgroup\@href {#1}{\urlprefix }}%
\providecommand \urlprefix  [0]{URL }%
\providecommand \Eprint [0]{\href }%
\providecommand \doibase [0]{https://doi.org/}%
\providecommand \selectlanguage [0]{\@gobble}%
\providecommand \bibinfo  [0]{\@secondoftwo}%
\providecommand \bibfield  [0]{\@secondoftwo}%
\providecommand \translation [1]{[#1]}%
\providecommand \BibitemOpen [0]{}%
\providecommand \bibitemStop [0]{}%
\providecommand \bibitemNoStop [0]{.\EOS\space}%
\providecommand \EOS [0]{\spacefactor3000\relax}%
\providecommand \BibitemShut  [1]{\csname bibitem#1\endcsname}%
\let\auto@bib@innerbib\@empty
%</preamble>
\bibitem [{\citenamefont {{LIGO Scientific Collaboration}}\ and\ \citenamefont
  {{Virgo Collaboration}}(2016{\natexlab{a}})}]{Abbott2016}%
  \BibitemOpen
  \bibfield  {author} {\bibinfo {author} {\bibnamefont {{LIGO Scientific
  Collaboration}}}\ and\ \bibinfo {author} {\bibnamefont {{Virgo
  Collaboration}}},\ }\href {https://doi.org/10.1103/PhysRevLett.116.061102}
  {\bibfield  {journal} {\bibinfo  {journal} {\prl}\ }\textbf {\bibinfo
  {volume} {116}},\ \bibinfo {eid} {061102} (\bibinfo {year}
  {2016}{\natexlab{a}})},\ \Eprint {https://arxiv.org/abs/1602.03837}
  {arXiv:1602.03837 [gr-qc]} \BibitemShut {NoStop}%
\bibitem [{\citenamefont {{LIGO Scientific Collaboration}}\ \emph
  {et~al.}(2017)\citenamefont {{LIGO Scientific Collaboration}}, \citenamefont
  {{Virgo Collaboration}}, \citenamefont {{IceCube Collaboration}},
  \citenamefont {{ANTARES Collaboration}}, \citenamefont {{Swift
  Collaboration}}, \citenamefont {{Dark Energy Camera GW-EM Collaboration}},
  \citenamefont {{DES Collaboration}}, \citenamefont {{DLT40 Collaboration}},
  \citenamefont {{CAASTRO Collaborations}}, \citenamefont {{VINROUGE
  Collaboration}}, \citenamefont {{BOOTES Collaboration}}, \citenamefont
  {{CALET Collaboration}}, \citenamefont {{IKI-GW Follow-up Collaboration}},
  \citenamefont {{LOFAR Collaboration}}, \citenamefont {{HAWC Collaboration}},
  \citenamefont {{Pierre Auger Collaboration}}, \citenamefont {{ALMA
  Collaboration}}, \citenamefont {{Pi of Sky Collaboration}},\ and\
  \citenamefont {{South Africa/MeerKAT}}}]{GW170817}%
  \BibitemOpen
  \bibfield  {author} {\bibinfo {author} {\bibnamefont {{LIGO Scientific
  Collaboration}}}, \bibinfo {author} {\bibnamefont {{Virgo Collaboration}}},
  \bibinfo {author} {\bibnamefont {{IceCube Collaboration}}}, \bibinfo {author}
  {\bibnamefont {{ANTARES Collaboration}}}, \bibinfo {author} {\bibnamefont
  {{Swift Collaboration}}}, \bibinfo {author} {\bibnamefont {{Dark Energy
  Camera GW-EM Collaboration}}}, \bibinfo {author} {\bibnamefont {{DES
  Collaboration}}}, \bibinfo {author} {\bibnamefont {{DLT40 Collaboration}}},
  \bibinfo {author} {\bibnamefont {{CAASTRO Collaborations}}}, \bibinfo
  {author} {\bibnamefont {{VINROUGE Collaboration}}}, \bibinfo {author}
  {\bibnamefont {{BOOTES Collaboration}}}, \bibinfo {author} {\bibnamefont
  {{CALET Collaboration}}}, \bibinfo {author} {\bibnamefont {{IKI-GW Follow-up
  Collaboration}}}, \bibinfo {author} {\bibnamefont {{LOFAR Collaboration}}},
  \bibinfo {author} {\bibnamefont {{HAWC Collaboration}}}, \bibinfo {author}
  {\bibnamefont {{Pierre Auger Collaboration}}}, \bibinfo {author}
  {\bibnamefont {{ALMA Collaboration}}}, \bibinfo {author} {\bibnamefont {{Pi
  of Sky Collaboration}}},\ and\ \bibinfo {author} {\bibfnamefont
  {S.}~\bibnamefont {{South Africa/MeerKAT}}},\ }\href
  {https://doi.org/10.3847/2041-8213/aa91c9} {\bibfield  {journal} {\bibinfo
  {journal} {\apjl}\ }\textbf {\bibinfo {volume} {848}},\ \bibinfo {eid} {L12}
  (\bibinfo {year} {2017})},\ \Eprint {https://arxiv.org/abs/1710.05833}
  {arXiv:1710.05833 [astro-ph.HE]} \BibitemShut {NoStop}%
\bibitem [{\citenamefont {{LIGO Scientific Collaboration}}\ and\ \citenamefont
  {{Virgo Collaboration}}(2016{\natexlab{b}})}]{Abbott2016b}%
  \BibitemOpen
  \bibfield  {author} {\bibinfo {author} {\bibnamefont {{LIGO Scientific
  Collaboration}}}\ and\ \bibinfo {author} {\bibnamefont {{Virgo
  Collaboration}}},\ }\href {https://doi.org/10.1103/PhysRevLett.116.131103}
  {\bibfield  {journal} {\bibinfo  {journal} {\prl}\ }\textbf {\bibinfo
  {volume} {116}},\ \bibinfo {eid} {131103} (\bibinfo {year}
  {2016}{\natexlab{b}})},\ \Eprint {https://arxiv.org/abs/1602.03838}
  {arXiv:1602.03838 [gr-qc]} \BibitemShut {NoStop}%
\bibitem [{\citenamefont {{Ezquiaga}}\ and\ \citenamefont
  {{Zumalac{\'a}rregui}}(2017)}]{Ezquiaga2017}%
  \BibitemOpen
  \bibfield  {author} {\bibinfo {author} {\bibfnamefont {J.~M.}\ \bibnamefont
  {{Ezquiaga}}}\ and\ \bibinfo {author} {\bibfnamefont {M.}~\bibnamefont
  {{Zumalac{\'a}rregui}}},\ }\href
  {https://doi.org/10.1103/PhysRevLett.119.251304} {\bibfield  {journal}
  {\bibinfo  {journal} {\prl}\ }\textbf {\bibinfo {volume} {119}},\ \bibinfo
  {eid} {251304} (\bibinfo {year} {2017})},\ \Eprint
  {https://arxiv.org/abs/1710.05901} {arXiv:1710.05901 [astro-ph.CO]}
  \BibitemShut {NoStop}%
\bibitem [{\citenamefont {{Schutz}}(1986)}]{Schutz1986}%
  \BibitemOpen
  \bibfield  {author} {\bibinfo {author} {\bibfnamefont {B.~F.}\ \bibnamefont
  {{Schutz}}},\ }\href {https://doi.org/10.1038/323310a0} {\bibfield  {journal}
  {\bibinfo  {journal} {\nat}\ }\textbf {\bibinfo {volume} {323}},\ \bibinfo
  {pages} {310} (\bibinfo {year} {1986})}\BibitemShut {NoStop}%
\bibitem [{\citenamefont {{Holz}}\ and\ \citenamefont
  {{Hughes}}(2005)}]{Holz2005}%
  \BibitemOpen
  \bibfield  {author} {\bibinfo {author} {\bibfnamefont {D.~E.}\ \bibnamefont
  {{Holz}}}\ and\ \bibinfo {author} {\bibfnamefont {S.~A.}\ \bibnamefont
  {{Hughes}}},\ }\href {https://doi.org/10.1086/431341} {\bibfield  {journal}
  {\bibinfo  {journal} {\apj}\ }\textbf {\bibinfo {volume} {629}},\ \bibinfo
  {pages} {15} (\bibinfo {year} {2005})},\ \Eprint
  {https://arxiv.org/abs/astro-ph/0504616} {arXiv:astro-ph/0504616 [astro-ph]}
  \BibitemShut {NoStop}%
\bibitem [{\citenamefont {Bogdanos}\ \emph {et~al.}(2010)\citenamefont
  {Bogdanos}, \citenamefont {Capozziello}, \citenamefont {De~Laurentis},\ and\
  \citenamefont {Nesseris}}]{Bogdanos:2009tn}%
  \BibitemOpen
  \bibfield  {author} {\bibinfo {author} {\bibfnamefont {C.}~\bibnamefont
  {Bogdanos}}, \bibinfo {author} {\bibfnamefont {S.}~\bibnamefont
  {Capozziello}}, \bibinfo {author} {\bibfnamefont {M.}~\bibnamefont
  {De~Laurentis}},\ and\ \bibinfo {author} {\bibfnamefont {S.}~\bibnamefont
  {Nesseris}},\ }\href {https://doi.org/10.1016/j.astropartphys.2010.08.001}
  {\bibfield  {journal} {\bibinfo  {journal} {Astropart. Phys.}\ }\textbf
  {\bibinfo {volume} {34}},\ \bibinfo {pages} {236} (\bibinfo {year} {2010})},\
  \Eprint {https://arxiv.org/abs/0911.3094} {arXiv:0911.3094 [gr-qc]}
  \BibitemShut {NoStop}%
\bibitem [{\citenamefont {Oikonomou}(2022)}]{Oikonomou:2022xoq}%
  \BibitemOpen
  \bibfield  {author} {\bibinfo {author} {\bibfnamefont {V.~K.}\ \bibnamefont
  {Oikonomou}},\ }\href {https://doi.org/10.1016/j.astropartphys.2022.102718}
  {\bibfield  {journal} {\bibinfo  {journal} {Astropart. Phys.}\ }\textbf
  {\bibinfo {volume} {141}},\ \bibinfo {pages} {102718} (\bibinfo {year}
  {2022})},\ \Eprint {https://arxiv.org/abs/2204.06304} {arXiv:2204.06304
  [gr-qc]} \BibitemShut {NoStop}%
\bibitem [{\citenamefont {Odintsov}\ \emph {et~al.}(2022)\citenamefont
  {Odintsov}, \citenamefont {Oikonomou},\ and\ \citenamefont
  {Myrzakulov}}]{Odintsov:2022cbm}%
  \BibitemOpen
  \bibfield  {author} {\bibinfo {author} {\bibfnamefont {S.~D.}\ \bibnamefont
  {Odintsov}}, \bibinfo {author} {\bibfnamefont {V.~K.}\ \bibnamefont
  {Oikonomou}},\ and\ \bibinfo {author} {\bibfnamefont {R.}~\bibnamefont
  {Myrzakulov}},\ }\href {https://doi.org/10.3390/sym14040729} {\bibfield
  {journal} {\bibinfo  {journal} {Symmetry}\ }\textbf {\bibinfo {volume}
  {14}},\ \bibinfo {pages} {729} (\bibinfo {year} {2022})},\ \Eprint
  {https://arxiv.org/abs/2204.00876} {arXiv:2204.00876 [gr-qc]} \BibitemShut
  {NoStop}%
\bibitem [{\citenamefont {{Chen}}\ \emph {et~al.}(2018)\citenamefont {{Chen}},
  \citenamefont {{Fishbach}},\ and\ \citenamefont {{Holz}}}]{Chen2018}%
  \BibitemOpen
  \bibfield  {author} {\bibinfo {author} {\bibfnamefont {H.-Y.}\ \bibnamefont
  {{Chen}}}, \bibinfo {author} {\bibfnamefont {M.}~\bibnamefont {{Fishbach}}},\
  and\ \bibinfo {author} {\bibfnamefont {D.~E.}\ \bibnamefont {{Holz}}},\
  }\href {https://doi.org/10.1038/s41586-018-0606-0} {\bibfield  {journal}
  {\bibinfo  {journal} {\nat}\ }\textbf {\bibinfo {volume} {562}},\ \bibinfo
  {pages} {545} (\bibinfo {year} {2018})},\ \Eprint
  {https://arxiv.org/abs/1712.06531} {arXiv:1712.06531 [astro-ph.CO]}
  \BibitemShut {NoStop}%
\bibitem [{\citenamefont {{Capozziello}}\ \emph {et~al.}(2011)\citenamefont
  {{Capozziello}}, \citenamefont {{de Laurentis}}, \citenamefont {{de
  Martino}},\ and\ \citenamefont {{Formisano}}}]{Capozziello2011}%
  \BibitemOpen
  \bibfield  {author} {\bibinfo {author} {\bibfnamefont {S.}~\bibnamefont
  {{Capozziello}}}, \bibinfo {author} {\bibfnamefont {M.}~\bibnamefont {{de
  Laurentis}}}, \bibinfo {author} {\bibfnamefont {I.}~\bibnamefont {{de
  Martino}}},\ and\ \bibinfo {author} {\bibfnamefont {M.}~\bibnamefont
  {{Formisano}}},\ }\href {https://doi.org/10.1007/s10509-010-0471-2}
  {\bibfield  {journal} {\bibinfo  {journal} {\apss}\ }\textbf {\bibinfo
  {volume} {332}},\ \bibinfo {pages} {31} (\bibinfo {year} {2011})},\ \Eprint
  {https://arxiv.org/abs/1004.4818} {arXiv:1004.4818 [astro-ph.CO]}
  \BibitemShut {NoStop}%
\bibitem [{\citenamefont {{Gray}}\ \emph {et~al.}(2020)\citenamefont {{Gray}},
  \citenamefont {{Hernandez}}, \citenamefont {{Qi}}, \citenamefont {{Sur}},
  \citenamefont {{Brady}}, \citenamefont {{Chen}}, \citenamefont {{Farr}},
  \citenamefont {{Fishbach}}, \citenamefont {{Gair}}, \citenamefont {{Ghosh}},
  \citenamefont {{Holz}}, \citenamefont {{Mastrogiovanni}}, \citenamefont
  {{Messenger}}, \citenamefont {{Steer}},\ and\ \citenamefont
  {{Veitch}}}]{Gray2020}%
  \BibitemOpen
  \bibfield  {author} {\bibinfo {author} {\bibfnamefont {R.}~\bibnamefont
  {{Gray}}}, \bibinfo {author} {\bibfnamefont {I.~M.}\ \bibnamefont
  {{Hernandez}}}, \bibinfo {author} {\bibfnamefont {H.}~\bibnamefont {{Qi}}},
  \bibinfo {author} {\bibfnamefont {A.}~\bibnamefont {{Sur}}}, \bibinfo
  {author} {\bibfnamefont {P.~R.}\ \bibnamefont {{Brady}}}, \bibinfo {author}
  {\bibfnamefont {H.-Y.}\ \bibnamefont {{Chen}}}, \bibinfo {author}
  {\bibfnamefont {W.~M.}\ \bibnamefont {{Farr}}}, \bibinfo {author}
  {\bibfnamefont {M.}~\bibnamefont {{Fishbach}}}, \bibinfo {author}
  {\bibfnamefont {J.~R.}\ \bibnamefont {{Gair}}}, \bibinfo {author}
  {\bibfnamefont {A.}~\bibnamefont {{Ghosh}}}, \bibinfo {author} {\bibfnamefont
  {D.~E.}\ \bibnamefont {{Holz}}}, \bibinfo {author} {\bibfnamefont
  {S.}~\bibnamefont {{Mastrogiovanni}}}, \bibinfo {author} {\bibfnamefont
  {C.}~\bibnamefont {{Messenger}}}, \bibinfo {author} {\bibfnamefont {D.~A.}\
  \bibnamefont {{Steer}}},\ and\ \bibinfo {author} {\bibfnamefont
  {J.}~\bibnamefont {{Veitch}}},\ }\href
  {https://doi.org/10.1103/PhysRevD.101.122001} {\bibfield  {journal} {\bibinfo
   {journal} {\prd}\ }\textbf {\bibinfo {volume} {101}},\ \bibinfo {eid}
  {122001} (\bibinfo {year} {2020})},\ \Eprint
  {https://arxiv.org/abs/1908.06050} {arXiv:1908.06050 [gr-qc]} \BibitemShut
  {NoStop}%
\bibitem [{\citenamefont {{Chase}}\ \emph {et~al.}(2022)\citenamefont
  {{Chase}}, \citenamefont {{O'Connor}}, \citenamefont {{Fryer}}, \citenamefont
  {{Troja}}, \citenamefont {{Korobkin}}, \citenamefont {{Wollaeger}},
  \citenamefont {{Ristic}}, \citenamefont {{Fontes}}, \citenamefont
  {{Hungerford}},\ and\ \citenamefont {{Herring}}}]{Chase2022}%
  \BibitemOpen
  \bibfield  {author} {\bibinfo {author} {\bibfnamefont {E.~A.}\ \bibnamefont
  {{Chase}}}, \bibinfo {author} {\bibfnamefont {B.}~\bibnamefont {{O'Connor}}},
  \bibinfo {author} {\bibfnamefont {C.~L.}\ \bibnamefont {{Fryer}}}, \bibinfo
  {author} {\bibfnamefont {E.}~\bibnamefont {{Troja}}}, \bibinfo {author}
  {\bibfnamefont {O.}~\bibnamefont {{Korobkin}}}, \bibinfo {author}
  {\bibfnamefont {R.~T.}\ \bibnamefont {{Wollaeger}}}, \bibinfo {author}
  {\bibfnamefont {M.}~\bibnamefont {{Ristic}}}, \bibinfo {author}
  {\bibfnamefont {C.~J.}\ \bibnamefont {{Fontes}}}, \bibinfo {author}
  {\bibfnamefont {A.~L.}\ \bibnamefont {{Hungerford}}},\ and\ \bibinfo {author}
  {\bibfnamefont {A.~M.}\ \bibnamefont {{Herring}}},\ }\href
  {https://doi.org/10.3847/1538-4357/ac3d25} {\bibfield  {journal} {\bibinfo
  {journal} {\apj}\ }\textbf {\bibinfo {volume} {927}},\ \bibinfo {eid} {163}
  (\bibinfo {year} {2022})},\ \Eprint {https://arxiv.org/abs/2105.12268}
  {arXiv:2105.12268 [astro-ph.HE]} \BibitemShut {NoStop}%
\bibitem [{\citenamefont {{Stratta}}\ \emph {et~al.}(2018)\citenamefont
  {{Stratta}} \emph {et~al.}}]{THESEUS:2017wvz}%
  \BibitemOpen
  \bibfield  {author} {\bibinfo {author} {\bibfnamefont {G.}~\bibnamefont
  {{Stratta}}} \emph {et~al.},\ }\href
  {https://doi.org/10.1016/j.asr.2018.04.013} {\bibfield  {journal} {\bibinfo
  {journal} {Advances in Space Research}\ }\textbf {\bibinfo {volume} {62}},\
  \bibinfo {pages} {662} (\bibinfo {year} {2018})},\ \Eprint
  {https://arxiv.org/abs/1712.08153} {arXiv:1712.08153 [astro-ph.HE]}
  \BibitemShut {NoStop}%
\bibitem [{\citenamefont {{Theseus Consortium}}(2021)}]{Amati2021}%
  \BibitemOpen
  \bibfield  {author} {\bibinfo {author} {\bibnamefont {{Theseus
  Consortium}}},\ }\href {https://doi.org/10.1007/s10686-021-09807-8}
  {\bibfield  {journal} {\bibinfo  {journal} {Experimental Astronomy}\ }\textbf
  {\bibinfo {volume} {52}},\ \bibinfo {pages} {183} (\bibinfo {year} {2021})},\
  \Eprint {https://arxiv.org/abs/2104.09531} {arXiv:2104.09531 [astro-ph.IM]}
  \BibitemShut {NoStop}%
\bibitem [{\citenamefont {{Stratta}}\ \emph {et~al.}(2022)\citenamefont
  {{Stratta}}, \citenamefont {{Amati}}, \citenamefont {{Branchesi}},
  \citenamefont {{Ciolfi}}, \citenamefont {{Tanvir}}, \citenamefont {{Bozzo}},
  \citenamefont {{G{\"o}tz}}, \citenamefont {{O'Brien}},\ and\ \citenamefont
  {{Santangelo}}}]{Stratta2022}%
  \BibitemOpen
  \bibfield  {author} {\bibinfo {author} {\bibfnamefont {G.}~\bibnamefont
  {{Stratta}}}, \bibinfo {author} {\bibfnamefont {L.}~\bibnamefont {{Amati}}},
  \bibinfo {author} {\bibfnamefont {M.}~\bibnamefont {{Branchesi}}}, \bibinfo
  {author} {\bibfnamefont {R.}~\bibnamefont {{Ciolfi}}}, \bibinfo {author}
  {\bibfnamefont {N.}~\bibnamefont {{Tanvir}}}, \bibinfo {author}
  {\bibfnamefont {E.}~\bibnamefont {{Bozzo}}}, \bibinfo {author} {\bibfnamefont
  {D.}~\bibnamefont {{G{\"o}tz}}}, \bibinfo {author} {\bibfnamefont
  {P.}~\bibnamefont {{O'Brien}}},\ and\ \bibinfo {author} {\bibfnamefont
  {A.}~\bibnamefont {{Santangelo}}},\ }\href
  {https://doi.org/10.3390/galaxies10030060} {\bibfield  {journal} {\bibinfo
  {journal} {Galaxies}\ }\textbf {\bibinfo {volume} {10}},\ \bibinfo {pages}
  {60} (\bibinfo {year} {2022})}\BibitemShut {NoStop}%
\bibitem [{\citenamefont {{Rosati}}\ \emph {et~al.}(2021)\citenamefont
  {{Rosati}} \emph {et~al.}}]{Rosati2021}%
  \BibitemOpen
  \bibfield  {author} {\bibinfo {author} {\bibfnamefont {P.}~\bibnamefont
  {{Rosati}}} \emph {et~al.},\ }\href
  {https://doi.org/10.1007/s10686-021-09764-2} {\bibfield  {journal} {\bibinfo
  {journal} {Experimental Astronomy}\ }\textbf {\bibinfo {volume} {52}},\
  \bibinfo {pages} {407} (\bibinfo {year} {2021})},\ \Eprint
  {https://arxiv.org/abs/2104.09535} {arXiv:2104.09535 [astro-ph.IM]}
  \BibitemShut {NoStop}%
\bibitem [{\citenamefont {{Tanvir}}\ \emph {et~al.}(2021)\citenamefont
  {{Tanvir}} \emph {et~al.}}]{Tanvir2021}%
  \BibitemOpen
  \bibfield  {author} {\bibinfo {author} {\bibfnamefont {N.~R.}\ \bibnamefont
  {{Tanvir}}} \emph {et~al.},\ }\href
  {https://doi.org/10.1007/s10686-021-09778-w} {\bibfield  {journal} {\bibinfo
  {journal} {Experimental Astronomy}\ }\textbf {\bibinfo {volume} {52}},\
  \bibinfo {pages} {219} (\bibinfo {year} {2021})},\ \Eprint
  {https://arxiv.org/abs/2104.09532} {arXiv:2104.09532 [astro-ph.IM]}
  \BibitemShut {NoStop}%
\bibitem [{\citenamefont {{Dainotti}}\ \emph {et~al.}(2021)\citenamefont
  {{Dainotti}}, \citenamefont {{De Simone}}, \citenamefont {{Schiavone}},
  \citenamefont {{Montani}}, \citenamefont {{Rinaldi}},\ and\ \citenamefont
  {{Lambiase}}}]{Dainotti2021}%
  \BibitemOpen
  \bibfield  {author} {\bibinfo {author} {\bibfnamefont {M.~G.}\ \bibnamefont
  {{Dainotti}}}, \bibinfo {author} {\bibfnamefont {B.}~\bibnamefont {{De
  Simone}}}, \bibinfo {author} {\bibfnamefont {T.}~\bibnamefont {{Schiavone}}},
  \bibinfo {author} {\bibfnamefont {G.}~\bibnamefont {{Montani}}}, \bibinfo
  {author} {\bibfnamefont {E.}~\bibnamefont {{Rinaldi}}},\ and\ \bibinfo
  {author} {\bibfnamefont {G.}~\bibnamefont {{Lambiase}}},\ }\href
  {https://doi.org/10.3847/1538-4357/abeb73} {\bibfield  {journal} {\bibinfo
  {journal} {\apj}\ }\textbf {\bibinfo {volume} {912}},\ \bibinfo {eid} {150}
  (\bibinfo {year} {2021})},\ \Eprint {https://arxiv.org/abs/2103.02117}
  {arXiv:2103.02117 [astro-ph.CO]} \BibitemShut {NoStop}%
\bibitem [{\citenamefont {{Dainotti}}\ \emph {et~al.}(2022)\citenamefont
  {{Dainotti}}, \citenamefont {{De Simone}}, \citenamefont {{Schiavone}},
  \citenamefont {{Montani}}, \citenamefont {{Rinaldi}}, \citenamefont
  {{Lambiase}}, \citenamefont {{Bogdan}},\ and\ \citenamefont
  {{Ugale}}}]{Dainotti2022}%
  \BibitemOpen
  \bibfield  {author} {\bibinfo {author} {\bibfnamefont {M.~G.}\ \bibnamefont
  {{Dainotti}}}, \bibinfo {author} {\bibfnamefont {B.~D.}\ \bibnamefont {{De
  Simone}}}, \bibinfo {author} {\bibfnamefont {T.}~\bibnamefont {{Schiavone}}},
  \bibinfo {author} {\bibfnamefont {G.}~\bibnamefont {{Montani}}}, \bibinfo
  {author} {\bibfnamefont {E.}~\bibnamefont {{Rinaldi}}}, \bibinfo {author}
  {\bibfnamefont {G.}~\bibnamefont {{Lambiase}}}, \bibinfo {author}
  {\bibfnamefont {M.}~\bibnamefont {{Bogdan}}},\ and\ \bibinfo {author}
  {\bibfnamefont {S.}~\bibnamefont {{Ugale}}},\ }\href
  {https://doi.org/10.3390/galaxies10010024} {\bibfield  {journal} {\bibinfo
  {journal} {Galaxies}\ }\textbf {\bibinfo {volume} {10}},\ \bibinfo {pages}
  {24} (\bibinfo {year} {2022})},\ \Eprint {https://arxiv.org/abs/2201.09848}
  {arXiv:2201.09848 [astro-ph.CO]} \BibitemShut {NoStop}%
\bibitem [{\citenamefont {{Messenger}}\ and\ \citenamefont
  {{Read}}(2012)}]{messenger:Read}%
  \BibitemOpen
  \bibfield  {author} {\bibinfo {author} {\bibfnamefont {C.}~\bibnamefont
  {{Messenger}}}\ and\ \bibinfo {author} {\bibfnamefont {J.}~\bibnamefont
  {{Read}}},\ }\href {https://doi.org/10.1103/PhysRevLett.108.091101}
  {\bibfield  {journal} {\bibinfo  {journal} {\prl}\ }\textbf {\bibinfo
  {volume} {108}},\ \bibinfo {eid} {091101} (\bibinfo {year} {2012})},\ \Eprint
  {https://arxiv.org/abs/1107.5725} {arXiv:1107.5725 [gr-qc]} \BibitemShut
  {NoStop}%
\bibitem [{\citenamefont {{Chatterjee}}\ \emph {et~al.}(2021)\citenamefont
  {{Chatterjee}}, \citenamefont {{Hegade K.~R.}}, \citenamefont {{Holder}},
  \citenamefont {{Holz}}, \citenamefont {{Perkins}}, \citenamefont {{Yagi}},\
  and\ \citenamefont {{Yunes}}}]{Chatterjee2021}%
  \BibitemOpen
  \bibfield  {author} {\bibinfo {author} {\bibfnamefont {D.}~\bibnamefont
  {{Chatterjee}}}, \bibinfo {author} {\bibfnamefont {A.}~\bibnamefont {{Hegade
  K.~R.}}}, \bibinfo {author} {\bibfnamefont {G.}~\bibnamefont {{Holder}}},
  \bibinfo {author} {\bibfnamefont {D.~E.}\ \bibnamefont {{Holz}}}, \bibinfo
  {author} {\bibfnamefont {S.}~\bibnamefont {{Perkins}}}, \bibinfo {author}
  {\bibfnamefont {K.}~\bibnamefont {{Yagi}}},\ and\ \bibinfo {author}
  {\bibfnamefont {N.}~\bibnamefont {{Yunes}}},\ }\href
  {https://doi.org/10.1103/PhysRevD.104.083528} {\bibfield  {journal} {\bibinfo
   {journal} {\prd}\ }\textbf {\bibinfo {volume} {104}},\ \bibinfo {eid}
  {083528} (\bibinfo {year} {2021})},\ \Eprint
  {https://arxiv.org/abs/2106.06589} {arXiv:2106.06589 [gr-qc]} \BibitemShut
  {NoStop}%
\bibitem [{\citenamefont {{The LIGO Scientific Collaboration}}\ \emph
  {et~al.}(2017)\citenamefont {{The LIGO Scientific Collaboration}},
  \citenamefont {{the Virgo Collaboration}}, \citenamefont {{the Dark Energy
  Camera GW-EM Collaboration}}, \citenamefont {{the DES Collaboration}},
  \citenamefont {{the 1M2H Collaboration}}, \citenamefont {{the DLT40
  Collaboration}}, \citenamefont {{the Las Cumbres Observatory Collaboration}},
  \citenamefont {{the Vinrouge Collaboration}},\ and\ \citenamefont {{the
  Master Collaboration}}}]{LIGO_H0_2017}%
  \BibitemOpen
  \bibfield  {author} {\bibinfo {author} {\bibnamefont {{The LIGO Scientific
  Collaboration}}}, \bibinfo {author} {\bibnamefont {{the Virgo
  Collaboration}}}, \bibinfo {author} {\bibnamefont {{the Dark Energy Camera
  GW-EM Collaboration}}}, \bibinfo {author} {\bibnamefont {{the DES
  Collaboration}}}, \bibinfo {author} {\bibnamefont {{the 1M2H
  Collaboration}}}, \bibinfo {author} {\bibnamefont {{the DLT40
  Collaboration}}}, \bibinfo {author} {\bibnamefont {{the Las Cumbres
  Observatory Collaboration}}}, \bibinfo {author} {\bibnamefont {{the Vinrouge
  Collaboration}}},\ and\ \bibinfo {author} {\bibnamefont {{the Master
  Collaboration}}},\ }\href {https://doi.org/10.1038/nature24471} {\bibfield
  {journal} {\bibinfo  {journal} {\nat}\ }\textbf {\bibinfo {volume} {551}},\
  \bibinfo {pages} {85} (\bibinfo {year} {2017})},\ \Eprint
  {https://arxiv.org/abs/1710.05835} {arXiv:1710.05835 [astro-ph.CO]}
  \BibitemShut {NoStop}%
\bibitem [{\citenamefont {{The LIGO Scientific Collaboration}}\ \emph
  {et~al.}(2021{\natexlab{a}})\citenamefont {{The LIGO Scientific
  Collaboration}}, \citenamefont {{the Virgo Collaboration}},\ and\
  \citenamefont {{the KAGRA Collaboration}}}]{LVK_H0_2021}%
  \BibitemOpen
  \bibfield  {author} {\bibinfo {author} {\bibnamefont {{The LIGO Scientific
  Collaboration}}}, \bibinfo {author} {\bibnamefont {{the Virgo
  Collaboration}}},\ and\ \bibinfo {author} {\bibnamefont {{the KAGRA
  Collaboration}}},\ }\href@noop {} {\  (\bibinfo {year}
  {2021}{\natexlab{a}})},\ \Eprint {https://arxiv.org/abs/2111.03604}
  {arXiv:2111.03604 [astro-ph.CO]} \BibitemShut {NoStop}%
\bibitem [{\citenamefont {{Di Valentino}}\ \emph {et~al.}(2021)\citenamefont
  {{Di Valentino}}, \citenamefont {{Mena}}, \citenamefont {{Pan}},
  \citenamefont {{Visinelli}}, \citenamefont {{Yang}}, \citenamefont
  {{Melchiorri}}, \citenamefont {{Mota}}, \citenamefont {{Riess}},\ and\
  \citenamefont {{Silk}}}]{DiValentino2021}%
  \BibitemOpen
  \bibfield  {author} {\bibinfo {author} {\bibfnamefont {E.}~\bibnamefont {{Di
  Valentino}}}, \bibinfo {author} {\bibfnamefont {O.}~\bibnamefont {{Mena}}},
  \bibinfo {author} {\bibfnamefont {S.}~\bibnamefont {{Pan}}}, \bibinfo
  {author} {\bibfnamefont {L.}~\bibnamefont {{Visinelli}}}, \bibinfo {author}
  {\bibfnamefont {W.}~\bibnamefont {{Yang}}}, \bibinfo {author} {\bibfnamefont
  {A.}~\bibnamefont {{Melchiorri}}}, \bibinfo {author} {\bibfnamefont {D.~F.}\
  \bibnamefont {{Mota}}}, \bibinfo {author} {\bibfnamefont {A.~G.}\
  \bibnamefont {{Riess}}},\ and\ \bibinfo {author} {\bibfnamefont
  {J.}~\bibnamefont {{Silk}}},\ }\href
  {https://doi.org/10.1088/1361-6382/ac086d} {\bibfield  {journal} {\bibinfo
  {journal} {Classical and Quantum Gravity}\ }\textbf {\bibinfo {volume}
  {38}},\ \bibinfo {eid} {153001} (\bibinfo {year} {2021})},\ \Eprint
  {https://arxiv.org/abs/2103.01183} {arXiv:2103.01183 [astro-ph.CO]}
  \BibitemShut {NoStop}%
\bibitem [{\citenamefont {{Abdalla}}\ \emph {et~al.}(2022)\citenamefont
  {{Abdalla}} \emph {et~al.}}]{Abdalla2022}%
  \BibitemOpen
  \bibfield  {author} {\bibinfo {author} {\bibfnamefont {E.}~\bibnamefont
  {{Abdalla}}} \emph {et~al.},\ }\href
  {https://doi.org/10.1016/j.jheap.2022.04.002} {\bibfield  {journal} {\bibinfo
   {journal} {Journal of High Energy Astrophysics}\ }\textbf {\bibinfo {volume}
  {34}},\ \bibinfo {pages} {49} (\bibinfo {year} {2022})},\ \Eprint
  {https://arxiv.org/abs/2203.06142} {arXiv:2203.06142 [astro-ph.CO]}
  \BibitemShut {NoStop}%
\bibitem [{\citenamefont {{Krishnan}}\ \emph {et~al.}(2021)\citenamefont
  {{Krishnan}}, \citenamefont {{{\'O} Colg{\'a}in}}, \citenamefont
  {{Sheikh-Jabbari}},\ and\ \citenamefont {{Yang}}}]{Krishnan2021}%
  \BibitemOpen
  \bibfield  {author} {\bibinfo {author} {\bibfnamefont {C.}~\bibnamefont
  {{Krishnan}}}, \bibinfo {author} {\bibfnamefont {E.}~\bibnamefont {{{\'O}
  Colg{\'a}in}}}, \bibinfo {author} {\bibfnamefont {M.~M.}\ \bibnamefont
  {{Sheikh-Jabbari}}},\ and\ \bibinfo {author} {\bibfnamefont {T.}~\bibnamefont
  {{Yang}}},\ }\href {https://doi.org/10.1103/PhysRevD.103.103509} {\bibfield
  {journal} {\bibinfo  {journal} {\prd}\ }\textbf {\bibinfo {volume} {103}},\
  \bibinfo {eid} {103509} (\bibinfo {year} {2021})},\ \Eprint
  {https://arxiv.org/abs/2011.02858} {arXiv:2011.02858 [astro-ph.CO]}
  \BibitemShut {NoStop}%
\bibitem [{\citenamefont {{Colg{\'a}in}}\ \emph
  {et~al.}(2022{\natexlab{a}})\citenamefont {{Colg{\'a}in}}, \citenamefont
  {{Sheikh-Jabbari}}, \citenamefont {{Solomon}}, \citenamefont {{Bargiacchi}},
  \citenamefont {{Capozziello}}, \citenamefont {{Dainotti}},\ and\
  \citenamefont {{Stojkovic}}}]{Colgain2022}%
  \BibitemOpen
  \bibfield  {author} {\bibinfo {author} {\bibfnamefont {E.~{\'O}.}\
  \bibnamefont {{Colg{\'a}in}}}, \bibinfo {author} {\bibfnamefont {M.~M.}\
  \bibnamefont {{Sheikh-Jabbari}}}, \bibinfo {author} {\bibfnamefont
  {R.}~\bibnamefont {{Solomon}}}, \bibinfo {author} {\bibfnamefont
  {G.}~\bibnamefont {{Bargiacchi}}}, \bibinfo {author} {\bibfnamefont
  {S.}~\bibnamefont {{Capozziello}}}, \bibinfo {author} {\bibfnamefont {M.~G.}\
  \bibnamefont {{Dainotti}}},\ and\ \bibinfo {author} {\bibfnamefont
  {D.}~\bibnamefont {{Stojkovic}}},\ }\href@noop {} {\bibfield  {journal}
  {\bibinfo  {journal} {\prd}\ } (\bibinfo {year} {2022}{\natexlab{a}})},\
  \Eprint {https://arxiv.org/abs/2203.10558} {arXiv:2203.10558 [astro-ph.CO]}
  \BibitemShut {NoStop}%
\bibitem [{\citenamefont {{Colg{\'a}in}}\ \emph
  {et~al.}(2022{\natexlab{b}})\citenamefont {{Colg{\'a}in}}, \citenamefont
  {{Sheikh-Jabbari}}, \citenamefont {{Solomon}}, \citenamefont {{Dainotti}},\
  and\ \citenamefont {{Stojkovic}}}]{Colgain2022b}%
  \BibitemOpen
  \bibfield  {author} {\bibinfo {author} {\bibfnamefont {E.~{\'O}.}\
  \bibnamefont {{Colg{\'a}in}}}, \bibinfo {author} {\bibfnamefont {M.~M.}\
  \bibnamefont {{Sheikh-Jabbari}}}, \bibinfo {author} {\bibfnamefont
  {R.}~\bibnamefont {{Solomon}}}, \bibinfo {author} {\bibfnamefont {M.~G.}\
  \bibnamefont {{Dainotti}}},\ and\ \bibinfo {author} {\bibfnamefont
  {D.}~\bibnamefont {{Stojkovic}}},\ }\href@noop {} {\  (\bibinfo {year}
  {2022}{\natexlab{b}})},\ \Eprint {https://arxiv.org/abs/2206.11447}
  {arXiv:2206.11447 [astro-ph.CO]} \BibitemShut {NoStop}%
\bibitem [{\citenamefont {{L{\'o}pez-Corredoira}}(2022)}]{Lopez2022}%
  \BibitemOpen
  \bibfield  {author} {\bibinfo {author} {\bibfnamefont {M.}~\bibnamefont
  {{L{\'o}pez-Corredoira}}},\ }\href {https://doi.org/10.1093/mnras/stac2567}
  {\bibfield  {journal} {\bibinfo  {journal} {\mnras}\ }\textbf {\bibinfo
  {volume} {517}},\ \bibinfo {pages} {5805} (\bibinfo {year} {2022})},\ \Eprint
  {https://arxiv.org/abs/2210.07078} {arXiv:2210.07078 [astro-ph.CO]}
  \BibitemShut {NoStop}%
\bibitem [{\citenamefont {Capozziello}\ \emph {et~al.}(2020)\citenamefont
  {Capozziello}, \citenamefont {Benetti},\ and\ \citenamefont
  {Spallicci}}]{Capozziello:2020nyq}%
  \BibitemOpen
  \bibfield  {author} {\bibinfo {author} {\bibfnamefont {S.}~\bibnamefont
  {Capozziello}}, \bibinfo {author} {\bibfnamefont {M.}~\bibnamefont
  {Benetti}},\ and\ \bibinfo {author} {\bibfnamefont {A.~D. A.~M.}\
  \bibnamefont {Spallicci}},\ }\href
  {https://doi.org/10.1007/s10701-020-00356-2} {\bibfield  {journal} {\bibinfo
  {journal} {Found. Phys.}\ }\textbf {\bibinfo {volume} {50}},\ \bibinfo
  {pages} {893} (\bibinfo {year} {2020})},\ \Eprint
  {https://arxiv.org/abs/2007.00462} {arXiv:2007.00462 [gr-qc]} \BibitemShut
  {NoStop}%
\bibitem [{\citenamefont {Spallicci}\ \emph {et~al.}(2022)\citenamefont
  {Spallicci}, \citenamefont {Benetti},\ and\ \citenamefont
  {Capozziello}}]{Spallicci:2021kye}%
  \BibitemOpen
  \bibfield  {author} {\bibinfo {author} {\bibfnamefont {A.~D. A.~M.}\
  \bibnamefont {Spallicci}}, \bibinfo {author} {\bibfnamefont {M.}~\bibnamefont
  {Benetti}},\ and\ \bibinfo {author} {\bibfnamefont {S.}~\bibnamefont
  {Capozziello}},\ }\href {https://doi.org/10.1007/s10701-021-00531-z}
  {\bibfield  {journal} {\bibinfo  {journal} {Found. Phys.}\ }\textbf {\bibinfo
  {volume} {52}},\ \bibinfo {pages} {23} (\bibinfo {year} {2022})},\ \Eprint
  {https://arxiv.org/abs/2112.07359} {arXiv:2112.07359 [physics.gen-ph]}
  \BibitemShut {NoStop}%
\bibitem [{\citenamefont {{Capozziello}}\ \emph {et~al.}(2023)\citenamefont
  {{Capozziello}}, \citenamefont {{Sarracino}},\ and\ \citenamefont
  {{Spallicci}}}]{Capozziello2023PDU}%
  \BibitemOpen
  \bibfield  {author} {\bibinfo {author} {\bibfnamefont {S.}~\bibnamefont
  {{Capozziello}}}, \bibinfo {author} {\bibfnamefont {G.}~\bibnamefont
  {{Sarracino}}},\ and\ \bibinfo {author} {\bibfnamefont {A.~D.~A.~M.}\
  \bibnamefont {{Spallicci}}},\ }\href
  {https://doi.org/10.1016/j.dark.2023.101201} {\bibfield  {journal} {\bibinfo
  {journal} {Physics of the Dark Universe}\ }\textbf {\bibinfo {volume} {40}},\
  \bibinfo {eid} {101201} (\bibinfo {year} {2023})},\ \Eprint
  {https://arxiv.org/abs/2302.13671} {arXiv:2302.13671 [astro-ph.CO]}
  \BibitemShut {NoStop}%
\bibitem [{\citenamefont {{Maggiore}}\ \emph {et~al.}(2020)\citenamefont
  {{Maggiore}} \emph {et~al.}}]{Maggiore2020}%
  \BibitemOpen
  \bibfield  {author} {\bibinfo {author} {\bibfnamefont {M.}~\bibnamefont
  {{Maggiore}}} \emph {et~al.},\ }\href
  {https://doi.org/10.1088/1475-7516/2020/03/050} {\bibfield  {journal}
  {\bibinfo  {journal} {\jcap}\ }\textbf {\bibinfo {volume} {2020}},\ \bibinfo
  {eid} {050} (\bibinfo {year} {2020})},\ \Eprint
  {https://arxiv.org/abs/1912.02622} {arXiv:1912.02622 [astro-ph.CO]}
  \BibitemShut {NoStop}%
\bibitem [{\citenamefont {{Branchesi}}\ \emph {et~al.}(2023)\citenamefont
  {{Branchesi}} \emph {et~al.}}]{Branchesi2023}%
  \BibitemOpen
  \bibfield  {author} {\bibinfo {author} {\bibfnamefont {M.}~\bibnamefont
  {{Branchesi}}} \emph {et~al.},\ }\href
  {https://doi.org/10.48550/arXiv.2303.15923} {\bibfield  {journal} {\bibinfo
  {journal} {arXiv e-prints}\ ,\ \bibinfo {eid} {arXiv:2303.15923}} (\bibinfo
  {year} {2023})},\ \Eprint {https://arxiv.org/abs/2303.15923}
  {arXiv:2303.15923 [gr-qc]} \BibitemShut {NoStop}%
\bibitem [{\citenamefont {{Planck Collaboration}}(2020)}]{Planck2020}%
  \BibitemOpen
  \bibfield  {author} {\bibinfo {author} {\bibnamefont {{Planck
  Collaboration}}},\ }\href {https://doi.org/10.1051/0004-6361/201833910}
  {\bibfield  {journal} {\bibinfo  {journal} {\aap}\ }\textbf {\bibinfo
  {volume} {641}},\ \bibinfo {eid} {A6} (\bibinfo {year} {2020})},\ \Eprint
  {https://arxiv.org/abs/1807.06209} {arXiv:1807.06209 [astro-ph.CO]}
  \BibitemShut {NoStop}%
\bibitem [{\citenamefont {{Riess}}\ \emph {et~al.}(2019)\citenamefont
  {{Riess}}, \citenamefont {{Casertano}}, \citenamefont {{Yuan}}, \citenamefont
  {{Macri}},\ and\ \citenamefont {{Scolnic}}}]{Riess2019}%
  \BibitemOpen
  \bibfield  {author} {\bibinfo {author} {\bibfnamefont {A.~G.}\ \bibnamefont
  {{Riess}}}, \bibinfo {author} {\bibfnamefont {S.}~\bibnamefont
  {{Casertano}}}, \bibinfo {author} {\bibfnamefont {W.}~\bibnamefont {{Yuan}}},
  \bibinfo {author} {\bibfnamefont {L.~M.}\ \bibnamefont {{Macri}}},\ and\
  \bibinfo {author} {\bibfnamefont {D.}~\bibnamefont {{Scolnic}}},\ }\href
  {https://doi.org/10.3847/1538-4357/ab1422} {\bibfield  {journal} {\bibinfo
  {journal} {\apj}\ }\textbf {\bibinfo {volume} {876}},\ \bibinfo {eid} {85}
  (\bibinfo {year} {2019})},\ \Eprint {https://arxiv.org/abs/1903.07603}
  {arXiv:1903.07603 [astro-ph.CO]} \BibitemShut {NoStop}%
\bibitem [{\citenamefont {{Zhang}}\ \emph {et~al.}(2019)\citenamefont
  {{Zhang}}, \citenamefont {{Zhang}}, \citenamefont {{Jin}}, \citenamefont
  {{Qi}},\ and\ \citenamefont {{Zhang}}}]{Zhang2019}%
  \BibitemOpen
  \bibfield  {author} {\bibinfo {author} {\bibfnamefont {J.-F.}\ \bibnamefont
  {{Zhang}}}, \bibinfo {author} {\bibfnamefont {M.}~\bibnamefont {{Zhang}}},
  \bibinfo {author} {\bibfnamefont {S.-J.}\ \bibnamefont {{Jin}}}, \bibinfo
  {author} {\bibfnamefont {J.-Z.}\ \bibnamefont {{Qi}}},\ and\ \bibinfo
  {author} {\bibfnamefont {X.}~\bibnamefont {{Zhang}}},\ }\href
  {https://doi.org/10.1088/1475-7516/2019/09/068} {\bibfield  {journal}
  {\bibinfo  {journal} {\jcap}\ }\textbf {\bibinfo {volume} {2019}},\ \bibinfo
  {eid} {068} (\bibinfo {year} {2019})},\ \Eprint
  {https://arxiv.org/abs/1907.03238} {arXiv:1907.03238 [astro-ph.CO]}
  \BibitemShut {NoStop}%
\bibitem [{\citenamefont {{Belgacem}}\ \emph
  {et~al.}(2019{\natexlab{a}})\citenamefont {{Belgacem}}, \citenamefont
  {{Dirian}}, \citenamefont {{Foffa}}, \citenamefont {{Howell}}, \citenamefont
  {{Maggiore}},\ and\ \citenamefont {{Regimbau}}}]{Belgacem:2019tbw}%
  \BibitemOpen
  \bibfield  {author} {\bibinfo {author} {\bibfnamefont {E.}~\bibnamefont
  {{Belgacem}}}, \bibinfo {author} {\bibfnamefont {Y.}~\bibnamefont
  {{Dirian}}}, \bibinfo {author} {\bibfnamefont {S.}~\bibnamefont {{Foffa}}},
  \bibinfo {author} {\bibfnamefont {E.~J.}\ \bibnamefont {{Howell}}}, \bibinfo
  {author} {\bibfnamefont {M.}~\bibnamefont {{Maggiore}}},\ and\ \bibinfo
  {author} {\bibfnamefont {T.}~\bibnamefont {{Regimbau}}},\ }\href
  {https://doi.org/10.1088/1475-7516/2019/08/015} {\bibfield  {journal}
  {\bibinfo  {journal} {\jcap}\ }\textbf {\bibinfo {volume} {2019}},\ \bibinfo
  {eid} {015} (\bibinfo {year} {2019}{\natexlab{a}})},\ \Eprint
  {https://arxiv.org/abs/1907.01487} {arXiv:1907.01487 [astro-ph.CO]}
  \BibitemShut {NoStop}%
\bibitem [{\citenamefont {{Jin}}\ \emph
  {et~al.}(2022{\natexlab{a}})\citenamefont {{Jin}}, \citenamefont {{Zhu}},
  \citenamefont {{Wang}}, \citenamefont {{Li}}, \citenamefont {{Zhang}},\ and\
  \citenamefont {{Zhang}}}]{Jin2022b}%
  \BibitemOpen
  \bibfield  {author} {\bibinfo {author} {\bibfnamefont {S.-J.}\ \bibnamefont
  {{Jin}}}, \bibinfo {author} {\bibfnamefont {R.-Q.}\ \bibnamefont {{Zhu}}},
  \bibinfo {author} {\bibfnamefont {L.-F.}\ \bibnamefont {{Wang}}}, \bibinfo
  {author} {\bibfnamefont {H.-L.}\ \bibnamefont {{Li}}}, \bibinfo {author}
  {\bibfnamefont {J.-F.}\ \bibnamefont {{Zhang}}},\ and\ \bibinfo {author}
  {\bibfnamefont {X.}~\bibnamefont {{Zhang}}},\ }\href@noop {} {\  (\bibinfo
  {year} {2022}{\natexlab{a}})},\ \Eprint {https://arxiv.org/abs/2204.04689}
  {arXiv:2204.04689 [astro-ph.CO]} \BibitemShut {NoStop}%
\bibitem [{\citenamefont {{Yu}}\ \emph {et~al.}(2021)\citenamefont {{Yu}},
  \citenamefont {{Song}}, \citenamefont {{Ai}}, \citenamefont {{Gao}},
  \citenamefont {{Wang}}, \citenamefont {{Wang}}, \citenamefont {{Lu}},
  \citenamefont {{Fang}},\ and\ \citenamefont {{Zhao}}}]{Yu2021}%
  \BibitemOpen
  \bibfield  {author} {\bibinfo {author} {\bibfnamefont {J.}~\bibnamefont
  {{Yu}}}, \bibinfo {author} {\bibfnamefont {H.}~\bibnamefont {{Song}}},
  \bibinfo {author} {\bibfnamefont {S.}~\bibnamefont {{Ai}}}, \bibinfo {author}
  {\bibfnamefont {H.}~\bibnamefont {{Gao}}}, \bibinfo {author} {\bibfnamefont
  {F.}~\bibnamefont {{Wang}}}, \bibinfo {author} {\bibfnamefont
  {Y.}~\bibnamefont {{Wang}}}, \bibinfo {author} {\bibfnamefont
  {Y.}~\bibnamefont {{Lu}}}, \bibinfo {author} {\bibfnamefont {W.}~\bibnamefont
  {{Fang}}},\ and\ \bibinfo {author} {\bibfnamefont {W.}~\bibnamefont
  {{Zhao}}},\ }\href {https://doi.org/10.3847/1538-4357/ac0628} {\bibfield
  {journal} {\bibinfo  {journal} {\apj}\ }\textbf {\bibinfo {volume} {916}},\
  \bibinfo {eid} {54} (\bibinfo {year} {2021})},\ \Eprint
  {https://arxiv.org/abs/2104.12374} {arXiv:2104.12374 [astro-ph.HE]}
  \BibitemShut {NoStop}%
\bibitem [{\citenamefont {{Belgacem}}\ \emph
  {et~al.}(2019{\natexlab{b}})\citenamefont {{Belgacem}} \emph
  {et~al.}}]{Belgacem2019}%
  \BibitemOpen
  \bibfield  {author} {\bibinfo {author} {\bibfnamefont {E.}~\bibnamefont
  {{Belgacem}}} \emph {et~al.},\ }\href
  {https://doi.org/10.1088/1475-7516/2019/07/024} {\bibfield  {journal}
  {\bibinfo  {journal} {\jcap}\ }\textbf {\bibinfo {volume} {2019}},\ \bibinfo
  {eid} {024} (\bibinfo {year} {2019}{\natexlab{b}})},\ \Eprint
  {https://arxiv.org/abs/1906.01593} {arXiv:1906.01593 [astro-ph.CO]}
  \BibitemShut {NoStop}%
\bibitem [{\citenamefont {{Belgacem}}\ \emph
  {et~al.}(2019{\natexlab{c}})\citenamefont {{Belgacem}}, \citenamefont
  {{Dirian}}, \citenamefont {{Finke}}, \citenamefont {{Foffa}},\ and\
  \citenamefont {{Maggiore}}}]{Belgacem2019b}%
  \BibitemOpen
  \bibfield  {author} {\bibinfo {author} {\bibfnamefont {E.}~\bibnamefont
  {{Belgacem}}}, \bibinfo {author} {\bibfnamefont {Y.}~\bibnamefont
  {{Dirian}}}, \bibinfo {author} {\bibfnamefont {A.}~\bibnamefont {{Finke}}},
  \bibinfo {author} {\bibfnamefont {S.}~\bibnamefont {{Foffa}}},\ and\ \bibinfo
  {author} {\bibfnamefont {M.}~\bibnamefont {{Maggiore}}},\ }\href
  {https://doi.org/10.1088/1475-7516/2019/11/022} {\bibfield  {journal}
  {\bibinfo  {journal} {\jcap}\ }\textbf {\bibinfo {volume} {2019}},\ \bibinfo
  {eid} {022} (\bibinfo {year} {2019}{\natexlab{c}})},\ \Eprint
  {https://arxiv.org/abs/1907.02047} {arXiv:1907.02047 [astro-ph.CO]}
  \BibitemShut {NoStop}%
\bibitem [{\citenamefont {{Ferreira}}\ \emph {et~al.}(2022)\citenamefont
  {{Ferreira}}, \citenamefont {{Barreiro}}, \citenamefont {{Mimoso}},\ and\
  \citenamefont {{Nunes}}}]{Ferreira2022}%
  \BibitemOpen
  \bibfield  {author} {\bibinfo {author} {\bibfnamefont {J.}~\bibnamefont
  {{Ferreira}}}, \bibinfo {author} {\bibfnamefont {T.}~\bibnamefont
  {{Barreiro}}}, \bibinfo {author} {\bibfnamefont {J.}~\bibnamefont
  {{Mimoso}}},\ and\ \bibinfo {author} {\bibfnamefont {N.~J.}\ \bibnamefont
  {{Nunes}}},\ }\href {https://doi.org/10.1103/PhysRevD.105.123531} {\bibfield
  {journal} {\bibinfo  {journal} {\prd}\ }\textbf {\bibinfo {volume} {105}},\
  \bibinfo {eid} {123531} (\bibinfo {year} {2022})},\ \Eprint
  {https://arxiv.org/abs/2203.13788} {arXiv:2203.13788 [astro-ph.CO]}
  \BibitemShut {NoStop}%
\bibitem [{\citenamefont {{Bachega}}\ \emph {et~al.}(2020)\citenamefont
  {{Bachega}}, \citenamefont {{Costa}}, \citenamefont {{Abdalla}},\ and\
  \citenamefont {{Fornazier}}}]{Bachega2020}%
  \BibitemOpen
  \bibfield  {author} {\bibinfo {author} {\bibfnamefont {R.~R.~A.}\
  \bibnamefont {{Bachega}}}, \bibinfo {author} {\bibfnamefont {A.~A.}\
  \bibnamefont {{Costa}}}, \bibinfo {author} {\bibfnamefont {E.}~\bibnamefont
  {{Abdalla}}},\ and\ \bibinfo {author} {\bibfnamefont {K.~S.~F.}\ \bibnamefont
  {{Fornazier}}},\ }\href {https://doi.org/10.1088/1475-7516/2020/05/021}
  {\bibfield  {journal} {\bibinfo  {journal} {\jcap}\ }\textbf {\bibinfo
  {volume} {2020}},\ \bibinfo {eid} {021} (\bibinfo {year} {2020})},\ \Eprint
  {https://arxiv.org/abs/1906.08909} {arXiv:1906.08909 [astro-ph.CO]}
  \BibitemShut {NoStop}%
\bibitem [{\citenamefont {{Caprini}}\ and\ \citenamefont
  {{Tamanini}}(2016)}]{Caprini2016}%
  \BibitemOpen
  \bibfield  {author} {\bibinfo {author} {\bibfnamefont {C.}~\bibnamefont
  {{Caprini}}}\ and\ \bibinfo {author} {\bibfnamefont {N.}~\bibnamefont
  {{Tamanini}}},\ }\href {https://doi.org/10.1088/1475-7516/2016/10/006}
  {\bibfield  {journal} {\bibinfo  {journal} {\jcap}\ }\textbf {\bibinfo
  {volume} {2016}},\ \bibinfo {eid} {006} (\bibinfo {year} {2016})},\ \Eprint
  {https://arxiv.org/abs/1607.08755} {arXiv:1607.08755 [astro-ph.CO]}
  \BibitemShut {NoStop}%
\bibitem [{\citenamefont {{Matos}}\ \emph {et~al.}(2022)\citenamefont
  {{Matos}}, \citenamefont {{Bellini}}, \citenamefont {{Calv{\~a}o}},\ and\
  \citenamefont {{Kunz}}}]{Matos2022}%
  \BibitemOpen
  \bibfield  {author} {\bibinfo {author} {\bibfnamefont {I.~S.}\ \bibnamefont
  {{Matos}}}, \bibinfo {author} {\bibfnamefont {E.}~\bibnamefont {{Bellini}}},
  \bibinfo {author} {\bibfnamefont {M.~O.}\ \bibnamefont {{Calv{\~a}o}}},\ and\
  \bibinfo {author} {\bibfnamefont {M.}~\bibnamefont {{Kunz}}},\ }\href@noop {}
  {\bibfield  {journal} {\bibinfo  {journal} {\jcap}\ } (\bibinfo {year}
  {2022})},\ \Eprint {https://arxiv.org/abs/2210.12174} {arXiv:2210.12174
  [astro-ph.CO]} \BibitemShut {NoStop}%
\bibitem [{\citenamefont {{Mukherjee}}\ \emph
  {et~al.}(2021{\natexlab{a}})\citenamefont {{Mukherjee}}, \citenamefont
  {{Wandelt}},\ and\ \citenamefont {{Silk}}}]{Mukherjee2021MNRAS}%
  \BibitemOpen
  \bibfield  {author} {\bibinfo {author} {\bibfnamefont {S.}~\bibnamefont
  {{Mukherjee}}}, \bibinfo {author} {\bibfnamefont {B.~D.}\ \bibnamefont
  {{Wandelt}}},\ and\ \bibinfo {author} {\bibfnamefont {J.}~\bibnamefont
  {{Silk}}},\ }\href {https://doi.org/10.1093/mnras/stab001} {\bibfield
  {journal} {\bibinfo  {journal} {\mnras}\ }\textbf {\bibinfo {volume} {502}},\
  \bibinfo {pages} {1136} (\bibinfo {year} {2021}{\natexlab{a}})},\ \Eprint
  {https://arxiv.org/abs/2012.15316} {arXiv:2012.15316 [astro-ph.CO]}
  \BibitemShut {NoStop}%
\bibitem [{\citenamefont {{Laghi}}\ \emph {et~al.}(2021)\citenamefont
  {{Laghi}}, \citenamefont {{Tamanini}}, \citenamefont {{Del Pozzo}},
  \citenamefont {{Sesana}}, \citenamefont {{Gair}}, \citenamefont {{Babak}},\
  and\ \citenamefont {{Izquierdo-Villalba}}}]{Laghi2021}%
  \BibitemOpen
  \bibfield  {author} {\bibinfo {author} {\bibfnamefont {D.}~\bibnamefont
  {{Laghi}}}, \bibinfo {author} {\bibfnamefont {N.}~\bibnamefont {{Tamanini}}},
  \bibinfo {author} {\bibfnamefont {W.}~\bibnamefont {{Del Pozzo}}}, \bibinfo
  {author} {\bibfnamefont {A.}~\bibnamefont {{Sesana}}}, \bibinfo {author}
  {\bibfnamefont {J.}~\bibnamefont {{Gair}}}, \bibinfo {author} {\bibfnamefont
  {S.}~\bibnamefont {{Babak}}},\ and\ \bibinfo {author} {\bibfnamefont
  {D.}~\bibnamefont {{Izquierdo-Villalba}}},\ }\href
  {https://doi.org/10.1093/mnras/stab2741} {\bibfield  {journal} {\bibinfo
  {journal} {\mnras}\ }\textbf {\bibinfo {volume} {508}},\ \bibinfo {pages}
  {4512} (\bibinfo {year} {2021})},\ \Eprint {https://arxiv.org/abs/2102.01708}
  {arXiv:2102.01708 [astro-ph.CO]} \BibitemShut {NoStop}%
\bibitem [{\citenamefont {{Muttoni}}\ \emph {et~al.}(2022)\citenamefont
  {{Muttoni}}, \citenamefont {{Mangiagli}}, \citenamefont {{Sesana}},
  \citenamefont {{Laghi}}, \citenamefont {{Del Pozzo}}, \citenamefont
  {{Izquierdo-Villalba}},\ and\ \citenamefont {{Rosati}}}]{Muttoni2022}%
  \BibitemOpen
  \bibfield  {author} {\bibinfo {author} {\bibfnamefont {N.}~\bibnamefont
  {{Muttoni}}}, \bibinfo {author} {\bibfnamefont {A.}~\bibnamefont
  {{Mangiagli}}}, \bibinfo {author} {\bibfnamefont {A.}~\bibnamefont
  {{Sesana}}}, \bibinfo {author} {\bibfnamefont {D.}~\bibnamefont {{Laghi}}},
  \bibinfo {author} {\bibfnamefont {W.}~\bibnamefont {{Del Pozzo}}}, \bibinfo
  {author} {\bibfnamefont {D.}~\bibnamefont {{Izquierdo-Villalba}}},\ and\
  \bibinfo {author} {\bibfnamefont {M.}~\bibnamefont {{Rosati}}},\ }\href
  {https://doi.org/10.1103/PhysRevD.105.043509} {\bibfield  {journal} {\bibinfo
   {journal} {\prd}\ }\textbf {\bibinfo {volume} {105}},\ \bibinfo {eid}
  {043509} (\bibinfo {year} {2022})},\ \Eprint
  {https://arxiv.org/abs/2109.13934} {arXiv:2109.13934 [astro-ph.CO]}
  \BibitemShut {NoStop}%
\bibitem [{\citenamefont {{Zhu}}\ \emph
  {et~al.}(2022{\natexlab{a}})\citenamefont {{Zhu}}, \citenamefont {{Xie}},
  \citenamefont {{Hu}}, \citenamefont {{Liu}}, \citenamefont {{Li}},
  \citenamefont {{Napolitano}}, \citenamefont {{Tang}}, \citenamefont
  {{Zhang}},\ and\ \citenamefont {{Mei}}}]{Zhu2022}%
  \BibitemOpen
  \bibfield  {author} {\bibinfo {author} {\bibfnamefont {L.-G.}\ \bibnamefont
  {{Zhu}}}, \bibinfo {author} {\bibfnamefont {L.-H.}\ \bibnamefont {{Xie}}},
  \bibinfo {author} {\bibfnamefont {Y.-M.}\ \bibnamefont {{Hu}}}, \bibinfo
  {author} {\bibfnamefont {S.}~\bibnamefont {{Liu}}}, \bibinfo {author}
  {\bibfnamefont {E.-K.}\ \bibnamefont {{Li}}}, \bibinfo {author}
  {\bibfnamefont {N.~R.}\ \bibnamefont {{Napolitano}}}, \bibinfo {author}
  {\bibfnamefont {B.-T.}\ \bibnamefont {{Tang}}}, \bibinfo {author}
  {\bibfnamefont {J.-D.}\ \bibnamefont {{Zhang}}},\ and\ \bibinfo {author}
  {\bibfnamefont {J.}~\bibnamefont {{Mei}}},\ }\href
  {https://doi.org/10.1007/s11433-021-1859-9} {\bibfield  {journal} {\bibinfo
  {journal} {Sci. China Phys. Mech.}\ }\textbf {\bibinfo {volume} {65}},\
  \bibinfo {eid} {259811} (\bibinfo {year} {2022}{\natexlab{a}})},\ \Eprint
  {https://arxiv.org/abs/2110.05224} {arXiv:2110.05224 [astro-ph.CO]}
  \BibitemShut {NoStop}%
\bibitem [{\citenamefont {{Zhu}}\ \emph
  {et~al.}(2022{\natexlab{b}})\citenamefont {{Zhu}}, \citenamefont {{Hu}},
  \citenamefont {{Wang}}, \citenamefont {{Zhang}}, \citenamefont {{Li}},
  \citenamefont {{Hendry}},\ and\ \citenamefont {{Mei}}}]{zhu2022b}%
  \BibitemOpen
  \bibfield  {author} {\bibinfo {author} {\bibfnamefont {L.-G.}\ \bibnamefont
  {{Zhu}}}, \bibinfo {author} {\bibfnamefont {Y.-M.}\ \bibnamefont {{Hu}}},
  \bibinfo {author} {\bibfnamefont {H.-T.}\ \bibnamefont {{Wang}}}, \bibinfo
  {author} {\bibfnamefont {J.-d.}\ \bibnamefont {{Zhang}}}, \bibinfo {author}
  {\bibfnamefont {X.-D.}\ \bibnamefont {{Li}}}, \bibinfo {author}
  {\bibfnamefont {M.}~\bibnamefont {{Hendry}}},\ and\ \bibinfo {author}
  {\bibfnamefont {J.}~\bibnamefont {{Mei}}},\ }\href
  {https://doi.org/10.1103/PhysRevResearch.4.013247} {\bibfield  {journal}
  {\bibinfo  {journal} {Phys. Rev. Res.}\ }\textbf {\bibinfo {volume} {4}},\
  \bibinfo {eid} {013247} (\bibinfo {year} {2022}{\natexlab{b}})},\ \Eprint
  {https://arxiv.org/abs/2104.11956} {arXiv:2104.11956 [astro-ph.CO]}
  \BibitemShut {NoStop}%
\bibitem [{\citenamefont {{Leandro}}\ \emph {et~al.}(2022)\citenamefont
  {{Leandro}}, \citenamefont {{Marra}},\ and\ \citenamefont
  {{Sturani}}}]{Sturani2022}%
  \BibitemOpen
  \bibfield  {author} {\bibinfo {author} {\bibfnamefont {H.}~\bibnamefont
  {{Leandro}}}, \bibinfo {author} {\bibfnamefont {V.}~\bibnamefont {{Marra}}},\
  and\ \bibinfo {author} {\bibfnamefont {R.}~\bibnamefont {{Sturani}}},\ }\href
  {https://doi.org/10.1103/PhysRevD.105.023523} {\bibfield  {journal} {\bibinfo
   {journal} {\prd}\ }\textbf {\bibinfo {volume} {105}},\ \bibinfo {eid}
  {023523} (\bibinfo {year} {2022})},\ \Eprint
  {https://arxiv.org/abs/2109.07537} {arXiv:2109.07537 [gr-qc]} \BibitemShut
  {NoStop}%
\bibitem [{\citenamefont {{Ye}}\ and\ \citenamefont
  {{Fishbach}}(2021)}]{Ye:2021klk}%
  \BibitemOpen
  \bibfield  {author} {\bibinfo {author} {\bibfnamefont {C.}~\bibnamefont
  {{Ye}}}\ and\ \bibinfo {author} {\bibfnamefont {M.}~\bibnamefont
  {{Fishbach}}},\ }\href {https://doi.org/10.1103/PhysRevD.104.043507}
  {\bibfield  {journal} {\bibinfo  {journal} {\prd}\ }\textbf {\bibinfo
  {volume} {104}},\ \bibinfo {eid} {043507} (\bibinfo {year} {2021})},\ \Eprint
  {https://arxiv.org/abs/2103.14038} {arXiv:2103.14038 [astro-ph.CO]}
  \BibitemShut {NoStop}%
\bibitem [{\citenamefont {{Ding}}\ \emph {et~al.}(2019)\citenamefont {{Ding}},
  \citenamefont {{Biesiada}}, \citenamefont {{Zheng}}, \citenamefont {{Liao}},
  \citenamefont {{Li}},\ and\ \citenamefont {{Zhu}}}]{Ding:2018zrk}%
  \BibitemOpen
  \bibfield  {author} {\bibinfo {author} {\bibfnamefont {X.}~\bibnamefont
  {{Ding}}}, \bibinfo {author} {\bibfnamefont {M.}~\bibnamefont {{Biesiada}}},
  \bibinfo {author} {\bibfnamefont {X.}~\bibnamefont {{Zheng}}}, \bibinfo
  {author} {\bibfnamefont {K.}~\bibnamefont {{Liao}}}, \bibinfo {author}
  {\bibfnamefont {Z.}~\bibnamefont {{Li}}},\ and\ \bibinfo {author}
  {\bibfnamefont {Z.-H.}\ \bibnamefont {{Zhu}}},\ }\href
  {https://doi.org/10.1088/1475-7516/2019/04/033} {\bibfield  {journal}
  {\bibinfo  {journal} {\jcap}\ }\textbf {\bibinfo {volume} {2019}},\ \bibinfo
  {eid} {033} (\bibinfo {year} {2019})},\ \Eprint
  {https://arxiv.org/abs/1801.05073} {arXiv:1801.05073 [astro-ph.CO]}
  \BibitemShut {NoStop}%
\bibitem [{\citenamefont {{Liu}}\ \emph {et~al.}(2019)\citenamefont {{Liu}},
  \citenamefont {{Li}},\ and\ \citenamefont {{Zhu}}}]{Liu2019}%
  \BibitemOpen
  \bibfield  {author} {\bibinfo {author} {\bibfnamefont {B.}~\bibnamefont
  {{Liu}}}, \bibinfo {author} {\bibfnamefont {Z.}~\bibnamefont {{Li}}},\ and\
  \bibinfo {author} {\bibfnamefont {Z.-H.}\ \bibnamefont {{Zhu}}},\ }\href
  {https://doi.org/10.1093/mnras/stz1179} {\bibfield  {journal} {\bibinfo
  {journal} {\mnras}\ }\textbf {\bibinfo {volume} {487}},\ \bibinfo {pages}
  {1980} (\bibinfo {year} {2019})},\ \Eprint {https://arxiv.org/abs/1904.11751}
  {arXiv:1904.11751 [astro-ph.CO]} \BibitemShut {NoStop}%
\bibitem [{\citenamefont {{Balaudo}}\ \emph {et~al.}(2022)\citenamefont
  {{Balaudo}}, \citenamefont {{Garoffolo}}, \citenamefont {{Martinelli}},
  \citenamefont {{Mukherjee}},\ and\ \citenamefont
  {{Silvestri}}}]{Balaudo2022}%
  \BibitemOpen
  \bibfield  {author} {\bibinfo {author} {\bibfnamefont {A.}~\bibnamefont
  {{Balaudo}}}, \bibinfo {author} {\bibfnamefont {A.}~\bibnamefont
  {{Garoffolo}}}, \bibinfo {author} {\bibfnamefont {M.}~\bibnamefont
  {{Martinelli}}}, \bibinfo {author} {\bibfnamefont {S.}~\bibnamefont
  {{Mukherjee}}},\ and\ \bibinfo {author} {\bibfnamefont {A.}~\bibnamefont
  {{Silvestri}}},\ }\href@noop {} {\  (\bibinfo {year} {2022})},\ \Eprint
  {https://arxiv.org/abs/2210.06398} {arXiv:2210.06398 [astro-ph.CO]}
  \BibitemShut {NoStop}%
\bibitem [{\citenamefont {{Del Pozzo}}\ \emph {et~al.}(2017)\citenamefont {{Del
  Pozzo}}, \citenamefont {{Li}},\ and\ \citenamefont
  {{Messenger}}}]{Delpozzo2017}%
  \BibitemOpen
  \bibfield  {author} {\bibinfo {author} {\bibfnamefont {W.}~\bibnamefont {{Del
  Pozzo}}}, \bibinfo {author} {\bibfnamefont {T.~G.~F.}\ \bibnamefont {{Li}}},\
  and\ \bibinfo {author} {\bibfnamefont {C.}~\bibnamefont {{Messenger}}},\
  }\href {https://doi.org/10.1103/PhysRevD.95.043502} {\bibfield  {journal}
  {\bibinfo  {journal} {\prd}\ }\textbf {\bibinfo {volume} {95}},\ \bibinfo
  {eid} {043502} (\bibinfo {year} {2017})},\ \Eprint
  {https://arxiv.org/abs/1506.06590} {arXiv:1506.06590 [gr-qc]} \BibitemShut
  {NoStop}%
\bibitem [{\citenamefont {{Ghosh}}\ \emph {et~al.}(2022)\citenamefont
  {{Ghosh}}, \citenamefont {{Biswas}},\ and\ \citenamefont
  {{Bose}}}]{ghosh2022}%
  \BibitemOpen
  \bibfield  {author} {\bibinfo {author} {\bibfnamefont {T.}~\bibnamefont
  {{Ghosh}}}, \bibinfo {author} {\bibfnamefont {B.}~\bibnamefont {{Biswas}}},\
  and\ \bibinfo {author} {\bibfnamefont {S.}~\bibnamefont {{Bose}}},\
  }\href@noop {} {\bibfield  {journal} {\bibinfo  {journal} {\prd}\ } (\bibinfo
  {year} {2022})},\ \Eprint {https://arxiv.org/abs/2203.11756}
  {arXiv:2203.11756 [astro-ph.CO]} \BibitemShut {NoStop}%
\bibitem [{\citenamefont {{Jin}}\ \emph
  {et~al.}(2022{\natexlab{b}})\citenamefont {{Jin}}, \citenamefont {{Li}},
  \citenamefont {{Zhang}},\ and\ \citenamefont {{Zhang}}}]{Jin2022a}%
  \BibitemOpen
  \bibfield  {author} {\bibinfo {author} {\bibfnamefont {S.-J.}\ \bibnamefont
  {{Jin}}}, \bibinfo {author} {\bibfnamefont {T.-N.}\ \bibnamefont {{Li}}},
  \bibinfo {author} {\bibfnamefont {J.-F.}\ \bibnamefont {{Zhang}}},\ and\
  \bibinfo {author} {\bibfnamefont {X.}~\bibnamefont {{Zhang}}},\ }\href@noop
  {} {\  (\bibinfo {year} {2022}{\natexlab{b}})},\ \Eprint
  {https://arxiv.org/abs/2202.11882} {arXiv:2202.11882 [gr-qc]} \BibitemShut
  {NoStop}%
\bibitem [{\citenamefont {{de Martino}}\ \emph {et~al.}(2020)\citenamefont {{de
  Martino}}, \citenamefont {{Chakrabarty}}, \citenamefont {{Cesare}},
  \citenamefont {{Gallo}}, \citenamefont {{Ostorero}},\ and\ \citenamefont
  {{Diaferio}}}]{deMartino2020}%
  \BibitemOpen
  \bibfield  {author} {\bibinfo {author} {\bibfnamefont {I.}~\bibnamefont {{de
  Martino}}}, \bibinfo {author} {\bibfnamefont {S.~S.}\ \bibnamefont
  {{Chakrabarty}}}, \bibinfo {author} {\bibfnamefont {V.}~\bibnamefont
  {{Cesare}}}, \bibinfo {author} {\bibfnamefont {A.}~\bibnamefont {{Gallo}}},
  \bibinfo {author} {\bibfnamefont {L.}~\bibnamefont {{Ostorero}}},\ and\
  \bibinfo {author} {\bibfnamefont {A.}~\bibnamefont {{Diaferio}}},\ }\href
  {https://doi.org/10.3390/universe6080107} {\bibfield  {journal} {\bibinfo
  {journal} {Universe}\ }\textbf {\bibinfo {volume} {6}},\ \bibinfo {pages}
  {107} (\bibinfo {year} {2020})},\ \Eprint {https://arxiv.org/abs/2007.15539}
  {arXiv:2007.15539 [astro-ph.CO]} \BibitemShut {NoStop}%
\bibitem [{\citenamefont {{Sotiriou}}\ and\ \citenamefont
  {{Faraoni}}(2010)}]{Sotiriou2010}%
  \BibitemOpen
  \bibfield  {author} {\bibinfo {author} {\bibfnamefont {T.~P.}\ \bibnamefont
  {{Sotiriou}}}\ and\ \bibinfo {author} {\bibfnamefont {V.}~\bibnamefont
  {{Faraoni}}},\ }\href {https://doi.org/10.1103/RevModPhys.82.451} {\bibfield
  {journal} {\bibinfo  {journal} {Rev. Mod. Phys.}\ }\textbf {\bibinfo {volume}
  {82}},\ \bibinfo {pages} {451} (\bibinfo {year} {2010})},\ \Eprint
  {https://arxiv.org/abs/0805.1726} {arXiv:0805.1726 [gr-qc]} \BibitemShut
  {NoStop}%
\bibitem [{\citenamefont {{Kase}}\ and\ \citenamefont
  {{Tsujikawa}}(2019)}]{Kase2019}%
  \BibitemOpen
  \bibfield  {author} {\bibinfo {author} {\bibfnamefont {R.}~\bibnamefont
  {{Kase}}}\ and\ \bibinfo {author} {\bibfnamefont {S.}~\bibnamefont
  {{Tsujikawa}}},\ }\href {https://doi.org/10.1142/S0218271819420057}
  {\bibfield  {journal} {\bibinfo  {journal} {Int. J. Mod. Phys. D}\ }\textbf
  {\bibinfo {volume} {28}},\ \bibinfo {eid} {1942005} (\bibinfo {year}
  {2019})},\ \Eprint {https://arxiv.org/abs/1809.08735} {arXiv:1809.08735
  [gr-qc]} \BibitemShut {NoStop}%
\bibitem [{\citenamefont {{Dam}}\ \emph {et~al.}(2017)\citenamefont {{Dam}},
  \citenamefont {{Heinesen}},\ and\ \citenamefont {{Wiltshire}}}]{Dam2017}%
  \BibitemOpen
  \bibfield  {author} {\bibinfo {author} {\bibfnamefont {L.~H.}\ \bibnamefont
  {{Dam}}}, \bibinfo {author} {\bibfnamefont {A.}~\bibnamefont {{Heinesen}}},\
  and\ \bibinfo {author} {\bibfnamefont {D.~L.}\ \bibnamefont {{Wiltshire}}},\
  }\href {https://doi.org/10.1093/mnras/stx1858} {\bibfield  {journal}
  {\bibinfo  {journal} {\mnras}\ }\textbf {\bibinfo {volume} {472}},\ \bibinfo
  {pages} {835} (\bibinfo {year} {2017})},\ \Eprint
  {https://arxiv.org/abs/1706.07236} {arXiv:1706.07236 [astro-ph.CO]}
  \BibitemShut {NoStop}%
\bibitem [{\citenamefont {{Colin}}\ \emph {et~al.}(2019)\citenamefont
  {{Colin}}, \citenamefont {{Mohayaee}}, \citenamefont {{Rameez}},\ and\
  \citenamefont {{Sarkar}}}]{Colin2019}%
  \BibitemOpen
  \bibfield  {author} {\bibinfo {author} {\bibfnamefont {J.}~\bibnamefont
  {{Colin}}}, \bibinfo {author} {\bibfnamefont {R.}~\bibnamefont {{Mohayaee}}},
  \bibinfo {author} {\bibfnamefont {M.}~\bibnamefont {{Rameez}}},\ and\
  \bibinfo {author} {\bibfnamefont {S.}~\bibnamefont {{Sarkar}}},\ }\href
  {https://doi.org/10.1051/0004-6361/201936373} {\bibfield  {journal} {\bibinfo
   {journal} {\aap}\ }\textbf {\bibinfo {volume} {631}},\ \bibinfo {eid} {L13}
  (\bibinfo {year} {2019})},\ \Eprint {https://arxiv.org/abs/1808.04597}
  {arXiv:1808.04597 [astro-ph.CO]} \BibitemShut {NoStop}%
\bibitem [{\citenamefont {{Pardo}}\ and\ \citenamefont
  {{Spergel}}(2020)}]{Pardo2020}%
  \BibitemOpen
  \bibfield  {author} {\bibinfo {author} {\bibfnamefont {K.}~\bibnamefont
  {{Pardo}}}\ and\ \bibinfo {author} {\bibfnamefont {D.~N.}\ \bibnamefont
  {{Spergel}}},\ }\href {https://doi.org/10.1103/PhysRevLett.125.211101}
  {\bibfield  {journal} {\bibinfo  {journal} {\prl}\ }\textbf {\bibinfo
  {volume} {125}},\ \bibinfo {eid} {211101} (\bibinfo {year} {2020})},\ \Eprint
  {https://arxiv.org/abs/2007.00555} {arXiv:2007.00555 [astro-ph.CO]}
  \BibitemShut {NoStop}%
\bibitem [{\citenamefont {{D'Agostino}}\ and\ \citenamefont
  {{Nunes}}(2019)}]{Dagostino2019}%
  \BibitemOpen
  \bibfield  {author} {\bibinfo {author} {\bibfnamefont {R.}~\bibnamefont
  {{D'Agostino}}}\ and\ \bibinfo {author} {\bibfnamefont {R.~C.}\ \bibnamefont
  {{Nunes}}},\ }\href {https://doi.org/10.1103/PhysRevD.100.044041} {\bibfield
  {journal} {\bibinfo  {journal} {\prd}\ }\textbf {\bibinfo {volume} {100}},\
  \bibinfo {eid} {044041} (\bibinfo {year} {2019})},\ \Eprint
  {https://arxiv.org/abs/1907.05516} {arXiv:1907.05516 [gr-qc]} \BibitemShut
  {NoStop}%
\bibitem [{\citenamefont {Capozziello}\ \emph {et~al.}(2021)\citenamefont
  {Capozziello}, \citenamefont {Dunsby},\ and\ \citenamefont
  {Luongo}}]{Capozziello:2021xjw}%
  \BibitemOpen
  \bibfield  {author} {\bibinfo {author} {\bibfnamefont {S.}~\bibnamefont
  {Capozziello}}, \bibinfo {author} {\bibfnamefont {P.~K.~S.}\ \bibnamefont
  {Dunsby}},\ and\ \bibinfo {author} {\bibfnamefont {O.}~\bibnamefont
  {Luongo}},\ }\href {https://doi.org/10.1093/mnras/stab3187} {\bibfield
  {journal} {\bibinfo  {journal} {Mon. Not. Roy. Astron. Soc.}\ }\textbf
  {\bibinfo {volume} {509}},\ \bibinfo {pages} {5399} (\bibinfo {year}
  {2021})},\ \Eprint {https://arxiv.org/abs/2106.15579} {arXiv:2106.15579
  [astro-ph.CO]} \BibitemShut {NoStop}%
\bibitem [{\citenamefont {Capozziello}\ \emph {et~al.}(2019)\citenamefont
  {Capozziello}, \citenamefont {D'Agostino},\ and\ \citenamefont
  {Luongo}}]{Capozziello:2019cav}%
  \BibitemOpen
  \bibfield  {author} {\bibinfo {author} {\bibfnamefont {S.}~\bibnamefont
  {Capozziello}}, \bibinfo {author} {\bibfnamefont {R.}~\bibnamefont
  {D'Agostino}},\ and\ \bibinfo {author} {\bibfnamefont {O.}~\bibnamefont
  {Luongo}},\ }\href {https://doi.org/10.1142/S0218271819300167} {\bibfield
  {journal} {\bibinfo  {journal} {Int. J. Mod. Phys. D}\ }\textbf {\bibinfo
  {volume} {28}},\ \bibinfo {pages} {1930016} (\bibinfo {year} {2019})},\
  \Eprint {https://arxiv.org/abs/1904.01427} {arXiv:1904.01427 [gr-qc]}
  \BibitemShut {NoStop}%
\bibitem [{\citenamefont {{Chevallier}}\ and\ \citenamefont
  {{Polarski}}(2001)}]{Chevallier2001}%
  \BibitemOpen
  \bibfield  {author} {\bibinfo {author} {\bibfnamefont {M.}~\bibnamefont
  {{Chevallier}}}\ and\ \bibinfo {author} {\bibfnamefont {D.}~\bibnamefont
  {{Polarski}}},\ }\href {https://doi.org/10.1142/S0218271801000822} {\bibfield
   {journal} {\bibinfo  {journal} {Int. J. Mod. Phys. D}\ }\textbf {\bibinfo
  {volume} {10}},\ \bibinfo {pages} {213} (\bibinfo {year} {2001})},\ \Eprint
  {https://arxiv.org/abs/gr-qc/0009008} {arXiv:gr-qc/0009008 [gr-qc]}
  \BibitemShut {NoStop}%
\bibitem [{\citenamefont {{Linder}}(2003)}]{Linder2003}%
  \BibitemOpen
  \bibfield  {author} {\bibinfo {author} {\bibfnamefont {E.~V.}\ \bibnamefont
  {{Linder}}},\ }\href {https://doi.org/10.1103/PhysRevLett.90.091301}
  {\bibfield  {journal} {\bibinfo  {journal} {\prl}\ }\textbf {\bibinfo
  {volume} {90}},\ \bibinfo {eid} {091301} (\bibinfo {year} {2003})},\ \Eprint
  {https://arxiv.org/abs/astro-ph/0208512} {arXiv:astro-ph/0208512 [astro-ph]}
  \BibitemShut {NoStop}%
\bibitem [{\citenamefont {{Copeland}}\ \emph {et~al.}(2006)\citenamefont
  {{Copeland}}, \citenamefont {{Sami}},\ and\ \citenamefont
  {{Tsujikawa}}}]{Copeland2006}%
  \BibitemOpen
  \bibfield  {author} {\bibinfo {author} {\bibfnamefont {E.~J.}\ \bibnamefont
  {{Copeland}}}, \bibinfo {author} {\bibfnamefont {M.}~\bibnamefont {{Sami}}},\
  and\ \bibinfo {author} {\bibfnamefont {S.}~\bibnamefont {{Tsujikawa}}},\
  }\href {https://doi.org/10.1142/S021827180600942X} {\bibfield  {journal}
  {\bibinfo  {journal} {Int. J. Mod. Phys. D}\ }\textbf {\bibinfo {volume}
  {15}},\ \bibinfo {pages} {1753} (\bibinfo {year} {2006})},\ \Eprint
  {https://arxiv.org/abs/hep-th/0603057} {arXiv:hep-th/0603057 [hep-th]}
  \BibitemShut {NoStop}%
\bibitem [{\citenamefont {{Gao}}\ \emph {et~al.}(2021)\citenamefont {{Gao}},
  \citenamefont {{Zhao}}, \citenamefont {{Xue}},\ and\ \citenamefont
  {{Zhang}}}]{Gao:2021}%
  \BibitemOpen
  \bibfield  {author} {\bibinfo {author} {\bibfnamefont {L.-Y.}\ \bibnamefont
  {{Gao}}}, \bibinfo {author} {\bibfnamefont {Z.-W.}\ \bibnamefont {{Zhao}}},
  \bibinfo {author} {\bibfnamefont {S.-S.}\ \bibnamefont {{Xue}}},\ and\
  \bibinfo {author} {\bibfnamefont {X.}~\bibnamefont {{Zhang}}},\ }\href
  {https://doi.org/10.1088/1475-7516/2021/07/005} {\bibfield  {journal}
  {\bibinfo  {journal} {\jcap}\ }\textbf {\bibinfo {volume} {2021}},\ \bibinfo
  {eid} {005} (\bibinfo {year} {2021})},\ \Eprint
  {https://arxiv.org/abs/2101.10714} {arXiv:2101.10714 [astro-ph.CO]}
  \BibitemShut {NoStop}%
\bibitem [{\citenamefont {{Bamba}}\ \emph {et~al.}(2012)\citenamefont
  {{Bamba}}, \citenamefont {{Capozziello}}, \citenamefont {{Nojiri}},\ and\
  \citenamefont {{Odintsov}}}]{Bamba2012}%
  \BibitemOpen
  \bibfield  {author} {\bibinfo {author} {\bibfnamefont {K.}~\bibnamefont
  {{Bamba}}}, \bibinfo {author} {\bibfnamefont {S.}~\bibnamefont
  {{Capozziello}}}, \bibinfo {author} {\bibfnamefont {S.}~\bibnamefont
  {{Nojiri}}},\ and\ \bibinfo {author} {\bibfnamefont {S.~D.}\ \bibnamefont
  {{Odintsov}}},\ }\href {https://doi.org/10.1007/s10509-012-1181-8} {\bibfield
   {journal} {\bibinfo  {journal} {\apss}\ }\textbf {\bibinfo {volume} {342}},\
  \bibinfo {pages} {155} (\bibinfo {year} {2012})},\ \Eprint
  {https://arxiv.org/abs/1205.3421} {arXiv:1205.3421 [gr-qc]} \BibitemShut
  {NoStop}%
\bibitem [{\citenamefont {{V{\"a}liviita}}\ \emph {et~al.}(2008)\citenamefont
  {{V{\"a}liviita}}, \citenamefont {{Majerotto}},\ and\ \citenamefont
  {{Maartens}}}]{Valiviita2008}%
  \BibitemOpen
  \bibfield  {author} {\bibinfo {author} {\bibfnamefont {J.}~\bibnamefont
  {{V{\"a}liviita}}}, \bibinfo {author} {\bibfnamefont {E.}~\bibnamefont
  {{Majerotto}}},\ and\ \bibinfo {author} {\bibfnamefont {R.}~\bibnamefont
  {{Maartens}}},\ }\href {https://doi.org/10.1088/1475-7516/2008/07/020}
  {\bibfield  {journal} {\bibinfo  {journal} {\jcap}\ }\textbf {\bibinfo
  {volume} {2008}},\ \bibinfo {eid} {020} (\bibinfo {year} {2008})},\ \Eprint
  {https://arxiv.org/abs/0804.0232} {arXiv:0804.0232 [astro-ph]} \BibitemShut
  {NoStop}%
\bibitem [{\citenamefont {{Gavela}}\ \emph {et~al.}(2009)\citenamefont
  {{Gavela}}, \citenamefont {{Hern{\'a}ndez}}, \citenamefont {{Lopez Honorez}},
  \citenamefont {{Mena}},\ and\ \citenamefont {{Rigolin}}}]{Gavela2009}%
  \BibitemOpen
  \bibfield  {author} {\bibinfo {author} {\bibfnamefont {M.~B.}\ \bibnamefont
  {{Gavela}}}, \bibinfo {author} {\bibfnamefont {D.}~\bibnamefont
  {{Hern{\'a}ndez}}}, \bibinfo {author} {\bibfnamefont {L.}~\bibnamefont
  {{Lopez Honorez}}}, \bibinfo {author} {\bibfnamefont {O.}~\bibnamefont
  {{Mena}}},\ and\ \bibinfo {author} {\bibfnamefont {S.}~\bibnamefont
  {{Rigolin}}},\ }\href {https://doi.org/10.1088/1475-7516/2009/07/034}
  {\bibfield  {journal} {\bibinfo  {journal} {\jcap}\ }\textbf {\bibinfo
  {volume} {2009}},\ \bibinfo {eid} {034} (\bibinfo {year} {2009})},\ \Eprint
  {https://arxiv.org/abs/0901.1611} {arXiv:0901.1611 [astro-ph.CO]}
  \BibitemShut {NoStop}%
\bibitem [{\citenamefont {{Wang}}\ \emph {et~al.}(2016)\citenamefont {{Wang}},
  \citenamefont {{Abdalla}}, \citenamefont {{Atrio-Barandela}},\ and\
  \citenamefont {{Pav{\'o}n}}}]{Wang2016}%
  \BibitemOpen
  \bibfield  {author} {\bibinfo {author} {\bibfnamefont {B.}~\bibnamefont
  {{Wang}}}, \bibinfo {author} {\bibfnamefont {E.}~\bibnamefont {{Abdalla}}},
  \bibinfo {author} {\bibfnamefont {F.}~\bibnamefont {{Atrio-Barandela}}},\
  and\ \bibinfo {author} {\bibfnamefont {D.}~\bibnamefont {{Pav{\'o}n}}},\
  }\href {https://doi.org/10.1088/0034-4885/79/9/096901} {\bibfield  {journal}
  {\bibinfo  {journal} {Rep. Prog. Phys.}\ }\textbf {\bibinfo {volume} {79}},\
  \bibinfo {eid} {096901} (\bibinfo {year} {2016})},\ \Eprint
  {https://arxiv.org/abs/1603.08299} {arXiv:1603.08299 [astro-ph.CO]}
  \BibitemShut {NoStop}%
\bibitem [{\citenamefont {{Yang}}\ \emph {et~al.}(2019)\citenamefont {{Yang}},
  \citenamefont {{Vagnozzi}}, \citenamefont {{Di Valentino}}, \citenamefont
  {{Nunes}}, \citenamefont {{Pan}},\ and\ \citenamefont {{Mota}}}]{Yang2019}%
  \BibitemOpen
  \bibfield  {author} {\bibinfo {author} {\bibfnamefont {W.}~\bibnamefont
  {{Yang}}}, \bibinfo {author} {\bibfnamefont {S.}~\bibnamefont {{Vagnozzi}}},
  \bibinfo {author} {\bibfnamefont {E.}~\bibnamefont {{Di Valentino}}},
  \bibinfo {author} {\bibfnamefont {R.~C.}\ \bibnamefont {{Nunes}}}, \bibinfo
  {author} {\bibfnamefont {S.}~\bibnamefont {{Pan}}},\ and\ \bibinfo {author}
  {\bibfnamefont {D.~F.}\ \bibnamefont {{Mota}}},\ }\href
  {https://doi.org/10.1088/1475-7516/2019/07/037} {\bibfield  {journal}
  {\bibinfo  {journal} {\jcap}\ }\textbf {\bibinfo {volume} {2019}},\ \bibinfo
  {eid} {037} (\bibinfo {year} {2019})},\ \Eprint
  {https://arxiv.org/abs/1905.08286} {arXiv:1905.08286 [astro-ph.CO]}
  \BibitemShut {NoStop}%
\bibitem [{\citenamefont {{Di Valentino}}\ \emph
  {et~al.}(2020{\natexlab{a}})\citenamefont {{Di Valentino}}, \citenamefont
  {{Melchiorri}}, \citenamefont {{Mena}},\ and\ \citenamefont
  {{Vagnozzi}}}]{DiValentino2020a}%
  \BibitemOpen
  \bibfield  {author} {\bibinfo {author} {\bibfnamefont {E.}~\bibnamefont {{Di
  Valentino}}}, \bibinfo {author} {\bibfnamefont {A.}~\bibnamefont
  {{Melchiorri}}}, \bibinfo {author} {\bibfnamefont {O.}~\bibnamefont
  {{Mena}}},\ and\ \bibinfo {author} {\bibfnamefont {S.}~\bibnamefont
  {{Vagnozzi}}},\ }\href {https://doi.org/10.1016/j.dark.2020.100666}
  {\bibfield  {journal} {\bibinfo  {journal} {Phys. Dark Universe}\ }\textbf
  {\bibinfo {volume} {30}},\ \bibinfo {eid} {100666} (\bibinfo {year}
  {2020}{\natexlab{a}})},\ \Eprint {https://arxiv.org/abs/1908.04281}
  {arXiv:1908.04281 [astro-ph.CO]} \BibitemShut {NoStop}%
\bibitem [{\citenamefont {{Di Valentino}}\ \emph
  {et~al.}(2020{\natexlab{b}})\citenamefont {{Di Valentino}}, \citenamefont
  {{Melchiorri}}, \citenamefont {{Mena}},\ and\ \citenamefont
  {{Vagnozzi}}}]{Divalentino2020b}%
  \BibitemOpen
  \bibfield  {author} {\bibinfo {author} {\bibfnamefont {E.}~\bibnamefont {{Di
  Valentino}}}, \bibinfo {author} {\bibfnamefont {A.}~\bibnamefont
  {{Melchiorri}}}, \bibinfo {author} {\bibfnamefont {O.}~\bibnamefont
  {{Mena}}},\ and\ \bibinfo {author} {\bibfnamefont {S.}~\bibnamefont
  {{Vagnozzi}}},\ }\href {https://doi.org/10.1103/PhysRevD.101.063502}
  {\bibfield  {journal} {\bibinfo  {journal} {\prd}\ }\textbf {\bibinfo
  {volume} {101}},\ \bibinfo {eid} {063502} (\bibinfo {year}
  {2020}{\natexlab{b}})},\ \Eprint {https://arxiv.org/abs/1910.09853}
  {arXiv:1910.09853 [astro-ph.CO]} \BibitemShut {NoStop}%
\bibitem [{\citenamefont {{Pan}}\ \emph {et~al.}(2019)\citenamefont {{Pan}},
  \citenamefont {{Yang}}, \citenamefont {{Di Valentino}}, \citenamefont
  {{Saridakis}},\ and\ \citenamefont {{Chakraborty}}}]{Pan2019}%
  \BibitemOpen
  \bibfield  {author} {\bibinfo {author} {\bibfnamefont {S.}~\bibnamefont
  {{Pan}}}, \bibinfo {author} {\bibfnamefont {W.}~\bibnamefont {{Yang}}},
  \bibinfo {author} {\bibfnamefont {E.}~\bibnamefont {{Di Valentino}}},
  \bibinfo {author} {\bibfnamefont {E.~N.}\ \bibnamefont {{Saridakis}}},\ and\
  \bibinfo {author} {\bibfnamefont {S.}~\bibnamefont {{Chakraborty}}},\ }\href
  {https://doi.org/10.1103/PhysRevD.100.103520} {\bibfield  {journal} {\bibinfo
   {journal} {\prd}\ }\textbf {\bibinfo {volume} {100}},\ \bibinfo {eid}
  {103520} (\bibinfo {year} {2019})},\ \Eprint
  {https://arxiv.org/abs/1907.07540} {arXiv:1907.07540 [astro-ph.CO]}
  \BibitemShut {NoStop}%
\bibitem [{\citenamefont {{Li}}\ and\ \citenamefont
  {{Shafieloo}}(2019)}]{Li2019}%
  \BibitemOpen
  \bibfield  {author} {\bibinfo {author} {\bibfnamefont {X.}~\bibnamefont
  {{Li}}}\ and\ \bibinfo {author} {\bibfnamefont {A.}~\bibnamefont
  {{Shafieloo}}},\ }\href {https://doi.org/10.3847/2041-8213/ab3e09} {\bibfield
   {journal} {\bibinfo  {journal} {\apjl}\ }\textbf {\bibinfo {volume} {883}},\
  \bibinfo {eid} {L3} (\bibinfo {year} {2019})},\ \Eprint
  {https://arxiv.org/abs/1906.08275} {arXiv:1906.08275 [astro-ph.CO]}
  \BibitemShut {NoStop}%
\bibitem [{\citenamefont {{Pan}}\ \emph {et~al.}(2020)\citenamefont {{Pan}},
  \citenamefont {{Yang}}, \citenamefont {{Di Valentino}}, \citenamefont
  {{Shafieloo}},\ and\ \citenamefont {{Chakraborty}}}]{Pan2020}%
  \BibitemOpen
  \bibfield  {author} {\bibinfo {author} {\bibfnamefont {S.}~\bibnamefont
  {{Pan}}}, \bibinfo {author} {\bibfnamefont {W.}~\bibnamefont {{Yang}}},
  \bibinfo {author} {\bibfnamefont {E.}~\bibnamefont {{Di Valentino}}},
  \bibinfo {author} {\bibfnamefont {A.}~\bibnamefont {{Shafieloo}}},\ and\
  \bibinfo {author} {\bibfnamefont {S.}~\bibnamefont {{Chakraborty}}},\ }\href
  {https://doi.org/10.1088/1475-7516/2020/06/062} {\bibfield  {journal}
  {\bibinfo  {journal} {\jcap}\ }\textbf {\bibinfo {volume} {2020}},\ \bibinfo
  {eid} {062} (\bibinfo {year} {2020})},\ \Eprint
  {https://arxiv.org/abs/1907.12551} {arXiv:1907.12551 [astro-ph.CO]}
  \BibitemShut {NoStop}%
\bibitem [{\citenamefont {{Yang}}\ \emph {et~al.}(2020)\citenamefont {{Yang}},
  \citenamefont {{Di Valentino}}, \citenamefont {{Mena}},\ and\ \citenamefont
  {{Pan}}}]{Yang2020}%
  \BibitemOpen
  \bibfield  {author} {\bibinfo {author} {\bibfnamefont {W.}~\bibnamefont
  {{Yang}}}, \bibinfo {author} {\bibfnamefont {E.}~\bibnamefont {{Di
  Valentino}}}, \bibinfo {author} {\bibfnamefont {O.}~\bibnamefont {{Mena}}},\
  and\ \bibinfo {author} {\bibfnamefont {S.}~\bibnamefont {{Pan}}},\ }\href
  {https://doi.org/10.1103/PhysRevD.102.023535} {\bibfield  {journal} {\bibinfo
   {journal} {\prd}\ }\textbf {\bibinfo {volume} {102}},\ \bibinfo {eid}
  {023535} (\bibinfo {year} {2020})},\ \Eprint
  {https://arxiv.org/abs/2003.12552} {arXiv:2003.12552 [astro-ph.CO]}
  \BibitemShut {NoStop}%
\bibitem [{\citenamefont {{Li}}\ and\ \citenamefont
  {{Shafieloo}}(2020)}]{Li:2020}%
  \BibitemOpen
  \bibfield  {author} {\bibinfo {author} {\bibfnamefont {X.}~\bibnamefont
  {{Li}}}\ and\ \bibinfo {author} {\bibfnamefont {A.}~\bibnamefont
  {{Shafieloo}}},\ }\href {https://doi.org/10.3847/1538-4357/abb3d0} {\bibfield
   {journal} {\bibinfo  {journal} {\apj}\ }\textbf {\bibinfo {volume} {902}},\
  \bibinfo {eid} {58} (\bibinfo {year} {2020})},\ \Eprint
  {https://arxiv.org/abs/2001.05103} {arXiv:2001.05103 [astro-ph.CO]}
  \BibitemShut {NoStop}%
\bibitem [{\citenamefont {{Yang}}\ \emph {et~al.}(2021)\citenamefont {{Yang}},
  \citenamefont {{Di Valentino}}, \citenamefont {{Pan}}, \citenamefont
  {{Shafieloo}},\ and\ \citenamefont {{Li}}}]{Yang2021}%
  \BibitemOpen
  \bibfield  {author} {\bibinfo {author} {\bibfnamefont {W.}~\bibnamefont
  {{Yang}}}, \bibinfo {author} {\bibfnamefont {E.}~\bibnamefont {{Di
  Valentino}}}, \bibinfo {author} {\bibfnamefont {S.}~\bibnamefont {{Pan}}},
  \bibinfo {author} {\bibfnamefont {A.}~\bibnamefont {{Shafieloo}}},\ and\
  \bibinfo {author} {\bibfnamefont {X.}~\bibnamefont {{Li}}},\ }\href
  {https://doi.org/10.1103/PhysRevD.104.063521} {\bibfield  {journal} {\bibinfo
   {journal} {\prd}\ }\textbf {\bibinfo {volume} {104}},\ \bibinfo {eid}
  {063521} (\bibinfo {year} {2021})},\ \Eprint
  {https://arxiv.org/abs/2103.03815} {arXiv:2103.03815 [astro-ph.CO]}
  \BibitemShut {NoStop}%
\bibitem [{\citenamefont {Mota}\ \emph {et~al.}(2011)\citenamefont {Mota},
  \citenamefont {Salzano},\ and\ \citenamefont {Capozziello}}]{Mota:2011iw}%
  \BibitemOpen
  \bibfield  {author} {\bibinfo {author} {\bibfnamefont {D.~F.}\ \bibnamefont
  {Mota}}, \bibinfo {author} {\bibfnamefont {V.}~\bibnamefont {Salzano}},\ and\
  \bibinfo {author} {\bibfnamefont {S.}~\bibnamefont {Capozziello}},\ }\href
  {https://doi.org/10.1103/PhysRevD.83.084038} {\bibfield  {journal} {\bibinfo
  {journal} {Phys. Rev. D}\ }\textbf {\bibinfo {volume} {83}},\ \bibinfo
  {pages} {084038} (\bibinfo {year} {2011})},\ \Eprint
  {https://arxiv.org/abs/1103.4215} {arXiv:1103.4215 [astro-ph.CO]}
  \BibitemShut {NoStop}%
\bibitem [{\citenamefont {Weinberg}(1976)}]{Weinberg1976}%
  \BibitemOpen
  \bibfield  {author} {\bibinfo {author} {\bibfnamefont {S.}~\bibnamefont
  {Weinberg}},\ }in\ \href {https://doi.org/10.1007/978-1-4684-0931-4_1} {\emph
  {\bibinfo {booktitle} {{14th International School of Subnuclear Physics:
  Understanding the Fundamental Constitutents of Matter}}}}\ (\bibinfo {year}
  {1976})\BibitemShut {NoStop}%
\bibitem [{\citenamefont {{Weinberg}}(2010)}]{Weinberg2010}%
  \BibitemOpen
  \bibfield  {author} {\bibinfo {author} {\bibfnamefont {S.}~\bibnamefont
  {{Weinberg}}},\ }\href {https://doi.org/10.1103/PhysRevD.81.083535}
  {\bibfield  {journal} {\bibinfo  {journal} {\prd}\ }\textbf {\bibinfo
  {volume} {81}},\ \bibinfo {eid} {083535} (\bibinfo {year} {2010})},\ \Eprint
  {https://arxiv.org/abs/0911.3165} {arXiv:0911.3165 [hep-th]} \BibitemShut
  {NoStop}%
\bibitem [{\citenamefont {{Xue}}(2015)}]{Xue2015}%
  \BibitemOpen
  \bibfield  {author} {\bibinfo {author} {\bibfnamefont {S.-S.}\ \bibnamefont
  {{Xue}}},\ }\href {https://doi.org/10.1016/j.nuclphysb.2015.05.022}
  {\bibfield  {journal} {\bibinfo  {journal} {Nucl. Phys. B}\ }\textbf
  {\bibinfo {volume} {897}},\ \bibinfo {pages} {326} (\bibinfo {year}
  {2015})},\ \Eprint {https://arxiv.org/abs/1410.6152} {arXiv:1410.6152
  [gr-qc]} \BibitemShut {NoStop}%
\bibitem [{\citenamefont {{Califano}}\ \emph {et~al.}(2023)\citenamefont
  {{Califano}}, \citenamefont {{de Martino}}, \citenamefont {{Vernieri}},\ and\
  \citenamefont {{Capozziello}}}]{Califano2022}%
  \BibitemOpen
  \bibfield  {author} {\bibinfo {author} {\bibfnamefont {M.}~\bibnamefont
  {{Califano}}}, \bibinfo {author} {\bibfnamefont {I.}~\bibnamefont {{de
  Martino}}}, \bibinfo {author} {\bibfnamefont {D.}~\bibnamefont
  {{Vernieri}}},\ and\ \bibinfo {author} {\bibfnamefont {S.}~\bibnamefont
  {{Capozziello}}},\ }\href {https://doi.org/10.1093/mnras/stac3230} {\bibfield
   {journal} {\bibinfo  {journal} {\mnras}\ }\textbf {\bibinfo {volume}
  {518}},\ \bibinfo {pages} {3372} (\bibinfo {year} {2023})},\ \Eprint
  {https://arxiv.org/abs/2205.11221} {arXiv:2205.11221 [astro-ph.CO]}
  \BibitemShut {NoStop}%
\bibitem [{\citenamefont {{Regimbau}}\ \emph {et~al.}(2012)\citenamefont
  {{Regimbau}} \emph {et~al.}}]{Regimbau:2012}%
  \BibitemOpen
  \bibfield  {author} {\bibinfo {author} {\bibfnamefont {T.}~\bibnamefont
  {{Regimbau}}} \emph {et~al.},\ }\href
  {https://doi.org/10.1103/PhysRevD.86.122001} {\bibfield  {journal} {\bibinfo
  {journal} {\prd}\ }\textbf {\bibinfo {volume} {86}},\ \bibinfo {eid} {122001}
  (\bibinfo {year} {2012})},\ \Eprint {https://arxiv.org/abs/1201.3563}
  {arXiv:1201.3563 [gr-qc]} \BibitemShut {NoStop}%
\bibitem [{\citenamefont {{Cai}}\ and\ \citenamefont
  {{Yang}}(2017)}]{Cai:2017}%
  \BibitemOpen
  \bibfield  {author} {\bibinfo {author} {\bibfnamefont {R.-G.}\ \bibnamefont
  {{Cai}}}\ and\ \bibinfo {author} {\bibfnamefont {T.}~\bibnamefont {{Yang}}},\
  }\href {https://doi.org/10.1103/PhysRevD.95.044024} {\bibfield  {journal}
  {\bibinfo  {journal} {\prd}\ }\textbf {\bibinfo {volume} {95}},\ \bibinfo
  {eid} {044024} (\bibinfo {year} {2017})},\ \Eprint
  {https://arxiv.org/abs/1608.08008} {arXiv:1608.08008 [astro-ph.CO]}
  \BibitemShut {NoStop}%
\bibitem [{\citenamefont {{Regimbau}}\ and\ \citenamefont
  {{Hughes}}(2009)}]{RegimbauHughes:2009}%
  \BibitemOpen
  \bibfield  {author} {\bibinfo {author} {\bibfnamefont {T.}~\bibnamefont
  {{Regimbau}}}\ and\ \bibinfo {author} {\bibfnamefont {S.~A.}\ \bibnamefont
  {{Hughes}}},\ }\href {https://doi.org/10.1103/PhysRevD.79.062002} {\bibfield
  {journal} {\bibinfo  {journal} {\prd}\ }\textbf {\bibinfo {volume} {79}},\
  \bibinfo {eid} {062002} (\bibinfo {year} {2009})},\ \Eprint
  {https://arxiv.org/abs/0901.2958} {arXiv:0901.2958 [gr-qc]} \BibitemShut
  {NoStop}%
\bibitem [{\citenamefont {{Meacher}}\ \emph {et~al.}(2016)\citenamefont
  {{Meacher}}, \citenamefont {{Cannon}}, \citenamefont {{Hanna}}, \citenamefont
  {{Regimbau}},\ and\ \citenamefont {{Sathyaprakash}}}]{Meacher:2016}%
  \BibitemOpen
  \bibfield  {author} {\bibinfo {author} {\bibfnamefont {D.}~\bibnamefont
  {{Meacher}}}, \bibinfo {author} {\bibfnamefont {K.}~\bibnamefont {{Cannon}}},
  \bibinfo {author} {\bibfnamefont {C.}~\bibnamefont {{Hanna}}}, \bibinfo
  {author} {\bibfnamefont {T.}~\bibnamefont {{Regimbau}}},\ and\ \bibinfo
  {author} {\bibfnamefont {B.~S.}\ \bibnamefont {{Sathyaprakash}}},\ }\href
  {https://doi.org/10.1103/PhysRevD.93.024018} {\bibfield  {journal} {\bibinfo
  {journal} {\prd}\ }\textbf {\bibinfo {volume} {93}},\ \bibinfo {eid} {024018}
  (\bibinfo {year} {2016})},\ \Eprint {https://arxiv.org/abs/1511.01592}
  {arXiv:1511.01592 [gr-qc]} \BibitemShut {NoStop}%
\bibitem [{\citenamefont {{Regimbau}}\ \emph {et~al.}(2017)\citenamefont
  {{Regimbau}}, \citenamefont {{Evans}}, \citenamefont {{Christensen}},
  \citenamefont {{Katsavounidis}}, \citenamefont {{Sathyaprakash}},\ and\
  \citenamefont {{Vitale}}}]{Regimbau:2017}%
  \BibitemOpen
  \bibfield  {author} {\bibinfo {author} {\bibfnamefont {T.}~\bibnamefont
  {{Regimbau}}}, \bibinfo {author} {\bibfnamefont {M.}~\bibnamefont {{Evans}}},
  \bibinfo {author} {\bibfnamefont {N.}~\bibnamefont {{Christensen}}}, \bibinfo
  {author} {\bibfnamefont {E.}~\bibnamefont {{Katsavounidis}}}, \bibinfo
  {author} {\bibfnamefont {B.}~\bibnamefont {{Sathyaprakash}}},\ and\ \bibinfo
  {author} {\bibfnamefont {S.}~\bibnamefont {{Vitale}}},\ }\href
  {https://doi.org/10.1103/PhysRevLett.118.151105} {\bibfield  {journal}
  {\bibinfo  {journal} {\prl}\ }\textbf {\bibinfo {volume} {118}},\ \bibinfo
  {eid} {151105} (\bibinfo {year} {2017})},\ \Eprint
  {https://arxiv.org/abs/1611.08943} {arXiv:1611.08943 [astro-ph.CO]}
  \BibitemShut {NoStop}%
\bibitem [{\citenamefont {{Vangioni}}\ \emph {et~al.}(2015)\citenamefont
  {{Vangioni}}, \citenamefont {{Olive}}, \citenamefont {{Prestegard}},
  \citenamefont {{Silk}}, \citenamefont {{Petitjean}},\ and\ \citenamefont
  {{Mandic}}}]{Vangioni:2014}%
  \BibitemOpen
  \bibfield  {author} {\bibinfo {author} {\bibfnamefont {E.}~\bibnamefont
  {{Vangioni}}}, \bibinfo {author} {\bibfnamefont {K.~A.}\ \bibnamefont
  {{Olive}}}, \bibinfo {author} {\bibfnamefont {T.}~\bibnamefont
  {{Prestegard}}}, \bibinfo {author} {\bibfnamefont {J.}~\bibnamefont
  {{Silk}}}, \bibinfo {author} {\bibfnamefont {P.}~\bibnamefont
  {{Petitjean}}},\ and\ \bibinfo {author} {\bibfnamefont {V.}~\bibnamefont
  {{Mandic}}},\ }\href {https://doi.org/10.1093/mnras/stu2600} {\bibfield
  {journal} {\bibinfo  {journal} {\mnras}\ }\textbf {\bibinfo {volume} {447}},\
  \bibinfo {pages} {2575} (\bibinfo {year} {2015})},\ \Eprint
  {https://arxiv.org/abs/1409.2462} {arXiv:1409.2462 [astro-ph.GA]}
  \BibitemShut {NoStop}%
\bibitem [{\citenamefont {{Tutukov}}\ and\ \citenamefont
  {{Yungelson}}(1994)}]{Tutukov:1994}%
  \BibitemOpen
  \bibfield  {author} {\bibinfo {author} {\bibfnamefont {A.~V.}\ \bibnamefont
  {{Tutukov}}}\ and\ \bibinfo {author} {\bibfnamefont {L.~R.}\ \bibnamefont
  {{Yungelson}}},\ }\href {https://doi.org/10.1093/mnras/268.4.871} {\bibfield
  {journal} {\bibinfo  {journal} {\mnras}\ }\textbf {\bibinfo {volume} {268}},\
  \bibinfo {pages} {871} (\bibinfo {year} {1994})}\BibitemShut {NoStop}%
\bibitem [{\citenamefont {{Lipunov}}\ \emph {et~al.}(1995)\citenamefont
  {{Lipunov}}, \citenamefont {{Postnov}}, \citenamefont {{Prokhorov}},
  \citenamefont {{Panchenko}},\ and\ \citenamefont
  {{Jorgensen}}}]{Lipunov:1995}%
  \BibitemOpen
  \bibfield  {author} {\bibinfo {author} {\bibfnamefont {V.~M.}\ \bibnamefont
  {{Lipunov}}}, \bibinfo {author} {\bibfnamefont {K.~A.}\ \bibnamefont
  {{Postnov}}}, \bibinfo {author} {\bibfnamefont {M.~E.}\ \bibnamefont
  {{Prokhorov}}}, \bibinfo {author} {\bibfnamefont {I.~E.}\ \bibnamefont
  {{Panchenko}}},\ and\ \bibinfo {author} {\bibfnamefont {H.~E.}\ \bibnamefont
  {{Jorgensen}}},\ }\href {https://doi.org/10.1086/176512} {\bibfield
  {journal} {\bibinfo  {journal} {\apj}\ }\textbf {\bibinfo {volume} {454}},\
  \bibinfo {pages} {593} (\bibinfo {year} {1995})},\ \Eprint
  {https://arxiv.org/abs/astro-ph/9504045} {arXiv:astro-ph/9504045 [astro-ph]}
  \BibitemShut {NoStop}%
\bibitem [{\citenamefont {{de Freitas Pacheco}}\ \emph
  {et~al.}(2006)\citenamefont {{de Freitas Pacheco}}, \citenamefont
  {{Regimbau}}, \citenamefont {{Vincent}},\ and\ \citenamefont
  {{Spallicci}}}]{Pacheco:2006}%
  \BibitemOpen
  \bibfield  {author} {\bibinfo {author} {\bibfnamefont {J.~A.}\ \bibnamefont
  {{de Freitas Pacheco}}}, \bibinfo {author} {\bibfnamefont {T.}~\bibnamefont
  {{Regimbau}}}, \bibinfo {author} {\bibfnamefont {S.}~\bibnamefont
  {{Vincent}}},\ and\ \bibinfo {author} {\bibfnamefont {A.}~\bibnamefont
  {{Spallicci}}},\ }\href {https://doi.org/10.1142/S0218271806007699}
  {\bibfield  {journal} {\bibinfo  {journal} {Int. J. of Mod. Phys. D}\
  }\textbf {\bibinfo {volume} {15}},\ \bibinfo {pages} {235} (\bibinfo {year}
  {2006})},\ \Eprint {https://arxiv.org/abs/astro-ph/0510727}
  {arXiv:astro-ph/0510727 [astro-ph]} \BibitemShut {NoStop}%
\bibitem [{\citenamefont {{Belczynski}}\ \emph {et~al.}(2006)\citenamefont
  {{Belczynski}}, \citenamefont {{Perna}}, \citenamefont {{Bulik}},
  \citenamefont {{Kalogera}}, \citenamefont {{Ivanova}},\ and\ \citenamefont
  {{Lamb}}}]{Belczynski:2006}%
  \BibitemOpen
  \bibfield  {author} {\bibinfo {author} {\bibfnamefont {K.}~\bibnamefont
  {{Belczynski}}}, \bibinfo {author} {\bibfnamefont {R.}~\bibnamefont
  {{Perna}}}, \bibinfo {author} {\bibfnamefont {T.}~\bibnamefont {{Bulik}}},
  \bibinfo {author} {\bibfnamefont {V.}~\bibnamefont {{Kalogera}}}, \bibinfo
  {author} {\bibfnamefont {N.}~\bibnamefont {{Ivanova}}},\ and\ \bibinfo
  {author} {\bibfnamefont {D.~Q.}\ \bibnamefont {{Lamb}}},\ }\href
  {https://doi.org/10.1086/505169} {\bibfield  {journal} {\bibinfo  {journal}
  {\apj}\ }\textbf {\bibinfo {volume} {648}},\ \bibinfo {pages} {1110}
  (\bibinfo {year} {2006})},\ \Eprint {https://arxiv.org/abs/astro-ph/0601458}
  {arXiv:astro-ph/0601458 [astro-ph]} \BibitemShut {NoStop}%
\bibitem [{\citenamefont {{O'Shaughnessy}}\ \emph {et~al.}(2008)\citenamefont
  {{O'Shaughnessy}}, \citenamefont {{Belczynski}},\ and\ \citenamefont
  {{Kalogera}}}]{Shaughnessy:2008}%
  \BibitemOpen
  \bibfield  {author} {\bibinfo {author} {\bibfnamefont {R.}~\bibnamefont
  {{O'Shaughnessy}}}, \bibinfo {author} {\bibfnamefont {K.}~\bibnamefont
  {{Belczynski}}},\ and\ \bibinfo {author} {\bibfnamefont {V.}~\bibnamefont
  {{Kalogera}}},\ }\href {https://doi.org/10.1086/526334} {\bibfield  {journal}
  {\bibinfo  {journal} {\apj}\ }\textbf {\bibinfo {volume} {675}},\ \bibinfo
  {pages} {566} (\bibinfo {year} {2008})},\ \Eprint
  {https://arxiv.org/abs/0706.4139} {arXiv:0706.4139 [astro-ph]} \BibitemShut
  {NoStop}%
\bibitem [{\citenamefont {{The LIGO Scientific Collaboration}}\ \emph
  {et~al.}(2021{\natexlab{b}})\citenamefont {{The LIGO Scientific
  Collaboration}}, \citenamefont {{the Virgo Collaboration}},\ and\
  \citenamefont {{the KAGRA Collaboration}}}]{LIGO2021:population}%
  \BibitemOpen
  \bibfield  {author} {\bibinfo {author} {\bibnamefont {{The LIGO Scientific
  Collaboration}}}, \bibinfo {author} {\bibnamefont {{the Virgo
  Collaboration}}},\ and\ \bibinfo {author} {\bibnamefont {{the KAGRA
  Collaboration}}},\ }\href@noop {} {\  (\bibinfo {year}
  {2021}{\natexlab{b}})},\ \Eprint {https://arxiv.org/abs/2111.03634}
  {arXiv:2111.03634 [astro-ph.HE]} \BibitemShut {NoStop}%
\bibitem [{\citenamefont {{Finn}}\ and\ \citenamefont
  {{Chernoff}}(1993)}]{FinnChernoff:1993}%
  \BibitemOpen
  \bibfield  {author} {\bibinfo {author} {\bibfnamefont {L.~S.}\ \bibnamefont
  {{Finn}}}\ and\ \bibinfo {author} {\bibfnamefont {D.~F.}\ \bibnamefont
  {{Chernoff}}},\ }\href {https://doi.org/10.1103/PhysRevD.47.2198} {\bibfield
  {journal} {\bibinfo  {journal} {\prd}\ }\textbf {\bibinfo {volume} {47}},\
  \bibinfo {pages} {2198} (\bibinfo {year} {1993})},\ \Eprint
  {https://arxiv.org/abs/gr-qc/9301003} {arXiv:gr-qc/9301003 [gr-qc]}
  \BibitemShut {NoStop}%
\bibitem [{\citenamefont {{Hild}}\ \emph {et~al.}(2011)\citenamefont {{Hild}}
  \emph {et~al.}}]{Sensitivity:2011}%
  \BibitemOpen
  \bibfield  {author} {\bibinfo {author} {\bibfnamefont {S.}~\bibnamefont
  {{Hild}}} \emph {et~al.},\ }\href
  {https://doi.org/10.1088/0264-9381/28/9/094013} {\bibfield  {journal}
  {\bibinfo  {journal} {Classical and Quantum Gravity}\ }\textbf {\bibinfo
  {volume} {28}},\ \bibinfo {eid} {094013} (\bibinfo {year} {2011})},\ \Eprint
  {https://arxiv.org/abs/1012.0908} {arXiv:1012.0908 [gr-qc]} \BibitemShut
  {NoStop}%
\bibitem [{\citenamefont {{Cutler}}\ and\ \citenamefont
  {{Flanagan}}(1994)}]{Cutler1994}%
  \BibitemOpen
  \bibfield  {author} {\bibinfo {author} {\bibfnamefont {C.}~\bibnamefont
  {{Cutler}}}\ and\ \bibinfo {author} {\bibfnamefont {{\'E}.~E.}\ \bibnamefont
  {{Flanagan}}},\ }\href {https://doi.org/10.1103/PhysRevD.49.2658} {\bibfield
  {journal} {\bibinfo  {journal} {\prd}\ }\textbf {\bibinfo {volume} {49}},\
  \bibinfo {pages} {2658} (\bibinfo {year} {1994})},\ \Eprint
  {https://arxiv.org/abs/gr-qc/9402014} {arXiv:gr-qc/9402014 [gr-qc]}
  \BibitemShut {NoStop}%
\bibitem [{\citenamefont {{Dalal}}\ \emph {et~al.}(2006)\citenamefont
  {{Dalal}}, \citenamefont {{Holz}}, \citenamefont {{Hughes}},\ and\
  \citenamefont {{Jain}}}]{Dalal:2006}%
  \BibitemOpen
  \bibfield  {author} {\bibinfo {author} {\bibfnamefont {N.}~\bibnamefont
  {{Dalal}}}, \bibinfo {author} {\bibfnamefont {D.~E.}\ \bibnamefont {{Holz}}},
  \bibinfo {author} {\bibfnamefont {S.~A.}\ \bibnamefont {{Hughes}}},\ and\
  \bibinfo {author} {\bibfnamefont {B.}~\bibnamefont {{Jain}}},\ }\href
  {https://doi.org/10.1103/PhysRevD.74.063006} {\bibfield  {journal} {\bibinfo
  {journal} {\prd}\ }\textbf {\bibinfo {volume} {74}},\ \bibinfo {eid} {063006}
  (\bibinfo {year} {2006})},\ \Eprint {https://arxiv.org/abs/astro-ph/0601275}
  {arXiv:astro-ph/0601275 [astro-ph]} \BibitemShut {NoStop}%
\bibitem [{\citenamefont {{Hirata}}\ \emph {et~al.}(2010)\citenamefont
  {{Hirata}}, \citenamefont {{Holz}},\ and\ \citenamefont
  {{Cutler}}}]{Hirata:2010}%
  \BibitemOpen
  \bibfield  {author} {\bibinfo {author} {\bibfnamefont {C.~M.}\ \bibnamefont
  {{Hirata}}}, \bibinfo {author} {\bibfnamefont {D.~E.}\ \bibnamefont
  {{Holz}}},\ and\ \bibinfo {author} {\bibfnamefont {C.}~\bibnamefont
  {{Cutler}}},\ }\href {https://doi.org/10.1103/PhysRevD.81.124046} {\bibfield
  {journal} {\bibinfo  {journal} {\prd}\ }\textbf {\bibinfo {volume} {81}},\
  \bibinfo {eid} {124046} (\bibinfo {year} {2010})},\ \Eprint
  {https://arxiv.org/abs/1004.3988} {arXiv:1004.3988 [astro-ph.CO]}
  \BibitemShut {NoStop}%
\bibitem [{\citenamefont {{Tamanini}}\ \emph {et~al.}(2016)\citenamefont
  {{Tamanini}}, \citenamefont {{Caprini}}, \citenamefont {{Barausse}},
  \citenamefont {{Sesana}}, \citenamefont {{Klein}},\ and\ \citenamefont
  {{Petiteau}}}]{Tamanini:2016}%
  \BibitemOpen
  \bibfield  {author} {\bibinfo {author} {\bibfnamefont {N.}~\bibnamefont
  {{Tamanini}}}, \bibinfo {author} {\bibfnamefont {C.}~\bibnamefont
  {{Caprini}}}, \bibinfo {author} {\bibfnamefont {E.}~\bibnamefont
  {{Barausse}}}, \bibinfo {author} {\bibfnamefont {A.}~\bibnamefont
  {{Sesana}}}, \bibinfo {author} {\bibfnamefont {A.}~\bibnamefont {{Klein}}},\
  and\ \bibinfo {author} {\bibfnamefont {A.}~\bibnamefont {{Petiteau}}},\
  }\href {https://doi.org/10.1088/1475-7516/2016/04/002} {\bibfield  {journal}
  {\bibinfo  {journal} {\jcap}\ }\textbf {\bibinfo {volume} {2016}},\ \bibinfo
  {eid} {002} (\bibinfo {year} {2016})},\ \Eprint
  {https://arxiv.org/abs/1601.07112} {arXiv:1601.07112 [astro-ph.CO]}
  \BibitemShut {NoStop}%
\bibitem [{\citenamefont {{Speri}}\ \emph {et~al.}(2021)\citenamefont
  {{Speri}}, \citenamefont {{Tamanini}}, \citenamefont {{Caldwell}},
  \citenamefont {{Gair}},\ and\ \citenamefont {{Wang}}}]{Speri:2021}%
  \BibitemOpen
  \bibfield  {author} {\bibinfo {author} {\bibfnamefont {L.}~\bibnamefont
  {{Speri}}}, \bibinfo {author} {\bibfnamefont {N.}~\bibnamefont {{Tamanini}}},
  \bibinfo {author} {\bibfnamefont {R.~R.}\ \bibnamefont {{Caldwell}}},
  \bibinfo {author} {\bibfnamefont {J.~R.}\ \bibnamefont {{Gair}}},\ and\
  \bibinfo {author} {\bibfnamefont {B.}~\bibnamefont {{Wang}}},\ }\href
  {https://doi.org/10.1103/PhysRevD.103.083526} {\bibfield  {journal} {\bibinfo
   {journal} {\prd}\ }\textbf {\bibinfo {volume} {103}},\ \bibinfo {eid}
  {083526} (\bibinfo {year} {2021})},\ \Eprint
  {https://arxiv.org/abs/2010.09049} {arXiv:2010.09049 [astro-ph.CO]}
  \BibitemShut {NoStop}%
\bibitem [{\citenamefont {{Hjorth}}\ \emph {et~al.}(2017)\citenamefont
  {{Hjorth}} \emph {et~al.}}]{Hjorth2017}%
  \BibitemOpen
  \bibfield  {author} {\bibinfo {author} {\bibfnamefont {J.}~\bibnamefont
  {{Hjorth}}} \emph {et~al.},\ }\href
  {https://doi.org/10.3847/2041-8213/aa9110} {\bibfield  {journal} {\bibinfo
  {journal} {\apjl}\ }\textbf {\bibinfo {volume} {848}},\ \bibinfo {eid} {L31}
  (\bibinfo {year} {2017})},\ \Eprint {https://arxiv.org/abs/1710.05856}
  {arXiv:1710.05856 [astro-ph.GA]} \BibitemShut {NoStop}%
\bibitem [{\citenamefont {{Howlett}}\ and\ \citenamefont
  {{Davis}}(2020)}]{Howlett2020}%
  \BibitemOpen
  \bibfield  {author} {\bibinfo {author} {\bibfnamefont {C.}~\bibnamefont
  {{Howlett}}}\ and\ \bibinfo {author} {\bibfnamefont {T.~M.}\ \bibnamefont
  {{Davis}}},\ }\href {https://doi.org/10.1093/mnras/staa049} {\bibfield
  {journal} {\bibinfo  {journal} {\mnras}\ }\textbf {\bibinfo {volume} {492}},\
  \bibinfo {pages} {3803} (\bibinfo {year} {2020})},\ \Eprint
  {https://arxiv.org/abs/1909.00587} {arXiv:1909.00587 [astro-ph.CO]}
  \BibitemShut {NoStop}%
\bibitem [{\citenamefont {{Mukherjee}}\ \emph
  {et~al.}(2021{\natexlab{b}})\citenamefont {{Mukherjee}} \emph
  {et~al.}}]{Mukherjee2021}%
  \BibitemOpen
  \bibfield  {author} {\bibinfo {author} {\bibfnamefont {S.}~\bibnamefont
  {{Mukherjee}}} \emph {et~al.},\ }\href
  {https://doi.org/10.1051/0004-6361/201936724} {\bibfield  {journal} {\bibinfo
   {journal} {\aap}\ }\textbf {\bibinfo {volume} {646}},\ \bibinfo {eid} {A65}
  (\bibinfo {year} {2021}{\natexlab{b}})},\ \Eprint
  {https://arxiv.org/abs/1909.08627} {arXiv:1909.08627 [astro-ph.CO]}
  \BibitemShut {NoStop}%
\bibitem [{\citenamefont {{Kocsis}}\ \emph {et~al.}(2006)\citenamefont
  {{Kocsis}}, \citenamefont {{Frei}}, \citenamefont {{Haiman}},\ and\
  \citenamefont {{Menou}}}]{Kocsis:2006}%
  \BibitemOpen
  \bibfield  {author} {\bibinfo {author} {\bibfnamefont {B.}~\bibnamefont
  {{Kocsis}}}, \bibinfo {author} {\bibfnamefont {Z.}~\bibnamefont {{Frei}}},
  \bibinfo {author} {\bibfnamefont {Z.}~\bibnamefont {{Haiman}}},\ and\
  \bibinfo {author} {\bibfnamefont {K.}~\bibnamefont {{Menou}}},\ }\href
  {https://doi.org/10.1086/498236} {\bibfield  {journal} {\bibinfo  {journal}
  {\apj}\ }\textbf {\bibinfo {volume} {637}},\ \bibinfo {pages} {27} (\bibinfo
  {year} {2006})},\ \Eprint {https://arxiv.org/abs/astro-ph/0505394}
  {arXiv:astro-ph/0505394 [astro-ph]} \BibitemShut {NoStop}%
\bibitem [{\citenamefont {{Cen}}\ and\ \citenamefont
  {{Ostriker}}(2000)}]{Cen2000}%
  \BibitemOpen
  \bibfield  {author} {\bibinfo {author} {\bibfnamefont {R.}~\bibnamefont
  {{Cen}}}\ and\ \bibinfo {author} {\bibfnamefont {J.~P.}\ \bibnamefont
  {{Ostriker}}},\ }\href {https://doi.org/10.1086/309090} {\bibfield  {journal}
  {\bibinfo  {journal} {\apj}\ }\textbf {\bibinfo {volume} {538}},\ \bibinfo
  {pages} {83} (\bibinfo {year} {2000})},\ \Eprint
  {https://arxiv.org/abs/astro-ph/9809370} {arXiv:astro-ph/9809370 [astro-ph]}
  \BibitemShut {NoStop}%
\bibitem [{\citenamefont {{Fu}}\ \emph {et~al.}(2020)\citenamefont {{Fu}},
  \citenamefont {{Yang}}, \citenamefont {{Chen}}, \citenamefont {{Zhou}},\ and\
  \citenamefont {{Chen}}}]{Fu2020}%
  \BibitemOpen
  \bibfield  {author} {\bibinfo {author} {\bibfnamefont {X.}~\bibnamefont
  {{Fu}}}, \bibinfo {author} {\bibfnamefont {J.}~\bibnamefont {{Yang}}},
  \bibinfo {author} {\bibfnamefont {Z.}~\bibnamefont {{Chen}}}, \bibinfo
  {author} {\bibfnamefont {L.}~\bibnamefont {{Zhou}}},\ and\ \bibinfo {author}
  {\bibfnamefont {J.}~\bibnamefont {{Chen}}},\ }\href
  {https://doi.org/10.1140/epjc/s10052-020-08479-6} {\bibfield  {journal}
  {\bibinfo  {journal} {Eur. Phys. J. C}\ }\textbf {\bibinfo {volume} {80}},\
  \bibinfo {eid} {893} (\bibinfo {year} {2020})},\ \Eprint
  {https://arxiv.org/abs/2009.03041} {arXiv:2009.03041 [astro-ph.CO]}
  \BibitemShut {NoStop}%
\bibitem [{\citenamefont {{Yang}}(2021)}]{Yang:2021qge}%
  \BibitemOpen
  \bibfield  {author} {\bibinfo {author} {\bibfnamefont {T.}~\bibnamefont
  {{Yang}}},\ }\href {https://doi.org/10.1088/1475-7516/2021/05/044} {\bibfield
   {journal} {\bibinfo  {journal} {\jcap}\ }\textbf {\bibinfo {volume}
  {2021}},\ \bibinfo {eid} {044} (\bibinfo {year} {2021})},\ \Eprint
  {https://arxiv.org/abs/2103.01923} {arXiv:2103.01923 [astro-ph.CO]}
  \BibitemShut {NoStop}%
\bibitem [{\citenamefont {{Mitra}}\ \emph {et~al.}(2021)\citenamefont
  {{Mitra}}, \citenamefont {{Mifsud}}, \citenamefont {{Mota}},\ and\
  \citenamefont {{Parkinson}}}]{Mitra2021}%
  \BibitemOpen
  \bibfield  {author} {\bibinfo {author} {\bibfnamefont {A.}~\bibnamefont
  {{Mitra}}}, \bibinfo {author} {\bibfnamefont {J.}~\bibnamefont {{Mifsud}}},
  \bibinfo {author} {\bibfnamefont {D.~F.}\ \bibnamefont {{Mota}}},\ and\
  \bibinfo {author} {\bibfnamefont {D.}~\bibnamefont {{Parkinson}}},\ }\href
  {https://doi.org/10.1093/mnras/stab165} {\bibfield  {journal} {\bibinfo
  {journal} {\mnras}\ }\textbf {\bibinfo {volume} {502}},\ \bibinfo {pages}
  {5563} (\bibinfo {year} {2021})},\ \Eprint {https://arxiv.org/abs/2010.00189}
  {arXiv:2010.00189 [astro-ph.CO]} \BibitemShut {NoStop}%
\bibitem [{\citenamefont {{Pian}}(2021)}]{Pian2021}%
  \BibitemOpen
  \bibfield  {author} {\bibinfo {author} {\bibfnamefont {E.}~\bibnamefont
  {{Pian}}},\ }\href {https://doi.org/10.3389/fspas.2020.609460} {\bibfield
  {journal} {\bibinfo  {journal} {Front. Astron. Space Sci.}\ }\textbf
  {\bibinfo {volume} {7}},\ \bibinfo {eid} {108} (\bibinfo {year} {2021})},\
  \Eprint {https://arxiv.org/abs/2009.12255} {arXiv:2009.12255 [astro-ph.HE]}
  \BibitemShut {NoStop}%
\bibitem [{\citenamefont {{Aartsen}}\ \emph {et~al.}(2017)\citenamefont
  {{Aartsen}} \emph {et~al.}}]{IceCube2017}%
  \BibitemOpen
  \bibfield  {author} {\bibinfo {author} {\bibfnamefont {M.~G.}\ \bibnamefont
  {{Aartsen}}} \emph {et~al.},\ }\href
  {https://doi.org/10.1088/1748-0221/12/03/P03012} {\bibfield  {journal}
  {\bibinfo  {journal} {J. of Inst.}\ }\textbf {\bibinfo {volume} {12}},\
  \bibinfo {pages} {P03012} (\bibinfo {year} {2017})},\ \Eprint
  {https://arxiv.org/abs/1612.05093} {arXiv:1612.05093 [astro-ph.IM]}
  \BibitemShut {NoStop}%
\bibitem [{\citenamefont {{Aartsen}}\ \emph {et~al.}(2020)\citenamefont
  {{Aartsen}} \emph {et~al.}}]{Aartsen2020}%
  \BibitemOpen
  \bibfield  {author} {\bibinfo {author} {\bibfnamefont {M.~G.}\ \bibnamefont
  {{Aartsen}}} \emph {et~al.},\ }\href
  {https://doi.org/10.1088/1475-7516/2020/07/042} {\bibfield  {journal}
  {\bibinfo  {journal} {\jcap}\ }\textbf {\bibinfo {volume} {2020}},\ \bibinfo
  {eid} {042} (\bibinfo {year} {2020})},\ \Eprint
  {https://arxiv.org/abs/1911.11809} {arXiv:1911.11809 [astro-ph.HE]}
  \BibitemShut {NoStop}%
\bibitem [{\citenamefont {{Spergel}}\ \emph {et~al.}(2015)\citenamefont
  {{Spergel}} \emph {et~al.}}]{Spergel2015}%
  \BibitemOpen
  \bibfield  {author} {\bibinfo {author} {\bibfnamefont {D.}~\bibnamefont
  {{Spergel}}} \emph {et~al.},\ }\href@noop {} {\  (\bibinfo {year} {2015})},\
  \Eprint {https://arxiv.org/abs/1503.03757} {arXiv:1503.03757 [astro-ph.IM]}
  \BibitemShut {NoStop}%
\bibitem [{\citenamefont {{Hounsell}}\ \emph {et~al.}(2018)\citenamefont
  {{Hounsell}} \emph {et~al.}}]{Hounsell2018}%
  \BibitemOpen
  \bibfield  {author} {\bibinfo {author} {\bibfnamefont {R.}~\bibnamefont
  {{Hounsell}}} \emph {et~al.},\ }\href
  {https://doi.org/10.3847/1538-4357/aac08b} {\bibfield  {journal} {\bibinfo
  {journal} {\apj}\ }\textbf {\bibinfo {volume} {867}},\ \bibinfo {eid} {23}
  (\bibinfo {year} {2018})},\ \Eprint {https://arxiv.org/abs/1702.01747}
  {arXiv:1702.01747 [astro-ph.IM]} \BibitemShut {NoStop}%
\bibitem [{\citenamefont {{Alfradique}}\ \emph {et~al.}(2022)\citenamefont
  {{Alfradique}}, \citenamefont {{Quartin}}, \citenamefont {{Amendola}},
  \citenamefont {{Castro}},\ and\ \citenamefont {{Toubiana}}}]{Alfradique2022}%
  \BibitemOpen
  \bibfield  {author} {\bibinfo {author} {\bibfnamefont {V.}~\bibnamefont
  {{Alfradique}}}, \bibinfo {author} {\bibfnamefont {M.}~\bibnamefont
  {{Quartin}}}, \bibinfo {author} {\bibfnamefont {L.}~\bibnamefont
  {{Amendola}}}, \bibinfo {author} {\bibfnamefont {T.}~\bibnamefont
  {{Castro}}},\ and\ \bibinfo {author} {\bibfnamefont {A.}~\bibnamefont
  {{Toubiana}}},\ }\href@noop {} {\  (\bibinfo {year} {2022})},\ \Eprint
  {https://arxiv.org/abs/2205.14034} {arXiv:2205.14034 [astro-ph.CO]}
  \BibitemShut {NoStop}%
\bibitem [{\citenamefont {{Amati}}\ \emph {et~al.}(2018)\citenamefont {{Amati}}
  \emph {et~al.}}]{THESEUS:2017qvx}%
  \BibitemOpen
  \bibfield  {author} {\bibinfo {author} {\bibfnamefont {L.}~\bibnamefont
  {{Amati}}} \emph {et~al.},\ }\href
  {https://doi.org/10.1016/j.asr.2018.03.010} {\bibfield  {journal} {\bibinfo
  {journal} {Adv. Space Res.}\ }\textbf {\bibinfo {volume} {62}},\ \bibinfo
  {pages} {191} (\bibinfo {year} {2018})},\ \Eprint
  {https://arxiv.org/abs/1710.04638} {arXiv:1710.04638 [astro-ph.IM]}
  \BibitemShut {NoStop}%
\bibitem [{\citenamefont {{Ciolfi}}\ \emph {et~al.}(2021)\citenamefont
  {{Ciolfi}} \emph {et~al.}}]{Ciolfi2021}%
  \BibitemOpen
  \bibfield  {author} {\bibinfo {author} {\bibfnamefont {R.}~\bibnamefont
  {{Ciolfi}}} \emph {et~al.},\ }\href
  {https://doi.org/10.1007/s10686-021-09795-9} {\bibfield  {journal} {\bibinfo
  {journal} {Exp. Astr.}\ }\textbf {\bibinfo {volume} {52}},\ \bibinfo {pages}
  {245} (\bibinfo {year} {2021})},\ \Eprint {https://arxiv.org/abs/2104.09534}
  {arXiv:2104.09534 [astro-ph.IM]} \BibitemShut {NoStop}%
\bibitem [{\citenamefont {{Ghirlanda}}\ \emph {et~al.}(2021)\citenamefont
  {{Ghirlanda}} \emph {et~al.}}]{Ghirlanda2021}%
  \BibitemOpen
  \bibfield  {author} {\bibinfo {author} {\bibfnamefont {G.}~\bibnamefont
  {{Ghirlanda}}} \emph {et~al.},\ }\href
  {https://doi.org/10.1007/s10686-021-09763-3} {\bibfield  {journal} {\bibinfo
  {journal} {Exp. Astr.}\ }\textbf {\bibinfo {volume} {52}},\ \bibinfo {pages}
  {277} (\bibinfo {year} {2021})},\ \Eprint {https://arxiv.org/abs/2104.10448}
  {arXiv:2104.10448 [astro-ph.IM]} \BibitemShut {NoStop}%
\bibitem [{\citenamefont {{Wanderman}}\ and\ \citenamefont
  {{Piran}}(2015)}]{Wanderman:2014eza}%
  \BibitemOpen
  \bibfield  {author} {\bibinfo {author} {\bibfnamefont {D.}~\bibnamefont
  {{Wanderman}}}\ and\ \bibinfo {author} {\bibfnamefont {T.}~\bibnamefont
  {{Piran}}},\ }\href {https://doi.org/10.1093/mnras/stv123} {\bibfield
  {journal} {\bibinfo  {journal} {\mnras}\ }\textbf {\bibinfo {volume} {448}},\
  \bibinfo {pages} {3026} (\bibinfo {year} {2015})},\ \Eprint
  {https://arxiv.org/abs/1405.5878} {arXiv:1405.5878 [astro-ph.HE]}
  \BibitemShut {NoStop}%
\bibitem [{\citenamefont {{Resmi}}\ \emph {et~al.}(2018)\citenamefont {{Resmi}}
  \emph {et~al.}}]{Resmi2018}%
  \BibitemOpen
  \bibfield  {author} {\bibinfo {author} {\bibfnamefont {L.}~\bibnamefont
  {{Resmi}}} \emph {et~al.},\ }\href {https://doi.org/10.3847/1538-4357/aae1a6}
  {\bibfield  {journal} {\bibinfo  {journal} {\apj}\ }\textbf {\bibinfo
  {volume} {867}},\ \bibinfo {eid} {57} (\bibinfo {year} {2018})},\ \Eprint
  {https://arxiv.org/abs/1803.02768} {arXiv:1803.02768 [astro-ph.HE]}
  \BibitemShut {NoStop}%
\bibitem [{\citenamefont {{Howell}}\ \emph {et~al.}(2019)\citenamefont
  {{Howell}}, \citenamefont {{Ackley}}, \citenamefont {{Rowlinson}},\ and\
  \citenamefont {{Coward}}}]{Howell:2019}%
  \BibitemOpen
  \bibfield  {author} {\bibinfo {author} {\bibfnamefont {E.~J.}\ \bibnamefont
  {{Howell}}}, \bibinfo {author} {\bibfnamefont {K.}~\bibnamefont {{Ackley}}},
  \bibinfo {author} {\bibfnamefont {A.}~\bibnamefont {{Rowlinson}}},\ and\
  \bibinfo {author} {\bibfnamefont {D.}~\bibnamefont {{Coward}}},\ }\href
  {https://doi.org/10.1093/mnras/stz455} {\bibfield  {journal} {\bibinfo
  {journal} {\mnras}\ }\textbf {\bibinfo {volume} {485}},\ \bibinfo {pages}
  {1435} (\bibinfo {year} {2019})},\ \Eprint {https://arxiv.org/abs/1811.09168}
  {arXiv:1811.09168 [astro-ph.HE]} \BibitemShut {NoStop}%
\bibitem [{\citenamefont {{Foreman-Mackey}}\ \emph {et~al.}(2013)\citenamefont
  {{Foreman-Mackey}}, \citenamefont {{Hogg}}, \citenamefont {{Lang}},\ and\
  \citenamefont {{Goodman}}}]{emcee}%
  \BibitemOpen
  \bibfield  {author} {\bibinfo {author} {\bibfnamefont {D.}~\bibnamefont
  {{Foreman-Mackey}}}, \bibinfo {author} {\bibfnamefont {D.~W.}\ \bibnamefont
  {{Hogg}}}, \bibinfo {author} {\bibfnamefont {D.}~\bibnamefont {{Lang}}},\
  and\ \bibinfo {author} {\bibfnamefont {J.}~\bibnamefont {{Goodman}}},\ }\href
  {https://doi.org/10.1086/670067} {\bibfield  {journal} {\bibinfo  {journal}
  {Publ. Astron. Soc. Pac.}\ }\textbf {\bibinfo {volume} {125}},\ \bibinfo
  {pages} {306} (\bibinfo {year} {2013})},\ \Eprint
  {https://arxiv.org/abs/1202.3665} {arXiv:1202.3665 [astro-ph.IM]}
  \BibitemShut {NoStop}%
\bibitem [{\citenamefont {{Mandel}}\ \emph {et~al.}(2019)\citenamefont
  {{Mandel}}, \citenamefont {{Farr}},\ and\ \citenamefont
  {{Gair}}}]{Mandel:2018mve}%
  \BibitemOpen
  \bibfield  {author} {\bibinfo {author} {\bibfnamefont {I.}~\bibnamefont
  {{Mandel}}}, \bibinfo {author} {\bibfnamefont {W.~M.}\ \bibnamefont
  {{Farr}}},\ and\ \bibinfo {author} {\bibfnamefont {J.~R.}\ \bibnamefont
  {{Gair}}},\ }\href {https://doi.org/10.1093/mnras/stz896} {\bibfield
  {journal} {\bibinfo  {journal} {\mnras}\ }\textbf {\bibinfo {volume} {486}},\
  \bibinfo {pages} {1086} (\bibinfo {year} {2019})},\ \Eprint
  {https://arxiv.org/abs/1809.02063} {arXiv:1809.02063 [physics.data-an]}
  \BibitemShut {NoStop}%
\bibitem [{\citenamefont {{Del Pozzo}}(2012)}]{DelPozzo:2011vcw}%
  \BibitemOpen
  \bibfield  {author} {\bibinfo {author} {\bibfnamefont {W.}~\bibnamefont {{Del
  Pozzo}}},\ }\href {https://doi.org/10.1103/PhysRevD.86.043011} {\bibfield
  {journal} {\bibinfo  {journal} {\prd}\ }\textbf {\bibinfo {volume} {86}},\
  \bibinfo {eid} {043011} (\bibinfo {year} {2012})},\ \Eprint
  {https://arxiv.org/abs/1108.1317} {arXiv:1108.1317 [astro-ph.CO]}
  \BibitemShut {NoStop}%
\bibitem [{\citenamefont {{Vitale}}\ \emph {et~al.}(2022)\citenamefont
  {{Vitale}}, \citenamefont {{Gerosa}}, \citenamefont {{Farr}},\ and\
  \citenamefont {{Taylor}}}]{Vitale:2020aaz}%
  \BibitemOpen
  \bibfield  {author} {\bibinfo {author} {\bibfnamefont {S.}~\bibnamefont
  {{Vitale}}}, \bibinfo {author} {\bibfnamefont {D.}~\bibnamefont {{Gerosa}}},
  \bibinfo {author} {\bibfnamefont {W.~M.}\ \bibnamefont {{Farr}}},\ and\
  \bibinfo {author} {\bibfnamefont {S.~R.}\ \bibnamefont {{Taylor}}},\ }in\
  \href {https://doi.org/10.1007/978-981-15-4702-7_45-1} {\emph {\bibinfo
  {booktitle} {Handbook of Gravitational Wave Astronomy}}}\ (\bibinfo {year}
  {2022})\ p.~\bibinfo {pages} {45}\BibitemShut {NoStop}%
\bibitem [{\citenamefont {{Poisson}}\ and\ \citenamefont
  {{Will}}(1995)}]{Poisson:1995ef}%
  \BibitemOpen
  \bibfield  {author} {\bibinfo {author} {\bibfnamefont {E.}~\bibnamefont
  {{Poisson}}}\ and\ \bibinfo {author} {\bibfnamefont {C.~M.}\ \bibnamefont
  {{Will}}},\ }\href {https://doi.org/10.1103/PhysRevD.52.848} {\bibfield
  {journal} {\bibinfo  {journal} {\prd}\ }\textbf {\bibinfo {volume} {52}},\
  \bibinfo {pages} {848} (\bibinfo {year} {1995})},\ \Eprint
  {https://arxiv.org/abs/gr-qc/9502040} {arXiv:gr-qc/9502040 [gr-qc]}
  \BibitemShut {NoStop}%
\bibitem [{\citenamefont {{Baird}}\ \emph {et~al.}(2013)\citenamefont
  {{Baird}}, \citenamefont {{Fairhurst}}, \citenamefont {{Hannam}},\ and\
  \citenamefont {{Murphy}}}]{Baird:2012cu}%
  \BibitemOpen
  \bibfield  {author} {\bibinfo {author} {\bibfnamefont {E.}~\bibnamefont
  {{Baird}}}, \bibinfo {author} {\bibfnamefont {S.}~\bibnamefont
  {{Fairhurst}}}, \bibinfo {author} {\bibfnamefont {M.}~\bibnamefont
  {{Hannam}}},\ and\ \bibinfo {author} {\bibfnamefont {P.}~\bibnamefont
  {{Murphy}}},\ }\href {https://doi.org/10.1103/PhysRevD.87.024035} {\bibfield
  {journal} {\bibinfo  {journal} {\prd}\ }\textbf {\bibinfo {volume} {87}},\
  \bibinfo {eid} {024035} (\bibinfo {year} {2013})},\ \Eprint
  {https://arxiv.org/abs/1211.0546} {arXiv:1211.0546 [gr-qc]} \BibitemShut
  {NoStop}%
\bibitem [{\citenamefont {{Bailes}}\ \emph {et~al.}(2021)\citenamefont
  {{Bailes}} \emph {et~al.}}]{Bailes2021}%
  \BibitemOpen
  \bibfield  {author} {\bibinfo {author} {\bibfnamefont {M.}~\bibnamefont
  {{Bailes}}} \emph {et~al.},\ }\href
  {https://doi.org/10.1038/s42254-021-00303-8} {\bibfield  {journal} {\bibinfo
  {journal} {Nat. Rev. Phys.}\ }\textbf {\bibinfo {volume} {3}},\ \bibinfo
  {pages} {344} (\bibinfo {year} {2021})}\BibitemShut {NoStop}%
\bibitem [{\citenamefont {{Arun}}\ \emph {et~al.}(2022)\citenamefont {{Arun}}
  \emph {et~al.}}]{Arun2022}%
  \BibitemOpen
  \bibfield  {author} {\bibinfo {author} {\bibfnamefont {K.~G.}\ \bibnamefont
  {{Arun}}} \emph {et~al.},\ }\href
  {https://doi.org/10.1007/s41114-022-00036-9} {\bibfield  {journal} {\bibinfo
  {journal} {Living Reviews in Relativity}\ }\textbf {\bibinfo {volume} {25}},\
  \bibinfo {eid} {4} (\bibinfo {year} {2022})},\ \Eprint
  {https://arxiv.org/abs/2205.01597} {arXiv:2205.01597 [gr-qc]} \BibitemShut
  {NoStop}%
\bibitem [{\citenamefont {{Zhan}}\ and\ \citenamefont
  {{Tyson}}(2018)}]{Zhan2018RPPh}%
  \BibitemOpen
  \bibfield  {author} {\bibinfo {author} {\bibfnamefont {H.}~\bibnamefont
  {{Zhan}}}\ and\ \bibinfo {author} {\bibfnamefont {J.~A.}\ \bibnamefont
  {{Tyson}}},\ }\href {https://doi.org/10.1088/1361-6633/aab1bd} {\bibfield
  {journal} {\bibinfo  {journal} {Rep. Prog. Phys.}\ }\textbf {\bibinfo
  {volume} {81}},\ \bibinfo {eid} {066901} (\bibinfo {year} {2018})},\ \Eprint
  {https://arxiv.org/abs/1707.06948} {arXiv:1707.06948 [astro-ph.CO]}
  \BibitemShut {NoStop}%
\bibitem [{\citenamefont {{Zhan}}\ \emph {et~al.}(2009)\citenamefont {{Zhan}}
  \emph {et~al.}}]{Zhan2009}%
  \BibitemOpen
  \bibfield  {author} {\bibinfo {author} {\bibfnamefont {H.}~\bibnamefont
  {{Zhan}}} \emph {et~al.},\ }in\ \href@noop {} {\emph {\bibinfo {booktitle}
  {astro2010: The Astronomy and Astrophysics Decadal Survey}}},\ Vol.\ \bibinfo
  {volume} {2010}\ (\bibinfo {year} {2009})\ p.\ \bibinfo {pages} {332},\
  \Eprint {https://arxiv.org/abs/0902.2599} {arXiv:0902.2599 [astro-ph.CO]}
  \BibitemShut {NoStop}%
\bibitem [{\citenamefont {{Square Kilometre Array Cosmology Science Working
  Group}}(2020)}]{ska2020}%
  \BibitemOpen
  \bibfield  {author} {\bibinfo {author} {\bibnamefont {{Square Kilometre Array
  Cosmology Science Working Group}}},\ }\href
  {https://doi.org/10.1017/pasa.2019.51} {\bibfield  {journal} {\bibinfo
  {journal} {\pasa}\ }\textbf {\bibinfo {volume} {37}},\ \bibinfo {eid} {e007}
  (\bibinfo {year} {2020})},\ \Eprint {https://arxiv.org/abs/1811.02743}
  {arXiv:1811.02743 [astro-ph.CO]} \BibitemShut {NoStop}%
\bibitem [{\citenamefont {{Xiao}}\ \emph {et~al.}(2022)\citenamefont {{Xiao}},
  \citenamefont {{Costa}},\ and\ \citenamefont {{Wang}}}]{Xiao2022}%
  \BibitemOpen
  \bibfield  {author} {\bibinfo {author} {\bibfnamefont {L.}~\bibnamefont
  {{Xiao}}}, \bibinfo {author} {\bibfnamefont {A.~A.}\ \bibnamefont
  {{Costa}}},\ and\ \bibinfo {author} {\bibfnamefont {B.}~\bibnamefont
  {{Wang}}},\ }\href {https://doi.org/10.1093/mnras/stab3256} {\bibfield
  {journal} {\bibinfo  {journal} {\mnras}\ }\textbf {\bibinfo {volume} {510}},\
  \bibinfo {pages} {1495} (\bibinfo {year} {2022})},\ \Eprint
  {https://arxiv.org/abs/2103.01796} {arXiv:2103.01796 [astro-ph.CO]}
  \BibitemShut {NoStop}%
\end{thebibliography}%

\end{document}